\documentclass[12pt,preprint]{aastex}
\usepackage{amssymb,amsmath,url}
\newcounter{address}
\newcounter{tableone}
\newcounter{tabletwo}
\usepackage{color,hyperref}
\definecolor{linkcolor}{rgb}{0,0,0.25}
\hypersetup{
  colorlinks=true,        
  linkcolor=linkcolor,    
  citecolor=linkcolor,    
  filecolor=linkcolor,    
  urlcolor=linkcolor      
}
\newcommand{\eg}{e.g.}
\newcommand{\ie}{i.e.}

\newcommand{\etal}{et~al.}
\newcommand{\cf}{cf.}
\newcommand{\Hipparcos}{\textsl{Hipparcos}}

\newcommand{\Tycho}{Tycho}
\newcommand{\gcs}{\textit{Geneva-Copenhagen Survey}}
\newcommand{\gcsabb}{GCS}
\newcommand{\normal}{{\cal N}}

\renewcommand{\vec}[1]{\mathbf{#1}} 

\newcommand{\mm}{\vec{m}}
\newcommand{\vv}{\vec{v}}

\newcommand{\ww}{\vec{w}}

\newcommand{\eex}{\vec{\hat{x}}}
\newcommand{\eey}{\vec{\hat{y}}}
\newcommand{\eez}{\vec{\hat{z}}}

\newcommand{\mmj}{\mm_j}
\newcommand{\mmk}{\mm_k}

\newcommand{\vvi}{\vv_i}

\newcommand{\wwi}{\ww_i}

\newcommand{\ten}[1]{\mathbf{#1}} 

\newcommand{\RR}{\ten{R}}
\renewcommand{\SS}{\ten{S}}

\newcommand{\VV}{\ten{V}}

\newcommand{\RRi}{\RR_i}

\newcommand{\SSi}{\SS_i}

\newcommand{\VVj}{\VV_{\!j}} 
\newcommand{\VVk}{\VV_{\!k}} 

\newcommand{\T}{^{\scriptscriptstyle \top}}   

\newcommand{\pij}{p_{ij}}

\newcommand{\gall}{{\it l}}
\newcommand{\galb}{{\it b}}

\newcommand{\parallax}{\ensuremath{\pi}}

\newcommand{\sectionname}{\S}
\newcommand{\vx}{\ensuremath{v_x}}
\newcommand{\vy}{\ensuremath{v_y}}
\newcommand{\vz}{\ensuremath{v_z}}

\newcommand{\eqnname}{equation}
\newcommand{\Eqnname}{Equation}

\newcommand{\bhr}{BHR}

\newcommand{\mhip}{\ensuremath{\mathrm{M}_{\mathrm{Hip}}}}
\newcommand{\bminusv}{\ensuremath{B-V}}
\newcommand{\nstars}{19,631}
\newcommand{\nstarsms}{15,023}

\newcommand{\ngcsstarsSSP}{7,682}
\newcommand{\ngcsstarscolorcutSSP}{7,577}
\newcommand{\ngcsstars}{9,575}
\newcommand{\ngcscolorcut}{7,577}
\newcommand{\nphotgcscolorcut}{9,330}
\newcommand{\nphotgcscolorcuthighz}{4,593}
\newcommand{\nphotgcscolorcutlowz}{4,737}

\newcommand{\kernelwidth}{\ensuremath{\lambda}}

\newcommand{\fiducialkernelwidth}{0.05}

\newcommand{\ssp}{SSP}
\newcommand{\lsrpijcut}{0.4}

\begin{document}

\title{The velocity distribution of nearby stars from \Hipparcos\ data\\
II. The nature of the low-velocity moving groups}
\author{
Jo~Bovy\altaffilmark{\ref{NYU},\ref{email}} \&
David~W.~Hogg\altaffilmark{\ref{NYU},\ref{MPIA}}
}
\altaffiltext{\theaddress}{\label{NYU}\stepcounter{address}
Center for Cosmology and Particle Physics, Department of Physics, New York 
University, 4 Washington Place, New York, NY 10003, USA}
\altaffiltext{\theaddress}{\stepcounter{address}\label{email}
To whom correspondence should be addressed: \texttt{jo.bovy@nyu.edu}}
\altaffiltext{\theaddress}{\label{MPIA}\refstepcounter{address}
  Max-Planck-Institut f\"ur Astronomie,
  K\"onigstuhl 17, D-69117 Heidelberg, Germany}

\begin{abstract}
The velocity distribution of nearby stars ($\lesssim 100$ pc) contains
many overdensities or ``moving groups'', clumps of comoving stars,
that are inconsistent with the standard assumption of an axisymmetric,
time-independent, and steady-state Galaxy. We study the age and
metallicity properties of the low-velocity moving groups based on the
reconstruction of the local velocity distribution in Paper I of this
series. We perform stringent, conservative hypothesis testing to
establish for each of these moving groups whether it could conceivably
consist of a coeval population of stars. We conclude that they do not:
the moving groups are not trivially associated with their eponymous
open clusters nor with any other inhomogeneous star formation
event. Concerning a possible dynamical origin of the moving groups, we
test whether any of the moving groups has a higher or lower
metallicity than the background population of thin disk stars, as
would generically be the case if the moving groups are associated with
resonances of the bar or spiral structure. We find clear evidence that
the Hyades moving group has higher than average metallicity and weak
evidence that the Sirius moving group has lower than average
metallicity, which could indicate that these two groups are related to
the inner Lindblad resonance of the spiral structure. Further we find
weak evidence that the Hercules moving group has higher than average
metallicity, as would be the case if it is associated with the bar's
outer Lindblad resonance. The Pleiades moving group shows no clear
metallicity anomaly, arguing against a common dynamical origin for the
Hyades and Pleiades groups. Overall, however, the moving groups are
barely distinguishable from the background population of stars,
raising the likelihood that the moving groups are associated with
transient perturbations.
\end{abstract}

\section{Introduction}\label{sec:intro}

Moving groups---clumps of stars in the Solar neighborhood sharing the
same space velocity---have been known for over a century
\citep{Maedler46a,Proctor69a} and their interpretation has touched on
some of the most basic facts about our Galaxy and the Universe. From
the location of the center of the Milky Way \citep{Maedler47a} over
the age and dynamical state of the Universe
\citep{Jeans15a,Jeans35a,Bok46a}, presently, the moving groups are
used to constrain the dynamical properties of the Galactic disk
\citep[\eg,][]{Dehnen01a,Quillen05a}. However, in order to
quantitatively constrain the fundamental properties of the Galaxy
using the presence of structure in the local velocity distribution,
the nature of the moving groups needs to be clarified. At present, the
evidence that the moving groups are not unmixed structure in phase
space consisting of the ghosts of past star-formation events, but are
instead dynamical effects arising from non-axisymmetric components of
the Galaxy's mass distribution, is by and large
circumstantial. Currently, any constraint on Galaxy dynamics arising
from the moving groups' existence or properties is subject to the
large uncertainty as to what the actual origin of the moving groups
is.

The structure of the local velocity distribution has received much
attention during the last century. While the simplest assumption is
that the distribution of velocities is a simple Gaussian distribution
\citep{Schwarzschild07a}, this assumption was untenable in the light
of observations that showed the presence of multiple ``streams'' in
the velocity distribution \citep{kapteyn05a,Eddington10a}. That these
streams are very prominent and make up a large part of the full
distribution is clear from the fact that their existence was so
readily established. Until the \Hipparcos\ mission, the actual
contribution of substructure in the velocity distribution was only
poorly characterized, but the rich \Hipparcos\ data set conclusively
showed that a large fraction of the local velocity distribution is in
the form of clumps \citep{Dehnen98b,Skuljan99a}; a quantitative
analysis shows that about 40\,percent of the stars in the Solar
neighborhood ($\lesssim 100$ pc) is part of a small number of moving
groups \citep{Bovy09a}. The velocity distribution with the moving
groups indicated is shown in \figurename~\ref{fig:veldist}.

The nature and origin of the moving groups has remained elusive all
this time, although considerable effort has been made both
observationally and theoretically to explain and interpret the
existence of the moving groups. For much of the last century the
consensus view was that the moving groups are the remnants of past
star formation events, coeval populations of stars that were once
closely associated in position as well as velocity but that have now
dispersed and spread out over vast regions of space into the loose
associations of stars that still retain a common motion. This view of
a dynamically unrelaxed Galaxy was first expressed by Jeans
\citep{Jeans15a} and its most vociferous proponent during the second
half of the century was Eggen \citep[\eg,][]{Eggen96a}. The Hyades and
Ursa Major moving groups seemed to fit into this framework as
disrupting clusters in a differentially rotating disk
\citep{Bok34a,Bok36a,Bok46a}. The inspection of the properties of
likely Hyades members showed that these followed a similar
color-luminosity relation as the Hyades and Praesepe open clusters
\citep{Eggen58a}, which seemed to vindicate the view of moving groups
as disrupting clusters. This explanation of the moving groups' origins
was contested, however,
\citep[\eg,][]{Breger68a,Wielen71a,Williams71a,Soderblom87a,Boesgaard88a}
and started to fall out of favor by the end of the century as
observational evidence started to appear that moving group members
were a much more varied population of stars than the open clusters
with which they were believed to be associated: \citet{Eggen93a} found
that the Hyades moving group has a different luminosity function than
the Hyades open cluster; \citet{Dehnen98b} found that moving groups
are present in various color subsamples of \Hipparcos\ stars and that
therefore, using color as a proxy for mean age, moving groups contain
stars of a wide range of ages. Nevertheless, the evaporating cluster
narrative still holds sway for (parts of) some moving groups
\citep{Asiain99a}, in particular for the small HR 1614 moving group
\citep{Feltzing00a,DeSilva07a}, which we do not study here because it
does not stand out as a kinematic overdensity in the overall velocity
distribution. In \sectionname\sectionname~\ref{sec:firstlook} and
\ref{sec:ssp} we ask whether the moving groups constitute a
single-burst stellar population.

In the last decade, there have been various indications that the
moving groups might have a dynamical origin. The Hercules moving group
in particular, an overdensity offset from the bulk of the velocity
distribution opposite the direction of Galactic rotation, displays a
wide range of metallicities \citep{Raboud98a,Bensby07a} and consists
mainly of old stars (\citealt{Caloi99a}; see also the earlier work by
\citealt{Blaauw70a}). The Hyades moving groups also seemed to contain
both old and young stars \citep{Chereul01a}, and soon all low-velocity
moving groups---excluding higher velocity features such as the
Arcturus moving group---were suspected to have a dynamical origin
\citep{Famaey05a,Famaey07a,Famaey08a}.

Theoretical considerations and simulations of orbits in
non-axisymmetric potentials such as that corresponding to the Galactic
bar or spiral structure also contributed to the belief that moving
groups might not be evaporating clusters of stars. The observed
pattern of moving groups can be thought of quite naturally as arising
from the bifurcation of orbits near resonances associated with the bar
\citep{Kalnajs91a,Dehnen01a,Fux01a} or steady-state spiral structure
\citep{Quillen05a}. Other simulations have shown that
moving-group-like structures also develop when considering transient
spiral structure \citep{deSimone04a}, recent bar growth
\citep{Minchev09a}, or the combined effect of spiral structure and
spiral arms \citep{Quillen03a,Chakrabarty07a,Antoja09a}. These
dynamical scenarios for the origin of the moving groups are discussed
in more detail in \sectionname~\ref{sec:dynamics}, in which we test
various of these dynamical scenarios.

Most of the non-axisymmetric perturbations that have been proposed to
create the moving groups are associated with stable, long-lived
perturbations, \eg, a long-lived density wave \citep{Lin64a}. However,
several pieces of evidence indicate that spiral structure might be
only short-lived and/or transient: spiral structure gradually heats
the disk \citep{Carlberg85a} such that it eventually becomes stable
against non-axisymmetric perturbations in the absence of a cooling
mechanism \citep{Sellwood84a}; spiral density waves tend to dissipate
within a few galactic revolutions if fresh waves are not continuously
created \citep{Toomre69a}; spiral structure is more common in high
density environments than in the field
\citep{Elmegreen82a,Elmegreen83a} where interactions between galaxies
that could induce transient spiral structure are more common; and
nearby galaxies show strong variations of the pattern speed with
galactocentric radius, which strongly constrains the lifetime of
grand-design spiral structure \citep{Merrifield09a,Meidt09a}. The
velocity distribution inferred from \Hipparcos\ data itself, with its
large amount of substructure, shows that spiral structure does not
operate on a smooth phase-space density and that spiral instabilities
that grow because of features in the phase-space distribution
\citep[\eg,][]{Sellwood89a,Sellwood91a} should therefore be expected
to be present.

One such instability driven by features in the angular-momentum
distribution such as grooves or ridges is the scenario proposed by
\citet{Sellwood91a} \citep[see also][]{Lovelace78a}. In this model for
the growth of spiral modes, an initial narrow groove in the
angular-momentum density grows into a well-defined large-scale spiral
pattern that dies off again after a few galactic rotations (at
corotation, which lies near the groove center). Since stars are
scattered at the inner Lindblad resonance (ILR of the spiral pattern,
an underdensity of stars in energy--angular-momentum space forms at
the Lindblad resonance, which could spur a new cycle of growth of a
spiral instability, albeit with a corotation radius near the ILR of
the previous pattern. Since the corotation radii of subsequent spiral
patterns move steadily inward, this recurrent cycle stops at a
certain point. In \sectionname~\ref{sec:sellwood}, we ask whether any
of the moving groups is a manifestation of this scenario.

Although the \Hipparcos\ data allowed the velocity distribution in the
Solar neighborhood to be studied in detail for the first time using
complete samples of stars, and theoretical work on the origin of the
moving groups has blossomed in recent years, little progress has been
made observationally to elucidate the nature of the moving groups. In
this paper, we use large samples of \Hipparcos\ stars---an order of
magnitude improvement over previous studies---to investigate the
origin of the kinematical substructures seen in
\figurename~\ref{fig:veldist}. We use the reconstruction of the local
velocity distribution from \citet{Bovy09a} to assign moving-group
membership probabilities to stars. We propagate the membership
uncertainty through all of the analyses of the properties of the
moving-group member stars. This avoids all of the biases that result
from making hard cuts on membership in investigations of this kind and
allows us to perform comprehensive tests to establish the origin of
each individual moving group.

Before we continue, it is worth pointing out that OB
associations---spatially localized associations of young stars
\citep[\eg,][]{deZeeuw99a}---are also sometimes referred to as moving
groups. The following does \emph{not} concern these OB associations.

The main parts of this paper are the following. In
\sectionname~\ref{sec:firstlook} we show that the moving groups are
not associated with their eponymous open clusters; in
\sectionname~\ref{sec:ssp}, we extend this result to show that the
moving groups are not associated with any single episode of star
formation; in \sectionname~\ref{sec:dynamics}, we test whether the
moving groups arise because of steady-state dynamical perturbations to
the axisymmetric disk potential; and in
\sectionname~\ref{sec:sellwood}, we look at whether the moving groups
are associated with the recurrent spiral structure scenario of
\citet{Sellwood91a}.

\section{Data}

We use the standard Galactic velocity coordinate system,
with the directions $x$, $y$, and $z$ (and associated unit vectors
$\eex$, $\eey$, and $\eez$) pointing toward the Galactic center, in
the direction of circular orbital motion, and toward the north
Galactic Pole, respectively. Vectors are everywhere taken to be column
vectors. The components of the velocity vector, $\eex\T\vv$,
$\eey\T\vv$, and $\eez\T\vv$, are conventionally referred to as $U$,
$V$, and $W$, respectively, but we will refer to them as \vx, \vy, and
\vz.

\subsection{Sample selection}

We follow the procedure of \citet{Dehnen98a} and \citet{Aumer09a} to
select a magnitude-limited, kinematically unbiased sample of single
main-sequence stars with accurate astrometry from the \Hipparcos\
catalog. We start by determining the magnitude to which the
\Hipparcos\ catalog is complete in 16 $\times$ 16 $\times$ 10 equal
width bins in $\sin \galb$, $\gall$, and color $B_\mathrm{T} -
V_\mathrm{T}$, the latter measured in the \Tycho\ passbands in the
interval (-0.3,1.5), by finding the $V_\mathrm{T}$ magnitude of the
second brightest star that is included in the \Tycho\ catalog
\citep{hog00a,hog00b}, but absent in the \Hipparcos\ catalog. We then
select in each bin all stars from the original \Hipparcos\ catalog
\citep{ESA97a} brighter in $V_\mathrm{T}$ than the limiting magnitude
in that bin. From this sample of stars we select single stars by using
the ``Solution type'' isol$_{n} < 10$ in the new reduction of the
\Hipparcos\ data \citep{vanLeeuwen07a}, stars with accurate astrometry
by selecting stars with relative parallax uncertainties smaller than
10\,percent (using the formal error on the parallax in the new
\Hipparcos\ catalog). Main-sequence stars are selected by using the
color--magnitude cuts from \citet{Aumer09a}
\begin{equation}
\begin{split}
\mhip &< \phantom{1}7.50\times ( \bminusv) -3.75\,,\phantom{11}\qquad \qquad \ \ \ \ \ \ \, \bminusv \leq 0.5\\
\mhip &< 15.33 \times (\bminusv) -7.665\,,\phantom{1}\qquad \quad 0.5\phantom{1} \leq \bminusv \leq 0.8\\
\mhip &< \phantom{1}4.43 \times (\bminusv) +1.055\,,\phantom{1}\qquad \quad 0.8\phantom{1} \leq \bminusv\\
\mhip &> \phantom{1}4.62 \times (\bminusv) +2.383\,,\phantom{1}\qquad \qquad \ \ \ \ \ \ \, \bminusv \leq 0.35\\\label{eq:cmdcuts}
\mhip &> \phantom{1}8.33 \times (\bminusv) +1.0845\,,\qquad \quad 0.35 \leq \bminusv \leq 0.65\\
\mhip &> \phantom{1}3.33 \times (\bminusv) +4.3375\,,\qquad \quad 0.65 \leq \bminusv \leq 1.25\\
\mhip &> \phantom{1}6.50 \times (\bminusv) + 0.375\,,\phantom{1}\qquad \quad 1.25 \leq \bminusv
\end{split}
\end{equation}
where \mhip\ is the absolute magnitude in \Hipparcos' own
passband.

This procedure selects \nstars\ stars from the \Hipparcos\ catalog,
\nstarsms\ of which are main-sequence stars. The color--magnitude
diagram of the full sample of \nstars\ stars is shown in
\figurename~\ref{fig:cmd}; the cuts defining the main-sequence are
also shown in this figure.

We refer the reader to \citet[][hereafter \bhr]{Bovy09a} for a
detailed explanation of how three-dimensional velocities are projected
onto the two-dimensional tangential plane observed by
\Hipparcos---since the \Hipparcos\ mission did not measure radial
velocities, this third velocity component is missing for all of the
stars in the sample---and how the uncertainties given in the
\Hipparcos\ catalog are propagated to the uncertainties in the
tangential velocity components. In what follows $\wwi$ will represent
the observed tangential velocity of star $i$, $\vvi$ its (unobserved)
three-dimensional velocity, $\RRi$ the projection matrix onto the
tangential plane for star $i$---\ie, $\RRi \vvi = \wwi$---and $\SSi$
the two-dimensional observational-uncertainty variance matrix in the
tangential-velocity plane.

\subsection{Probabilistic moving-group membership determination}\label{sec:member}

\bhr\ reconstructed the velocity distribution of nearby stars by
deconvolving the observed tangential velocity distribution of a
kinematically unbiased sample of 11,865 \Hipparcos\ stars. The
deconvolution algorithm \citep{Bovy09b} represents the underlying
velocity distribution as a sum, or mixture, of Gaussian components,
and can properly handle arbitrary uncertainties, including missing
data, provided that there are no significant star--to--star
correlations. These are believed to be insignificant in the most
recent release of the \Hipparcos\ data \citep{vanLeeuwen07a}. Model
selection, most notably the selection of the ``right' number of
components in the mixture, was based on predicting the radial
velocities in the \gcs\ \citep[\gcsabb;][]{Nordstroem04a}. \bhr\ found
that the underlying three-dimensional velocity distribution was best
represented by a ten-component mixture of Gaussians and found the 99
best-fit parameters of this decomposition.

Although in Gaussian-mixture deconvolution the individual components
do not necessarily have any meaningful interpretation---the Gaussians
are simply basis functions of an expansion---many of the Gaussian
components in the best-fit mixture could be identified unambiguously
with peaks in the velocity distribution, most of which correspond to
known moving groups. For the purposes of this paper we will use the
representation of the velocity distribution as a mixture of 10
Gaussian components with parameters given in Table 1 of \bhr; we will
come back to this choice in the discussion in
\sectionname~\ref{sec:discussion}. We will identify the main moving
groups in the velocity distribution with components in Table 1 of
\bhr\ as follows: component 2 corresponds to the NGC 1901 moving
group; component 4 to the Hercules moving group; component 5 to the
Sirius moving group; components 6 and 7 to the Pleiades moving group;
component 8 to the Hyades moving group; and component 10 to the
Arcturus moving group.

We can now probabilistically assign stars to Gaussian components or
moving groups. For each star $i$ we calculate the probability that it
is associated with component $j$ of the Gaussian mixture model for the
local velocity distribution
\begin{equation}\label{eq:pij}
p_{ij} = \frac{\alpha_j\,\normal\left(\wwi | \RRi \mmj, \RRi \VVj
\RRi\T + \SSi\right)}{\sum_k\alpha_k\,\normal\left(\wwi | \RRi \mmk, \RRi \VVk
\RRi\T+\SSi\right)}\,,
\end{equation}
where $\alpha_j,\mmj,\VVj$ are the amplitude, mean, and variance of
the $j$th Gaussian component, which are given in Table 1 of \bhr; see
\citet{Bovy09b} for a derivation of this formula. For the Pleiades
moving group for each star $i$ we add up the probabilities of it being
associated with component 6 or 7, \ie, $p_{i,\mathrm{Pleiades}} =
p_{i6}+p_{i7}$, since two of the components of the mixture are
associated with the Pleiades moving group (see \bhr\ for an extended
discussion of this).

\section{A first look: Are the low-velocity moving groups associated with their eponymous open clusters?}\label{sec:firstlook}

To get a first idea about the properties of the moving groups we can
look at ``probabilistic'' color--magnitude diagrams of the groups,
which will form the basis for everything we do in the remainder of
this paper. Using the probabilities $p_{ij}$ for each star $i$ to be
part of moving group $j$, we can create color--magnitude diagrams for
the different moving groups that are weighted by the probabilities of
each star to be part of that particular moving group. Such
color--magnitude diagrams are shown in \figurename~\ref{fig:groupcmd}
for the six moving groups unambiguously detected by \bhr. In these
color--magnitude diagrams each star is plotted as a dot with the
grayscale of that dot proportional to the probability of the star to
be part of the moving group. For clarity only those stars with $p_{ij}
> 0.1$ are plotted. It is clear from this figure that very few stars
can be associated with the Arcturus moving group. For this reason we
will not discuss the Arcturus moving group in this paper, instead we
will focus on the remaining, low-velocity, moving groups.

The color--magnitude relation of the low-velocity moving groups in
\figurename~\ref{fig:groupcmd} is very broad for each of the moving
groups. Care must be taken, however, in interpreting this fact, since
the effect of parallax uncertainties is not shown in this figure and
the observed scatter in the color--magnitude relation might well have
some contribution from this uncertainty propagated. We will come back
to this question below. More disturbing, therefore, is the systematic
offset between the color--magnitude relation of the moving group and
the isochrone of the open cluster associated with the moving group. No
open cluster is associated with the Hercules moving group, and we will
therefore ignore it for the remainder of this section.

Two isochrones in the BV photometric system
\citep{Marigo08a,Bertelli94a,Maiz06a}\footnote{Retrieved using the Web
interface provided by Leo Girardi at the Astronomical Observatory of
Padua \url{http://stev.oapd.inaf.it/cgi-bin/cmd_2.2}} are plotted for
each of the moving groups corresponding to the proposed ages and
metallicities for the associated open clusters found in the
literature; see the caption of \figurename~\ref{fig:groupcmd} for the
details on each open cluster. It is clear from this figure that the
isochrones of the open clusters do not represent the color--magnitude
relation of their associated moving groups well, although a caveat
remains about the effect of parallax uncertainties and the effect of
low-probability moving groups members, which is hard to gauge from
this figure. To make the comparison between the open clusters' and the
moving groups' age and metallicity more quantitative, we show in
\figurename~\ref{fig:iscluster} a comparison between the observed
parallaxes of the moving group members and the predicted parallaxes
based on the open clusters' isochrone and the observed photometry for
each main-sequence star in the sample; this comparison is similar to
the one performed by \citet{Famaey08a}. That is, using the observed
color of a star and the $M_V$ versus \bminusv\ relation corresponding
to the isochrone of the associated open cluster we predict the
absolute magnitude of the star and convert this to a model parallax
using the observed apparent $V$ magnitude. Conservatively, we do not
consider any star for which we cannot obtain a photometric parallax in
this way, for example because its color is inconsistent with the age
and metallicity of the cluster; if any such stars is a high
probability member of the moving group, this star alone rules out the
open-cluster origin of the moving group. In order to compute the
photometric parallax we use, for each moving group, the first of the
isochrones mentioned in the caption of \figurename~\ref{fig:groupcmd};
the results for the second set of isochrones are very similar to the
ones presented below. In each of the histograms all of the
main-sequence stars of the sample are plotted; their contributions to
each histogram are weighted by their probabilities of being members of
the moving group in question, as calculated in
\sectionname~\ref{sec:member}.

Two different comparisons between the observed parallaxes and the
model parallaxes are shown in \figurename~\ref{fig:iscluster}. The
left histogram shows the distribution of observed and model parallaxes
for each moving group. Although the effect of parallax uncertainties
(typically $\sim\!1$ mas) is not included in this comparison, a clear
offset between the observed and model parallaxes can be seen. For each
moving group, the hypothesis that the moving group originated from the
open cluster systematically underestimates the distance to each
star. This effect is made even more apparent in the right figure of
each of the panels. Shown here is the distribution of the difference
between the observed and the predicted parallaxes, normalized using
the observational uncertainty on the parallax. If the single-burst
stellar population corresponding to the open-cluster explanation for
the moving groups were correct, this histogram should be that of a
Gaussian distribution of mean zero and standard deviation
one. However, it is immediately clear that the distribution is much
broader than this expected Gaussian, and that it is significantly
skewed. This skewness corresponds to the systematic offsets between
model and observations discussed above. As we will argue now, it is
this excessive skewness, rather than the excessive width, of this
distribution that shows that the moving groups cannot be fully
explained as being part of the evaporation of their eponymous open
clusters.

\begin{figure}
\includegraphics{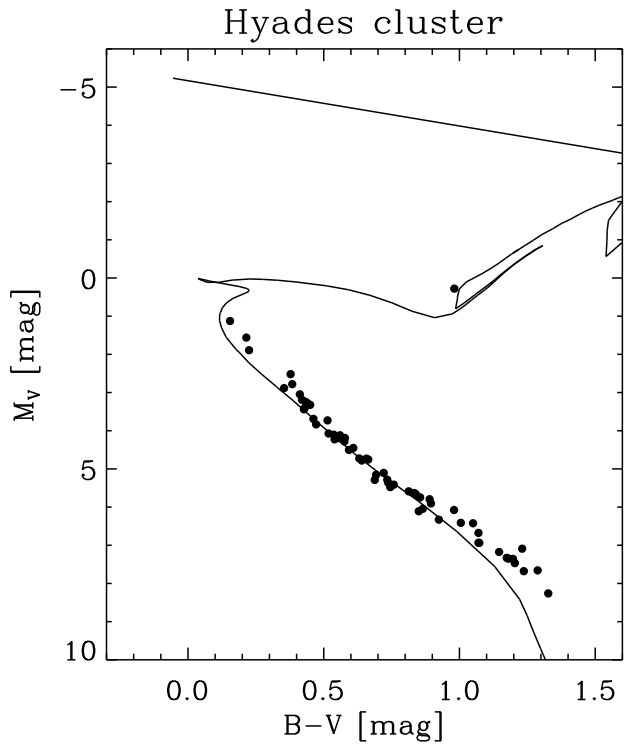}
\caption{Color--magnitude diagram of the Hyades cluster with the 625
  Myr, $Z=0.019$ isochrone overlaid. These members are selected from
  the catalog of Hyades members compiled by \citet{Perryman98a}: we
  selected those stars that have a final membership entry `1', that
  are single, and that lie within 10 pc of the center of the Hyades
  cluster.}\label{fig:hyadesclustercmd}
\end{figure}

The reason why it is dangerous to attach too much significance to the
much larger-than-expected width of the parallax residual distribution
is because even the associated open clusters have a small amount of
scatter in their age and metallicity properties that has not been
taken into account here and this scatter would have to be added to the
variance of the expected Gaussian to see whether the model is a good
fit. This fact is illustrated in \figurename
s~\ref{fig:hyadesclustercmd} and \ref{fig:hyadesclusterhist}. Shown
here are figures similar to \figurename s~\ref{fig:groupcmd} and
\ref{fig:iscluster} but for the Hyades cluster itself. We have taken a
list of probable Hyades cluster members in the \Hipparcos\ catalog
from Table 2 of \citet{Perryman98a}: We only select those stars that
have a final membership entry `1', that are single, and that lie
within 10 pc of the center of the Hyades cluster. This procedure
selects 61 Hyades members. The color--magnitude diagram of these stars
is shown in \figurename~\ref{fig:hyadesclustercmd} with the 625 Myr,
$Z= 0.019$ isochrone overlaid. The Hyades members hug the isochrone
closely, especially in the range 0.4 mag $< \bminusv <$ 0.9 mag. The
correspondence between the isochrone and the members' color and
magnitude becomes less good for redder stars at $\bminusv > 1.0$ mag;
the reason for this is unclear, since the color--magnitude diagrams
with best-fit isochrones do not extend this far redward in
\citet{Perryman98a}, but it might be related to subtle effects in the
calculation of the theoretical isochrones. As expected, there is no
sign of a large, systematic offset between the theoretical isochrone
and the observed color--magnitude relation. Note that most Hyades open
cluster members considered here are high-probability members of the
Hyades moving group: all but 15 of the stars are above the $\pij =
0.4$ threshold that gives an overall level of moving-groups structure
in the velocity distribution comparable to the observed level (see
below); 39 of the stars even have membership probabilities
larger than 0.7. Thus, if all high-probability members of the moving
groups studied in this section were as consistent with a single-burst
stellar population as the Hyades open cluster we would expect to
detect this by our procedure.

\figurename~\ref{fig:hyadesclusterhist} shows for the Hyades cluster
the same histograms as presented in \figurename~\ref{fig:iscluster}
for the moving groups. The observed distribution of parallaxes and the
distribution of model parallaxes for the Hyades members, again
calculated using the Hyades isochrone and the observed photometry for
the stars, are very similar and no systematic difference such as the
one observed for the moving groups in \figurename~\ref{fig:iscluster}
can be seen. This is confirmed in the right panel of
\figurename~\ref{fig:hyadesclusterhist} where the histogram of the
normalized difference between observed and model parallaxes is shown
for the 61 Hyades members. While this distribution peaks at zero,
indicating that there is no systematic bias in the model parallaxes,
the distribution is broader than the expected standard-deviation-one
Gaussian distribution. This indicates that the scatter in age and
metallicity of the Hyades members produces a scatter in the model
parallaxes that should be taken into account in the model comparison
above. The distribution in the right panel also exhibits a small
amount of skewness and heavy-tailed behavior. However, this skewness
is small and simply related to the departure from the isochrone of the
stars redward of $\bminusv = 1.0$ mag; there is no indication of the
smooth skewness and heavy tail seen in
\figurename~\ref{fig:iscluster}. Thus, while the breadth of the model
comparison histograms in the right panels of
\figurename~\ref{fig:iscluster} does not provide a convincing reason
to reject the open cluster origin of the moving groups, the large
amount of skewness and the heavy tails in these distributions does
clearly indicate that the model of the moving groups being evaporating
parts of their eponymous open clusters is not a good fit to the
moving-group photometric properties. We can safely say that the
kinematically identified, low-velocity, moving groups do not appear to
have the same stellar population as the open clusters after which they
are named.

One might worry about the influence of the very low probability
$(p_{ij} < 0.1)$ and/or the red ($\bminusv > 0.9$ mag)---because of
the discrepancy between the model isochrone and the observed absolute
magnitudes of these stars for the Hyades cluster discussed
above---moving group members on the discussion above. We have
therefore repeated the previous analysis leaving out stars from the
sample which satisfy either of these two criteria. The histograms
obtained in this way are barely distinguishable from the distributions
shown in \figurename~\ref{fig:iscluster} and the argument made in the
previous paragraphs continues to hold.

\section{Strategy for the second part of this paper}\label{sec:strategy}

Even though we have shown that the moving groups cannot be considered
to be the evaporating parts of their associated open clusters, the
question still remains whether they can be considered to be some
single-burst stellar population, perhaps originating from an open
cluster that has completely evaporated and thus has no presently
identifiable core. It might be that it is merely a coincidence that
the kinematically defined moving groups' space velocities roughly
coincide with those of prominent open clusters, while the moving
groups are actually remnants of older open clusters that are hard to
identify at the present day. We have also failed to explain the origin
of the Hercules moving group in the previous section because of the
lack of an associated open cluster. Therefore, in what follows we will
test the hypothesis that the low-velocity moving groups each comprise
\emph{some} single-burst stellar population, with an \emph{a priori}
unknown age and metallicity. If the moving groups fail to live up to
this hypothesis, we can confidently say that they are not remnants of
inhomogeneous star formation, but instead most likely have a dynamical
origin.

\begin{figure}
\includegraphics[width=0.5\textwidth]{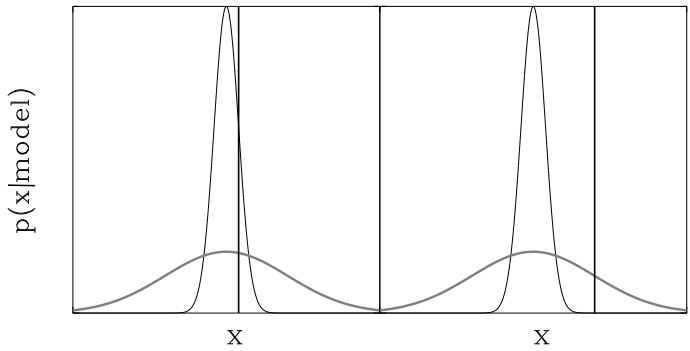}
\caption{Model selection: the $y$-axis represents the probability of
  measuring the value on the $x$-axis for a foreground model
  (\emph{thin, black curve}) and a background model (\emph{thick, gray
    curve}). The foreground model makes very informative predictions
  while the broader background model makes less informative
  predictions. Therefore, when both the foreground model and the
  background model predict the right observed value (\emph{vertical
    line}) the observed value has a larger probability for the
  foreground model (\emph{left panel}); when the foreground model
  fails to predict the observed value, the observed value is more
  probable under the background model \emph{right
    panel}).}\label{fig:model_selection}
\end{figure}

Hypothesis testing or model selection is at its strongest when two
mutually exclusive hypotheses can be pitted against each other as
opposed to merely testing whether a particular hypothesis provides a
good fit to the data. Fortunately, we are in a situation here in which
there is a well-specified background hypothesis: this hypothesis is
simply that the stars in the moving groups are nothing more than a
sparse sampling of the full locally observed disk population, that is,
that there is nothing special about their age and chemical composition
to distinguish them from local disk stars as a whole. We are also
lucky to have a non-parametric model at our disposal for this
background hypothesis: this model is nothing more than the observed
local population of disk stars. Thus, we can test whether the moving
groups' photometric properties are better described by the model in
which each contains just a single-burst stellar population or by the
model in which each contains just the same population as the
background stars. The single-burst stellar population model can make
very tight predictions for the photometric properties of the stars
while the background model can only make very broad statements about
the moving groups' member properties. If the tight predictions play
out, this will lead to clear evidence of the evaporating-cluster
nature of the moving groups because the photometric properties of the
member stars will be much more probable than they are under the
background hypothesis. However, if the single-burst stellar population
hypothesis fails to predict the photometric properties of the moving
group members, then the background model will be preferred. This
conceptual view of the model selection procedure which we will use in
the second part of this paper is illustrated in
\figurename~\ref{fig:model_selection}.

Coupled with an initial mass function, the age and metallicity of a
single-burst stellar population imply a density, or distribution, in
the color--magnitude plane, and testing whether a population of stars
consists of a single stellar population is equivalent to checking
whether the observed distribution of stars in the color--magnitude
plane is consistent with this density. This is a very strict test of
the coeval hypothesis that depends on choosing, or inferring, the
right initial mass function and having complete samples of stars at
one's disposal. A more conservative approach, which does not rely on
these two assumptions, would be to test whether the relation between
color and absolute magnitude predicted for a coeval population of a
certain age and metallicity is observed in the sample. That is, rather
than testing whether the predicted density is observed in the
color--magnitude plane of a sample of stars, we test whether the
predicted regression $M_V(\bminusv)$ is consistent with the data. In
practice, we use the predicted $M_V(\bminusv)$ relation combined with
the observed photometry of a star to predict a photometric parallax
for the star, in exactly the way described on the previous
section. This photometric parallax is then compared to the observed
parallax, taking into account the observational uncertainty on the
parallax.

For the hypothesis that we are interested in testing this is advisable
because mass segregation and selective evaporation have been shown to
affect the luminosity functions of open clusters, both in simulations
\citep{Aarseth72a,Terlevich87a,delafuentemarcos95a,Bonnell98a}---whether
primordial \citep{Bonnell98a,Hillenbrand98a} or dynamical
\citep{McMillan07a,Moeckel09a,Allison09a}---as well as observationally
in some of the open clusters associated with the moving groups studied
here (Hyades:
\citealt{Reid92a,Perryman98a,Reid99a,Dobbie02a,Bouvier08a}; Pleiades:
\citealt{Bouvier98a,Hambly99a,Adams01a}; NGC 1901:
\citealt{Carraro07a}). There is some debate about whether mass
segregation has actually been observed in massive open clusters
\citep{Ascenso09a}. We can expect low-mass stars to be preferentially
ejected from open clusters, although quantitative estimates of this
effect are still highly uncertain. It would be hard to predict the
complete two-dimensional model density in the color--magnitude
plane. However, whether or not selective evaporation plays a large
role in the evolution and evaporation of open clusters, the relation
$M_V(\bminusv)$ should always hold if the sample of stars originated
from a single star-formation event and the model selection test based
on it will not be affected.

The test will hinge on the existence of a background model that states
that the stars in the moving groups are similar to the local disk
population as a whole. In the next section we will refine this
background model and put it in such a form that we can use it
quantitatively in the model selection test. That is, we will turn the
bulk photometric properties of the local disk stars in \Hipparcos\
into a photometric parallax relation---predicted model parallax plus
model scatter---which can be compared to the photometric parallax
obtained for a single-burst stellar population for each star.

\section{The background model}\label{sec:background}

\begin{figure}
\includegraphics[]{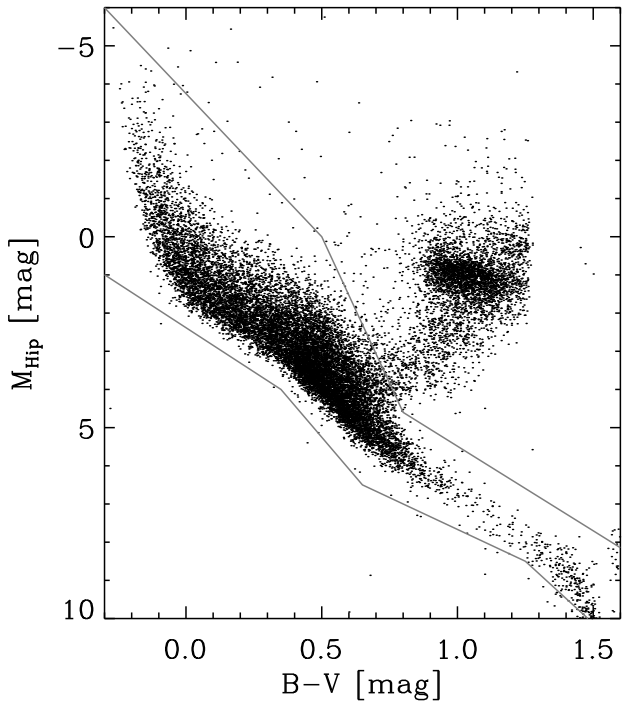}
\caption{Color--magnitude diagram of the full \Hipparcos\ sample of
  \nstars\ stars, selected to be kinematically unbiased and consist of
  single stars with relative parallax uncertainties $\lesssim$
  10\,percent. The \nstarsms\ main-sequence stars that we use in the
  hypothesis tests in \sectionname \sectionname \ref{sec:firstlook},
  \ref{sec:background}, and \ref{sec:ssp} lie between the gray lines.
  \mhip\ is the absolute magnitude in \Hipparcos' own
  passband.}\label{fig:cmd}
\end{figure}

\begin{figure}
\includegraphics[width=0.3\textwidth]{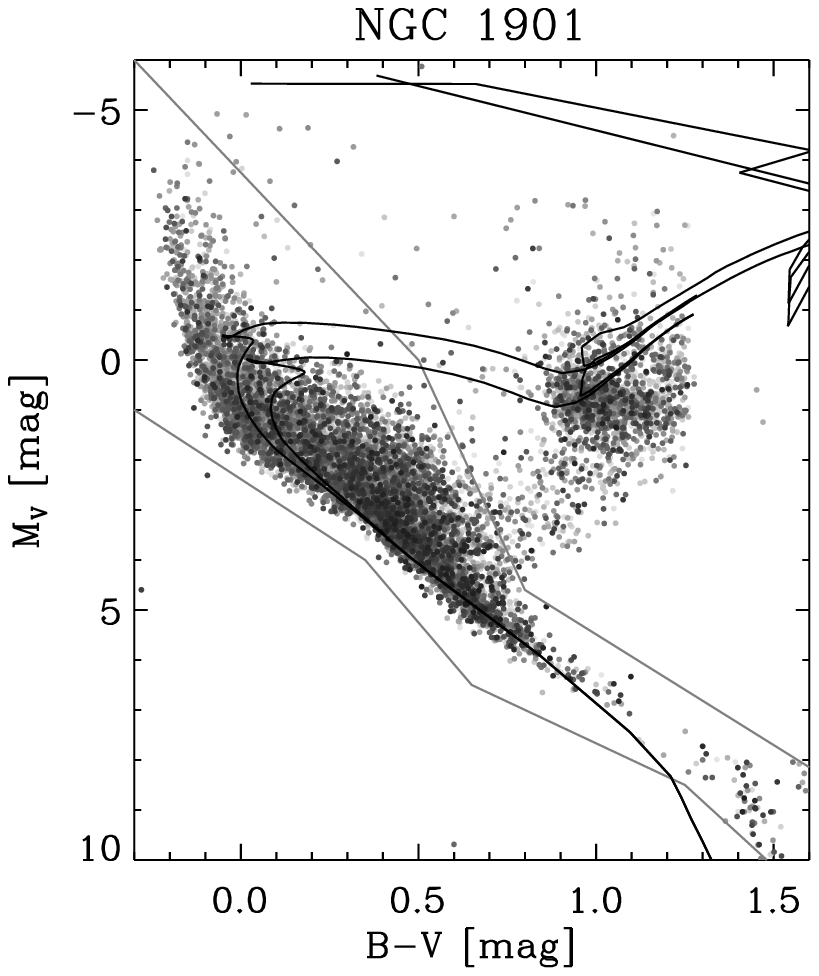}
\includegraphics[width=0.3\textwidth]{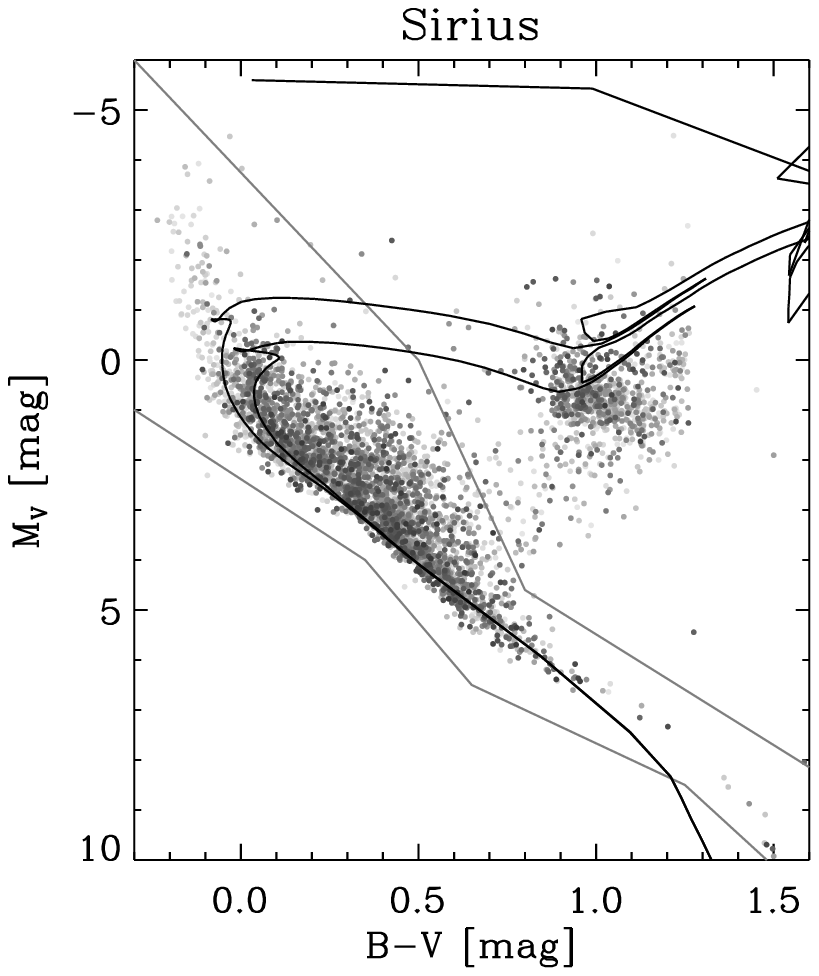}
\includegraphics[width=0.3\textwidth]{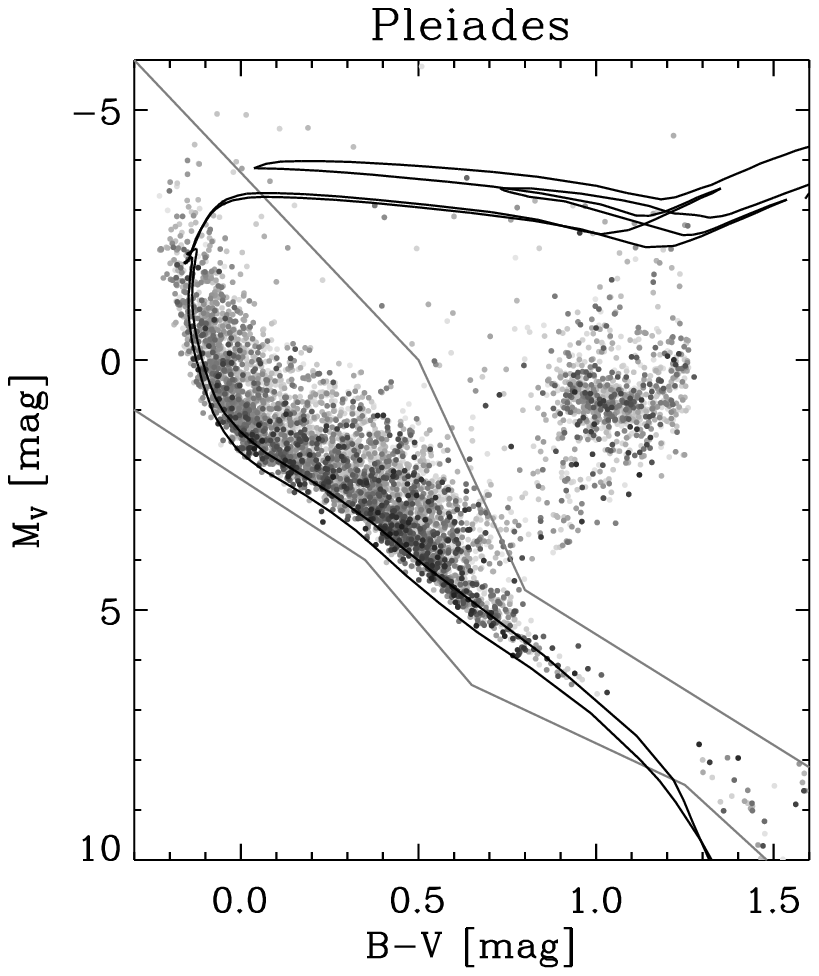}\\
\includegraphics[width=0.3\textwidth]{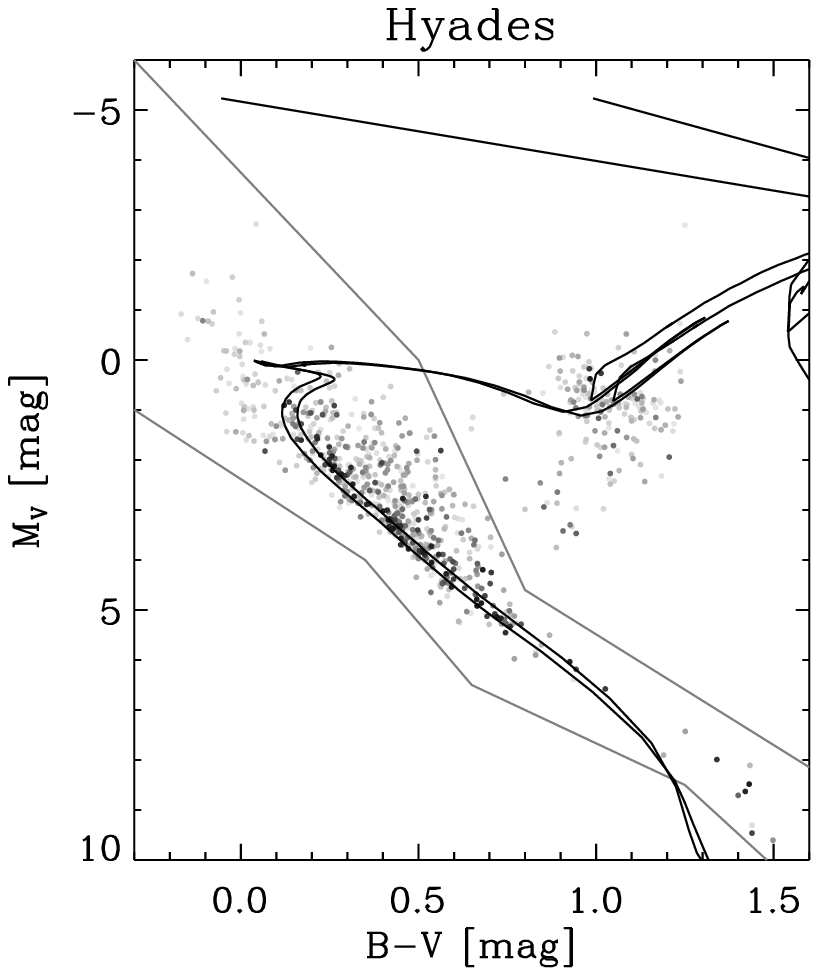}
\includegraphics[width=0.3\textwidth]{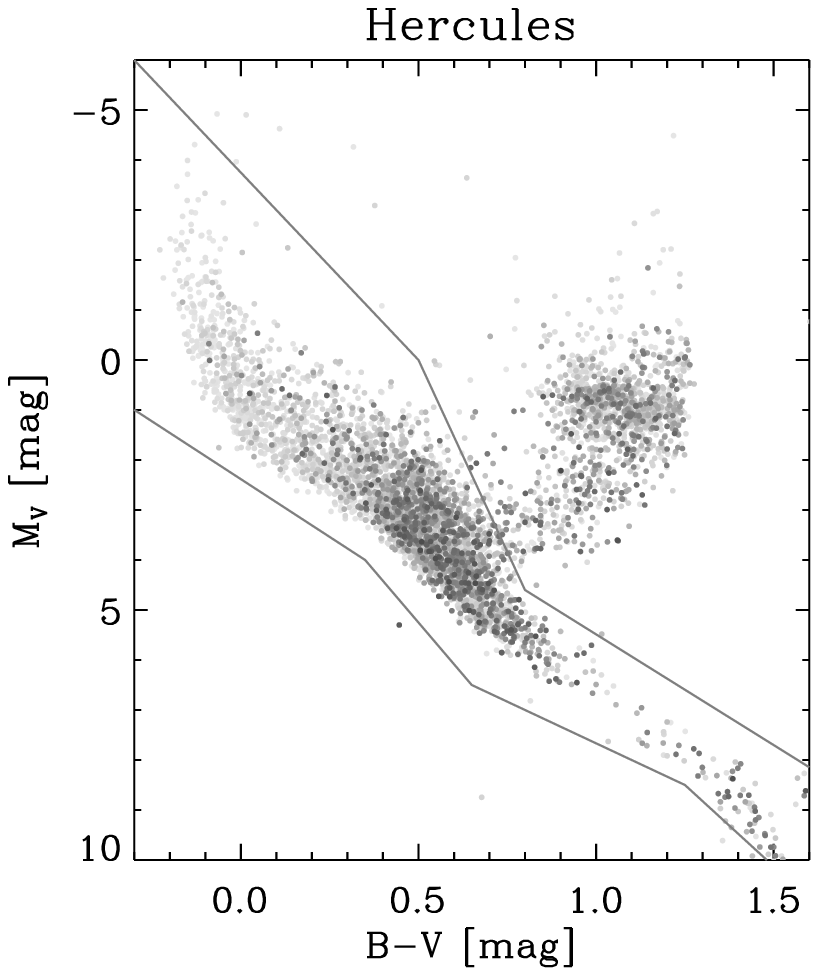}
\includegraphics[width=0.3\textwidth]{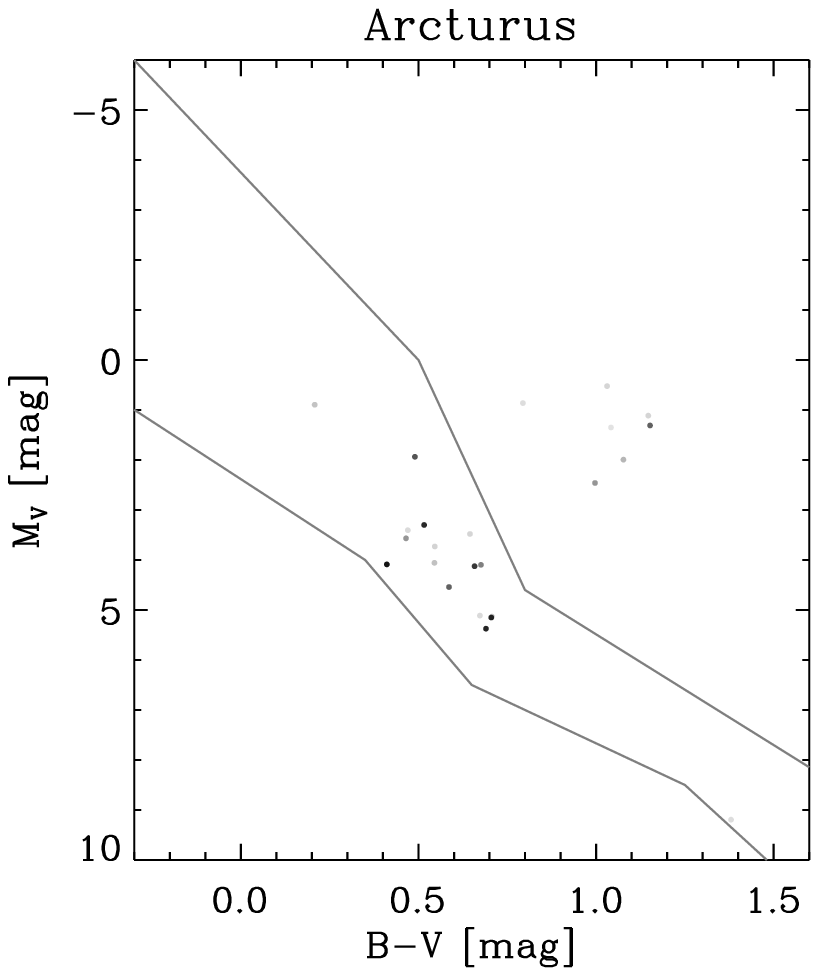}
\caption{Color--magnitude diagrams of the six moving groups detected in
\bhr. The points are grayscale-coded with the probability of each star
to be part of the moving group (see the text); only stars that have a
probability larger than 0.1 of being part of the moving group are
plotted. For those moving groups potentially associated with an open
cluster, theoretical isochrones \citep{Marigo08a,Bertelli94a} for the
open cluster are overlaid: the 400 Myr, $Z=0.016$ \citep{Carraro07a}
and the 600 Myr, $Z=0.016$ \citep{Pavani01a} isochrone for the NGC
1901 cluster; the 300 Myr, $Z=0.016$ \citep{Soderblom93a} and the 500
Myr, $Z=0.016$ \citep{King03a} isochrone for the Ursa Major (Sirius)
cluster; the 100 Myr, $Z= 0.018$ \citep{Boesgaard90a,Gratton00a} and
the 100 Myr, $Z=0.008$ \citep{Percival05a} isochrone for the Pleiades
cluster ; the 625 Myr, $Z = 0.026$ isochrone \citep{Perryman98a} and
the 625 Myr, $Z=0.019$ isochrone for the Hyades cluster. The
main-sequence cuts from \figurename~\ref{fig:cmd} are indicated in
gray.}\label{fig:groupcmd}
\end{figure}

\begin{figure}
\includegraphics[width=.5\textwidth]{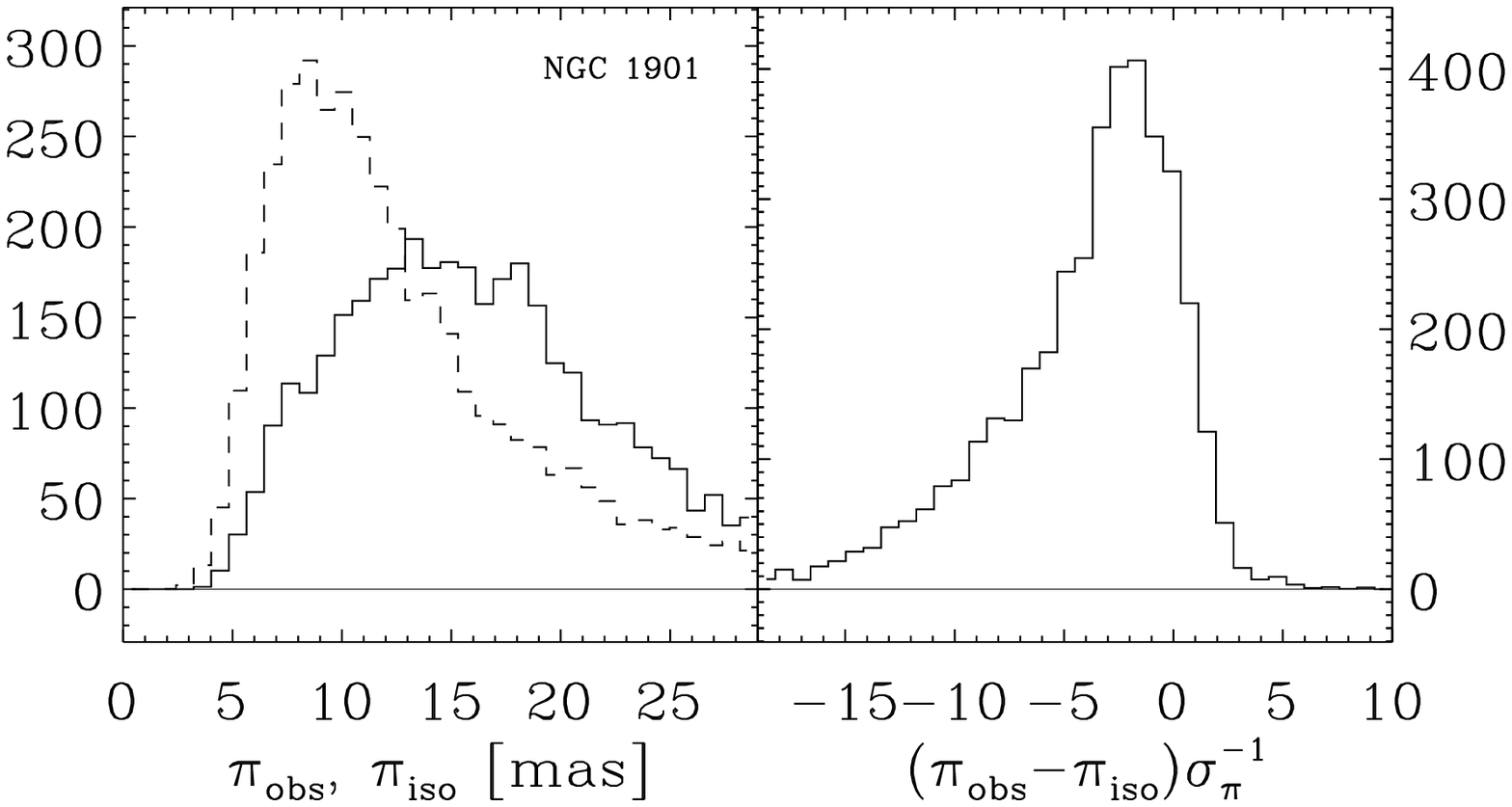}
\includegraphics[width=.5\textwidth]{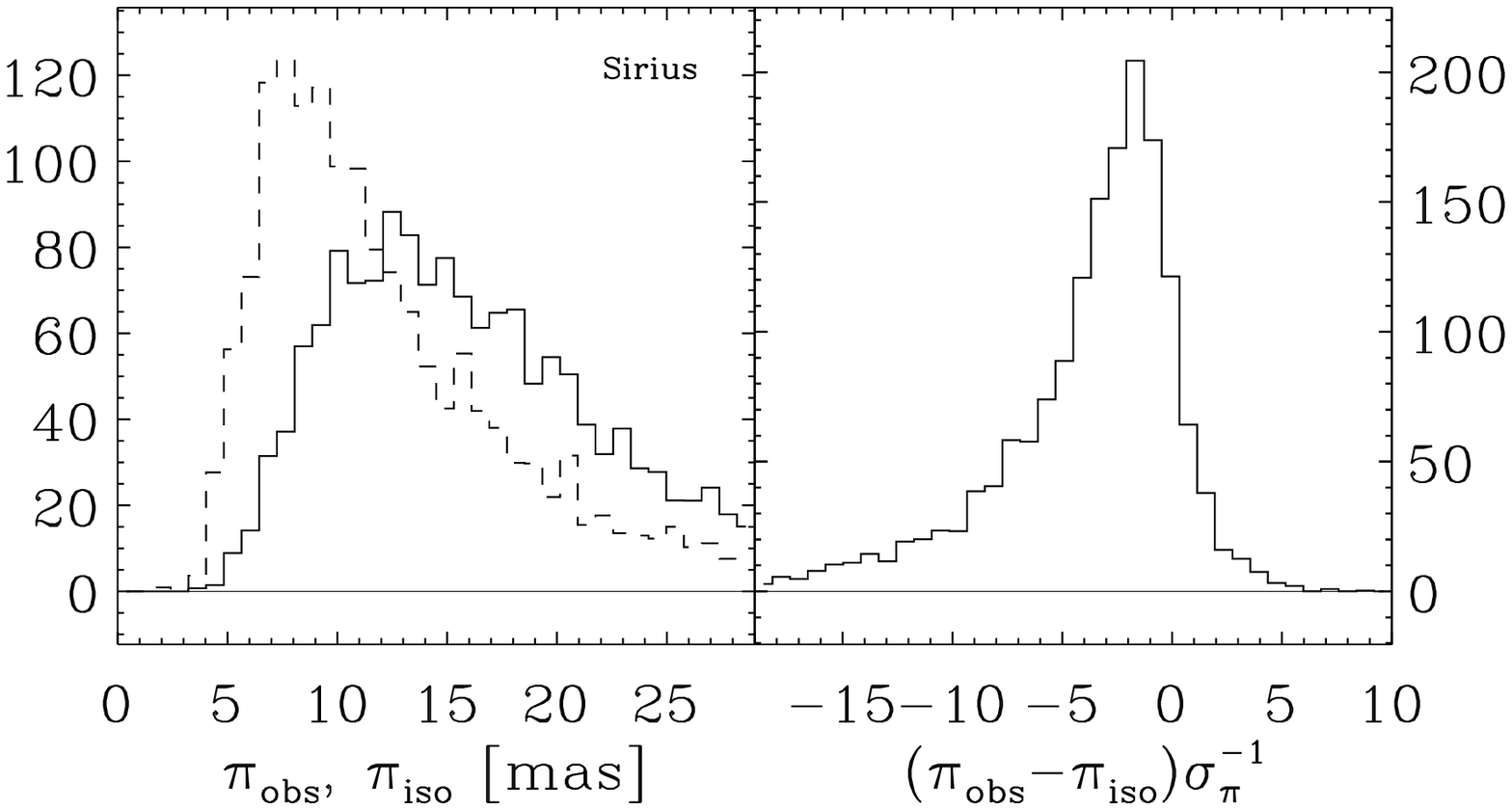}
\includegraphics[width=.5\textwidth]{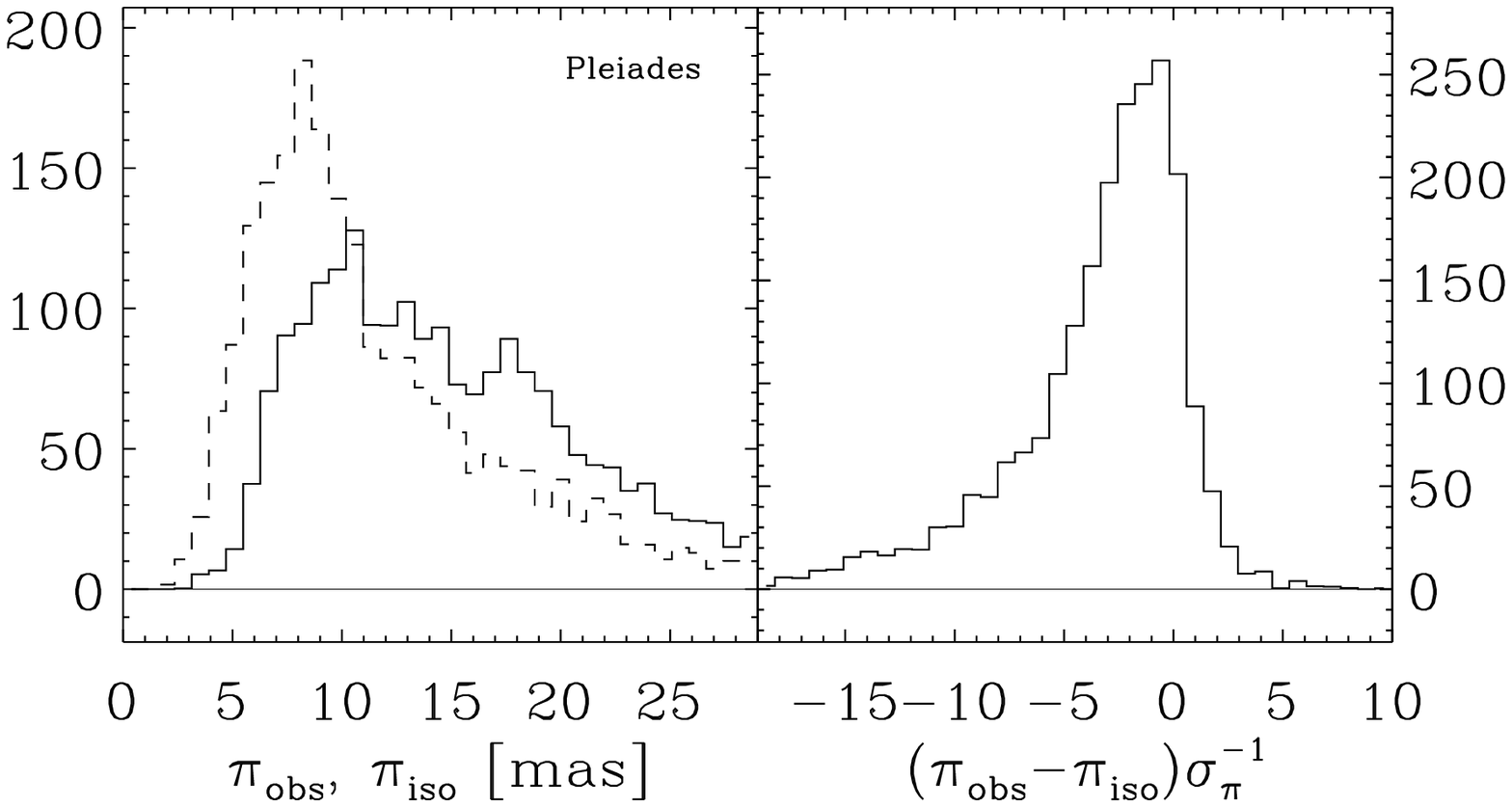}
\includegraphics[width=.5\textwidth]{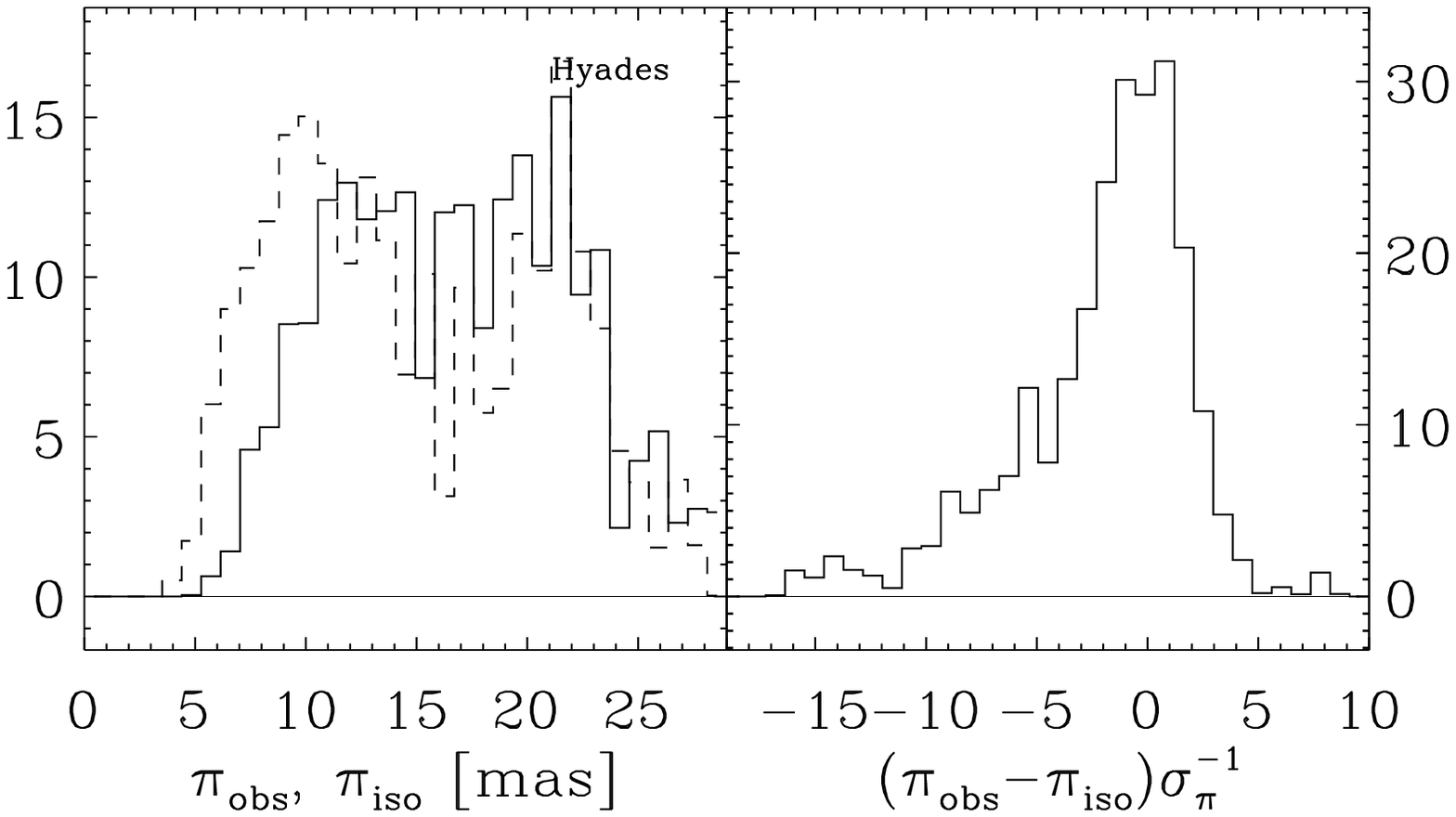}
\caption{Observed parallaxes vs. model parallaxes assuming a
  single-burst stellar population identical to that of the associated
  cluster of the moving groups: comparison of the distribution of
  observed parallaxes (\emph{dashed lines}) with that of the model
  parallaxes (\emph{solid lines}) in the left figure of each panel;
  histogram of the normalized difference between model and observed
  parallax in the right figure.  Each star is weighted by its
  probability of being part of the moving group in question. The
  isochrone used in this figure corresponds to the first age and
  metallicity pair mentioned in the caption of
  \figurename~\ref{fig:groupcmd} for each open
  cluster.}\label{fig:iscluster}
\end{figure}

\begin{figure}
\includegraphics[width=0.5\textwidth]{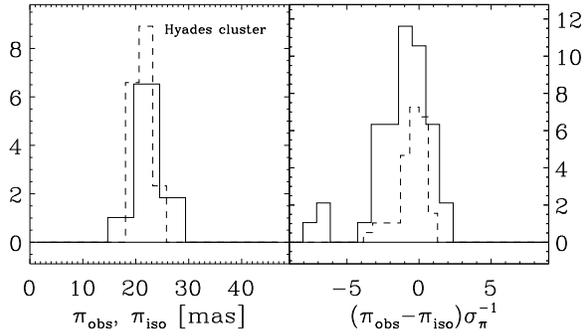}
\caption{Same as \figurename~\ref{fig:iscluster} but for the Hyades
cluster of \figurename~\ref{fig:hyadesclustercmd}. The dashed
histogram in the right panel is what one gets after adding a 0.2
mag spread in quadrature to the observational uncertainty in the
parallax (this histogram has been scaled down by a factor of three for
display purposes).}\label{fig:hyadesclusterhist}
\end{figure}

Given that we have at our disposal a large number of stars with
accurate photometry and parallaxes to estimate a one-dimensional
photometric parallax relation, it is unlikely that any parametric
model could capture the observed relation and its scatter in all of
its details. It is, therefore, advisable to use a non-parametric
approach to estimate the photometric parallax relation and its
intrinsic scatter for the background population. Principled
probabilistic approaches to this exist \citep[\eg, using Gaussian
Process regression:][]{rasmussen2005a} but given the low-dimensional
nature of the problem and the large amount of data---from
\figurename~\ref{fig:cmd} it is clear that at most points there are at
least dozens of stars with which to estimate the local relation---we
can expect simpler procedures to perform adequately.

To constrain the background model we use all of the \nstarsms\ stars
in our \Hipparcos\ sample. Strictly speaking, we are testing whether
one or more of the moving groups is distinct from the local disk
population of stars in that it consists solely of stars of a narrow
age and metallicity range, and therefore, including these moving group
members in the background model mixes into the background model the
stellar populations of the moving groups. This could complicate model
selection, since it will be harder to distinguish between the
background and the foreground models (for the purposes of this section
and the next, the foreground model for each moving group is that it is
a single-burst stellar population) when the background model is more
like the foreground model than it should be. In principle we should
test each combination of moving-group/not-a-moving-group for each of
the moving groups and build background models out of stars that are
not believed to be part of a single-burst stellar population in that
particular model selection test. This would be impractical, not in the
least because few of the stars can be confidently assigned to a
specific moving group or even background, and making subsamples would
necessarily involve making hard cuts on membership probabilities, with
all of the biases that would result from that. We therefore use all of
the stars to construct the background model and investigate each
moving group separately in the following by testing against this
background model. Given that more than 60\,percent of the stars are
believed to be part of the background (see \bhr) and that the
population of moving groups taken together would presumably span some
range of age and metallicity, the effect of including real moving
group members in the background sample should be small. It is
important to note, however, that even if the moving groups
significantly affect the background model, this will only bring the
foreground and the background model closer together, but the
foreground model should still be preferred over the background model
when the moving group is a single-burst stellar population.

\begin{figure}
\includegraphics[width=\textwidth]{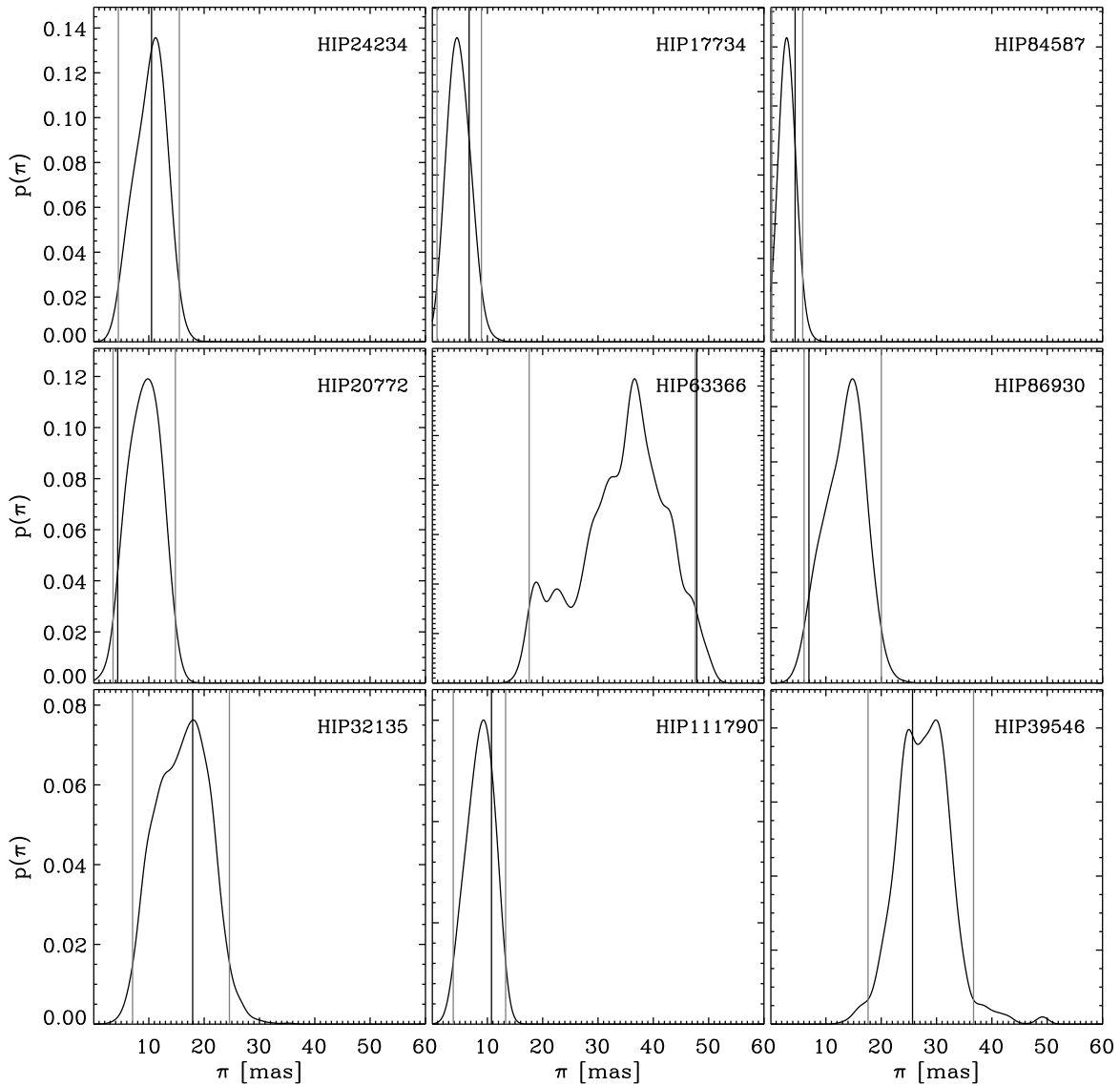}
\caption{Background model predictions for the parallax of 9 random
  stars in the basic \Hipparcos\ sample. The background model consists
  of a linear smoother with a Tricube kernel with width parameter
  \kernelwidth\ = \fiducialkernelwidth. In each panel the background
  model has been convolved with the observational parallax
  uncertainty. The observed parallax (\emph{thick, black line}) as
  well as 95\,percent confidence regions (\emph{thin, gray lines}) are
  indicated.}\label{fig:back_random}
\end{figure}

From our sample of \nstarsms\ main-sequence stars, we construct a
non-parametric photometric parallax relation: For each star $i$ we
take the stars in our \Hipparcos\ sample in a small color bin (see
below) around star $i$'s color and consider the absolute magnitudes of
the stars in this color bin to represent the complete set of absolute
magnitudes that a star in this color bin could have, that is, the
probability of the absolute magnitude of star $i$ is given by
\begin{equation}\label{eq:deltaabsmag1}
p(M_{V,i} | (\bminusv)_i) = \sum_{\substack{j\\(\bminusv)_j \approx
    (\bminusv)_i}} \delta (M_{V,i}-M_{V,j})\,,
\end{equation}
where $\delta(\cdot)$ is the Dirac delta function. The exact meaning
and implementation of $(\bminusv)_j \approx (\bminusv)_i$ are
discussed in detail below. Given this finite set of possible absolute
magnitudes for star $i$, we use the observed $V$ magnitude of star $i$
to derive a probability estimate for its parallax $\parallax_i$, that
is,
\begin{align}\label{eq:photplx1}
p(\parallax_i | V_i,M_{V}) &= \delta(\parallax_i -
\parallax[V_i,M_{V}] )\,,\\ \parallax[V,M_V] &= 10^{\left[(M_V -
    V)/5+2
    \right]}\,,\label{eq:photplx2}\\ p(\parallax_i|V_i,(\bminusv)_i)
&= \sum_{\substack{j\\(\bminusv)_j \approx (\bminusv)_i}} \delta
(\parallax_i - \parallax[V_i,M_{V,j}] )\,,
\end{align}
where $[\pi] = $ mas. To define the notion of ``nearness'' that is the
implementation of $(\bminusv)_j \approx (\bminusv)_i$ in the
expressions above we use the concept of a kernel (in this sense the
method described here is similar to that of a linear smoother in
non-parametric statistics; \citealt{Wasserman05a}). Using a kernel
$w(\cdot;\lambda)$ we define a distance between two colors $x_i \equiv
(\bminusv)_i$ and $x_j \equiv (\bminusv)_j$ as $w(x_i-x_j;\lambda)$,
where $\lambda$ is a width parameter of the kernel, and we use this
notion of distance to weight the contributions of the various stars in
the background sample. These weights are inserted into
\eqnname~(\ref{eq:deltaabsmag1}) as follows
\begin{equation}\label{eq:deltaabsmag2}
p(M_{V,i} | (\bminusv)_i) = \frac{1}{W}\,\sum_j
w(x_i-x_j;\lambda)\,\delta (M_{V,i}-M_{V,j})\,,
\end{equation}
where $x_i$ and $x_j$ are the colors of the stars and $W$ is a
normalization factor equal to $\sum_j w(x_i-x_j;\lambda)$. To compare
this photometric parallax with the observed trigonometric parallax
$\parallax_{\mathrm{obs},i}$ we convolve this distribution with the
observational uncertainty $\sigma_{\parallax,i}$, assumed Gaussian:
\begin{equation}\label{eq:backphotplx}
p(\parallax_{\mathrm{obs},i} | \sigma_{\parallax,i},V_i,(\bminusv)_i)
= \frac{1}{W}\,\sum_j
w(x_i-x_j;\lambda)\,\normal(\parallax_{\mathrm{obs},i} |
\parallax[V_i,M_{V,j}],\sigma_{\parallax,i}^2 )\,,\\
\end{equation}
where $\normal(\cdot)$ is the normalized, one-dimensional Gaussian
distribution and $\parallax[V,M_V]$ is given by
\eqnname~(\ref{eq:photplx2}). The probability distributions for the
observed parallax obtained in this way are shown for a random sample
of stars in \figurename~\ref{fig:back_random} together with the actual
observed value of the trigonometric parallax (the kernel and its width
used in this figure are the optimal ones for the background model as
discussed below).

Several considerations play a role in choosing a kernel function
$w(\cdot;\lambda)$. On the one hand one wants a kernel that is as
smooth as possible, smoothly going from giving high weights to points
that are close in color space to low weights for stars on the other
side of the main sequence. However, it is computationally advantageous
to use a kernel that has finite support such that in constructing the
photometric parallax prediction in \eqnname~(\ref{eq:backphotplx})
only a subset of the \nstarsms\ in the whole sample are used. For this
reason, a Gaussian kernel
\begin{equation}\label{eq:gaussiankernel}
w(u;\lambda) = \exp\left(-\frac{u^2}{2\lambda^2}\right)\,,
\end{equation}
while smooth, is unwieldy. Therefore, we have considered the following
finite-range kernels: the Tricube kernel
\begin{equation}\label{eq:tricube}
w(u;\lambda) =  \left(1-\left(\frac{u}{\lambda}\right)^3\right)^3\,,\qquad u \leq \lambda\,,
\end{equation}
and the Epanechnikov kernel
\begin{equation}\label{eq:epanechnikov}
w(u;\lambda) = (1-\left(\frac{u}{\lambda}\right)^2)\,, \qquad u \leq \lambda\,.
\end{equation}
Of these, the Tricube kernel is everywhere differentiable; it combines
the best of both worlds.

Each of the kernels has a width parameter $\lambda$ that is unknown
\emph{a priori}. We need to train the background model, \ie, establish
a good value of $\lambda$. We train the background model using
leave-one-out cross-validation \citep{Stone74a}: for each choice of
the width parameter $\lambda$ on a logarithmically spaced grid in
$\lambda$, we compute the probability of each of the observed
parallaxes in our sample using as the training $\{\bminusv,M_V\}$-set
all of the other stars in our sample. We then multiply the
probabilities thus obtained for all of the stars and take the
logarithm of this product to compute the ``score'' for that value of
$\lambda$; this quantity is also known as the
``pseudo-likelihood''. The value of $\lambda$ with the highest score
is the preferred value of $\lambda$.

\begin{figure}
\includegraphics{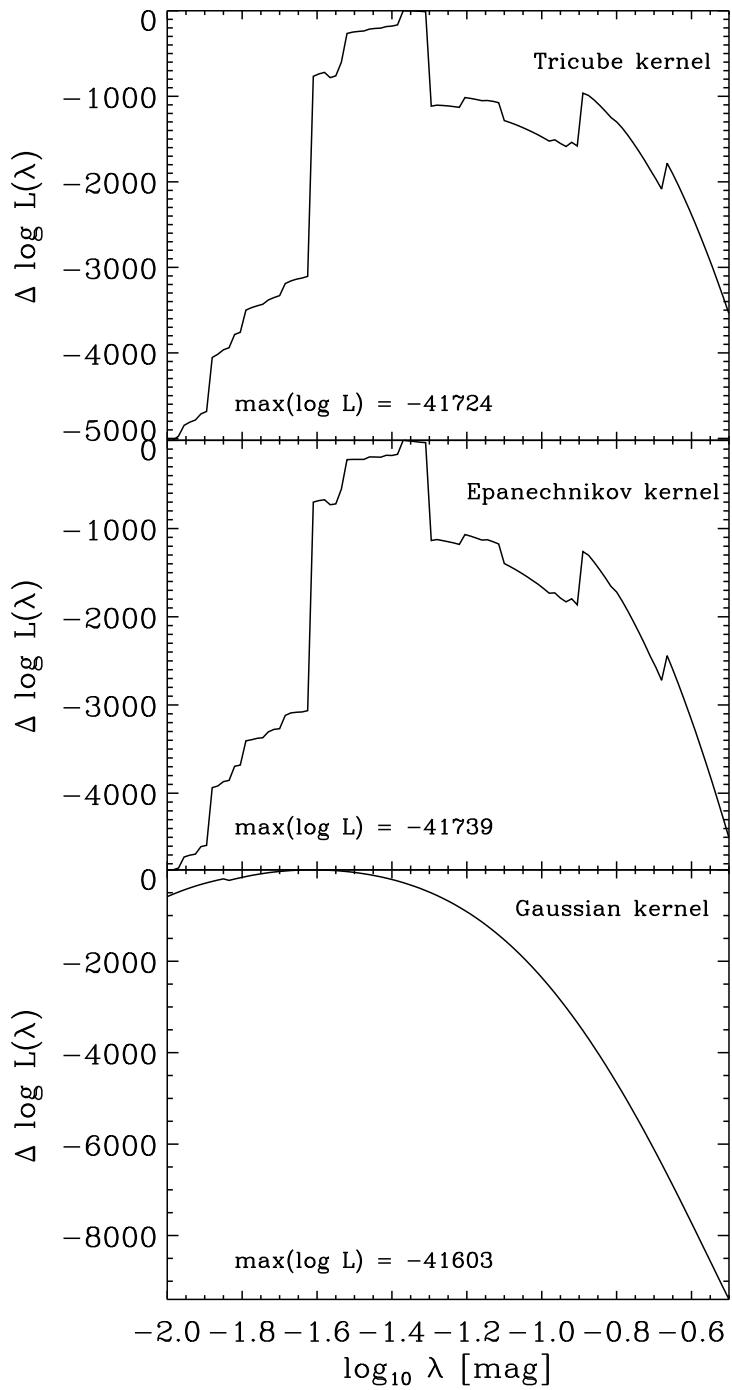}
\caption{Selection of the width parameter \kernelwidth\ of the kernel
  used in the kernel-regression background
  model.}\label{fig:kernelwidthselection}
\end{figure}

We computed the cross-validation score for a range of values of
$\lambda$ for each of the kernels; these are shown in
\figurename~\ref{fig:kernelwidthselection}. It is clear that all three
kernels agree on the best value of $\lambda$ (keeping in mind that the
Gaussian kernel has infinite range and only approaches zero for $u >
2\lambda$). As expected, the resulting cross-validation score curve
for the Gaussian kernel is much smoother than the corresponding curves
for the Tricube and Epanechnikov kernels and the maximum score for the
Gaussian kernel is somewhat higher than that for the Tricube kernel,
but because computations with the Gaussian kernel are much slower and
the gain in performance is small, we choose the Tricube kernel for the
background model. This is, again, a conservative choice, since a
slightly worse background fit can only make it easier for the
foreground model to be preferred. All three kernels agree that the
optimal width is approximately $\lambda$ = 0.05 mag and this is the
value used in the background model.

\begin{figure}
\includegraphics{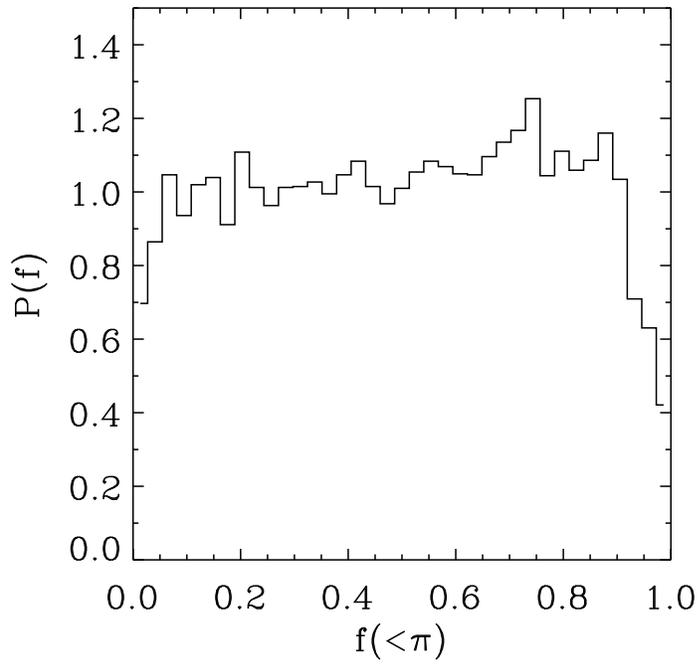}
\caption{Distribution of the quantiles at which the observed parallax
  is found of the background-model predictive distribution for the
  parallax. This curve should be flat for perfectly consistent
  predictive distributions---meaning that they correctly predict all
  of the quantiles of the distribution.}\label{fig:back_quants}
\end{figure}

To test whether the background model with the chosen kernel parameters
actually provides a good fit to the data or whether it is merely the
best possible fit among the possibilities explored (note that we do
not expect this to be the case as this is a non-parametric model), we
have checked whether the background-model parallax probability
distribution in \eqnname~(\ref{eq:backphotplx}) is a consistent
description of the parallax distribution in that all of the quantiles
of the distribution are correct. \figurename~\ref{fig:back_quants}
shows the distribution of the quantiles of the parallax distribution
at which the observed, trigonometric parallaxes are found. If the
background model is a good description of the observed parallaxes,
then this distribution should be uniform. That is, if
\eqnname~(\ref{eq:backphotplx}) correctly predicts the 95\,percent
confidence interval, the 90\,percent interval, and so on, then the
background model is a good fit to the data. The distribution in
\figurename~\ref{fig:back_quants} is flat over most of the range
between zero and one, with the only major deviations near the edges of
this interval, and we can therefore say that the background model
provides a good fit to the bulk of the data. That the background model
fails for stars at the edges of the parallax distribution is not
surprising as these are rare: nearby and faint or distant and bright
stars are sparsely sampled regions of the color--magnitude diagram in
a magnitude-limited sample as is clear from \figurename~\ref{fig:cmd}.

\section{The moving groups are not single-burst stellar populations}\label{sec:ssp}

The goal of this section is to establish whether the moving groups
could conceivably arise from an evaporating cluster, or whether their
stellar content is inconsistent with being produced by a single burst
of star formation. We will fit a model of a single-burst stellar
population to each of the five moving groups and test whether this
model is a better fit to the moving-groups data than the background
model described in the previous section. The foreground hypothesis for
the purpose of this section is therefore that the moving group is
characterized by a single age and metallicity.

Like the background model, the foreground model defines a photometric
parallax relation. While that defined by the background sample is a
broad model, roughly consisting of a mean photometric parallax
relation and a large amount of scatter around this mean, the
foreground model's photometric parallax relation is very narrow, or
informative, in that it is given by the single isochrone corresponding
to an assumed age and metallicity (single in the sense of being the
unique isochrone in the Padova database), smoothed by the
observational uncertainty. The probability of an observed,
trigonometric parallax $\parallax_{\mathrm{obs},i}$ assuming a certain
age and metallicity $Z$ and given the star's color $(\bminusv)_i$,
apparent magnitude $V_i$, and observational uncertainty
$\sigma_{\parallax,i}$ is given by
\begin{equation}\label{eq:photplxiso}
p(\parallax_{\mathrm{obs},i} | \mathrm{Age},Z,\sigma_{\parallax,i},V_i,(\bminusv)_i)
= \normal \left(\parallax_{\mathrm{obs},i}|\parallax[V_i,(\bminusv)_i,\mathrm{Age},Z],\sigma_{\parallax,i}^2\right)\,,
\end{equation}
where the photometric parallax
$\parallax[V_i,(\bminusv)_i,\mathrm{Age},Z]$ is derived from the
isochrone absolute magnitude as in \eqnname~(\ref{eq:photplx2}). The
absolute magnitude is derived from the isochrone by reading off the
absolute magnitude along the isochrone corresponding to the assumed
age and metallicity at the star's observed color $(\bminusv)_i$.

\Eqnname~(\ref{eq:photplxiso}) is not the whole story. First,
\figurename s~\ref{fig:hyadesclustercmd} and
\ref{fig:hyadesclusterhist} show that even an open cluster itself is
not perfectly fit by a single isochrone, that is, the right histogram
in \figurename~\ref{fig:hyadesclusterhist} is much broader than the
unit variance Gaussian distribution. We find that there is 0.2 mag of
root variance in absolute magnitude with respect to the isochrone
locus of the stars in \figurename~\ref{fig:hyadesclustercmd}. We
propagate this to a variance in the parallaxes of the Hyades cluster
members and add it in quadrature to the observational parallax
uncertainty. The resulting photometric parallax--observed parallax
comparison is also shown in \figurename~\ref{fig:hyadesclusterhist} as
the dashed histogram. This distribution has close to unit variance;
now the open-cluster scenario provides a good fit to the data (this
procedure is somewhat equivalent to adding a small amount of unmodeled
noise or ``jitter'').

Second, the assumption of a certain age and metallicity for a moving
group is too easily falsified. When we observe a star that is a member
of a moving group, but that has a color that is inconsistent with that
age and metallicity, \eg, because the star is too young to still be on
the main sequence of an old population of stars, this combination of
age and metallicity is ruled out by the existence of this single star
alone. As useful as the idea of falsification has been in
epistemology, and as helpful as it could be in this case if we had
high probability members of the moving groups in our sample, the ease
of falsification is actually problematic, since we cannot confidently
assign any of the stars in our sample to moving groups and we need to
take the odds that a star is in fact a background interloper into
account. The proper way to take this interloper probability into
account is to divide the probability of a star's properties among the
foreground and background hypotheses in a way that is proportional to
the probability that the star is part of the moving group or
not. Thus, we write the probability of the observed parallax of each
star as
\begin{equation}\label{eq:forebackground}
p(\parallax_{\mathrm{obs},i}) = p(\parallax_{\mathrm{obs},i} |
\mathrm{foreground})^{\pij} \,p(\parallax_{\mathrm{obs},i} |
\mathrm{background})^{1-\pij}\,,
\end{equation}
where $\pij$ is the probability that star $i$ is a member of moving
group $j$; see \eqnname~(\ref{eq:pij}). A low-probability member of a
moving group, one that is most likely \emph{not} a member, has the bad
property that it can rule out a certain age and metallicity due to its
color being inconsistent with it, since the first factor in
\eqnname~(\ref{eq:forebackground}) will be zero for any non-zero
$\pij$ and the probability of an age and metallicity of a moving group
is given by Bayes's theorem
\begin{equation}\label{eq:bayesSSP}
p(\mathrm{Age},Z | \{\parallax_{\mathrm{obs},i}\} ) \propto p(\mathrm{Age},Z) \prod_i p(\parallax_{\mathrm{obs},i} | \mathrm{Age},Z)\,,
\end{equation}
where we have implicitly assumed the other observational properties of
the star (\ie, its color, apparent magnitude, and observational
parallax uncertainty) in the conditional probabilities. A single star
with a color that is inconsistent with the age and metallicity under
investigation for the moving group therefore rules out this age and
metallicity, as it factors in a zero probability in the product in
\eqnname~(\ref{eq:bayesSSP}). 

Instead of just using the isochrone prediction in evaluating the
probability of an observed parallax under the foreground model in
\eqnname~(\ref{eq:photplxiso}), we add a small contribution from the
background into the probability, such that the first factor in
\eqnname~(\ref{eq:forebackground}) becomes
\begin{equation}\label{eq:photplxfore}
p(\parallax_{\mathrm{obs},i} | \mathrm{foreground}) = (1-\alpha)\,p(\parallax_{\mathrm{obs},i} | \mathrm{Age},Z) + \alpha \,p(\parallax_{\mathrm{obs},i} | \mathrm{background})\,,
\end{equation}
where the background probability is given by
\eqnname~(\ref{eq:backphotplx}). The parameter $\alpha$ is, in
general, a free parameter and is a measure of the amount of background
contamination. The total foreground probability is obtained by
substituting this equation into \eqnname~(\ref{eq:forebackground}). A
star whose color is inconsistent with an assumed age and metallicity
will now automatically resort to its background probability, since
then $p(\parallax_{\mathrm{obs},i} | \mathrm{Age},Z)$ is zero and
\begin{equation}
\begin{split}
p(\parallax_{\mathrm{obs},i} | \mathrm{foreground}) &= \left[(1-\alpha)\,p(\parallax_{\mathrm{obs},i} | \mathrm{Age},Z) + \alpha \,p(\parallax_{\mathrm{obs},i} | \mathrm{background})
\right]^{\pij}\\
& \qquad \times p(\parallax_{\mathrm{obs},i} | \mathrm{background})^{1-\pij} \\
&= \alpha^{\pij} p(\parallax_{\mathrm{obs},i} | \mathrm{background})\,.
\end{split}
\end{equation}
Low probability members have $\pij \approx 0$ such that $\alpha^{\pij}
\approx 1$. This expression shows that when $\alpha$ is a free
parameter, it will be advantageous to make it large when
high-probability members are inconsistent with the assumed age and
metallicity, to make the fit at least as good as the background fit.

When $\alpha$ is allowed to take any value between zero and one, it is
clear that if the fit prefers a value of $\alpha$ that is close to
one, this will be an indication that the single-burst stellar
population model is not a very good fit to the moving-group data. But
what value of $\alpha$ do we expect if the data \emph{are} consistent
with the moving group having originated from a single burst of star
formation? In order to answer this question, we look at the overall
properties of the velocity distribution. We look at the fraction of
stars in one of the moving groups as a function of a hard cut on the
membership probabilities $\pij$ to assign moving-group members. We
find that a hard cut of $\pij > $ \lsrpijcut\ gives rise to a fraction
of moving-group substructure consistent with the overall fraction of
substructure observed in the velocity distribution, \ie,
40\,percent. We can then ask: What is the accumulated fraction of
membership probability of stars with membership probabilities less
than this hard cut? This gives an estimate of the background
contamination for each moving group, that is, it gives an indication
of the influence of the background stars on inferences using the
membership probabilities. These background contamination levels
$\alpha$ are given in \tablename~\ref{table:bestfitssp} on the first
line for each moving group.

If we allow the fit to vary the background contamination level
$\alpha$ and we find that the fit prefers values of $\alpha$ that are
much larger than the value of $\alpha$ estimated for each moving group
from the global analysis described above, that is a strong indication
that the moving groups are not single-burst stellar populations. This
does not rule out that certain parts of the moving group are
consistent with being created in a single burst---a preferred value of
$\alpha$ that is close to but not equal to unity could suggest
this. Therefore, we perform two fits: one in which we fix $\alpha$ at
the value determined in the last paragraph and the other in which we
allow $\alpha$ to take on any value between zero and one. In both
cases we vary the age and metallicity of the underlying isochrone on a
grid in log age and metallicity space. The best fit is then given by
the combination of age, metallicity, and---if left free---$\alpha$
that maximizes the probability of the foreground model in
\begin{equation}\label{eq:maxlikeSSP}
p(\log \mathrm{Age},Z(, \alpha)) \propto \prod_i p(\parallax_{\mathrm{obs},i} | \log \mathrm{Age}, Z(, \alpha))\,,
\end{equation}
where the individual conditional probabilities are given by
\eqnname~(\ref{eq:forebackground}) and the parentheses around $\alpha$
indicate that we either fix $\alpha$ or vary it. This is a
maximum-likelihood fit or, equivalently but relevant in what follows,
the maximum of the posterior probability distribution for age,
metallicity, and $\alpha$ with uniform priors on log age, metallicity,
and $\alpha$. The latter attitude permits marginalization over subsets
of the parameters. In performing the fit, the last factor in
\eqnname~(\ref{eq:forebackground}) is irrelevant, as it does not
depend on any of the fit parameters, but in the model selection it
does need to be taken into account.

\begin{figure}
\begin{center}
\includegraphics[width=0.3\textwidth]{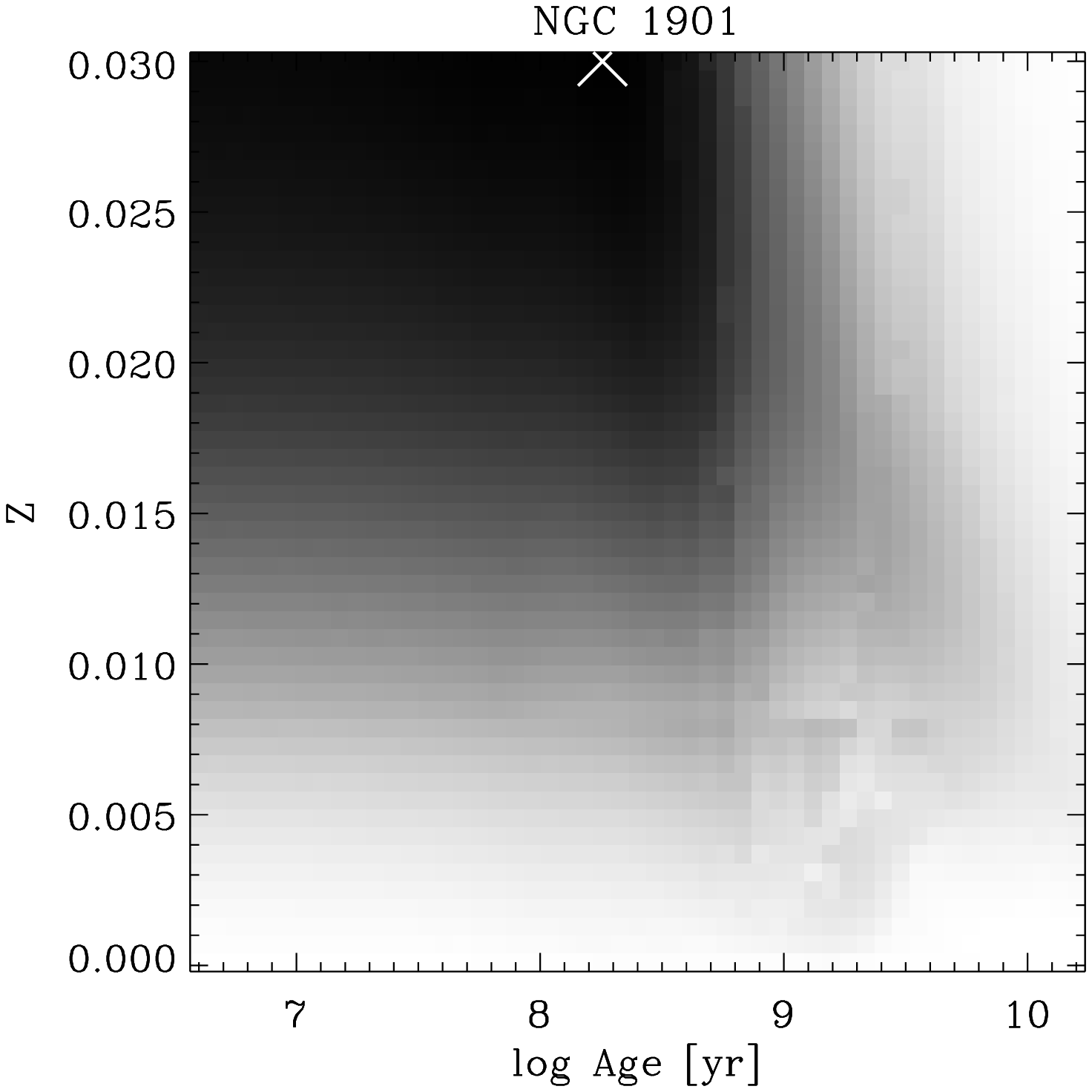}
\includegraphics[width=0.3\textwidth]{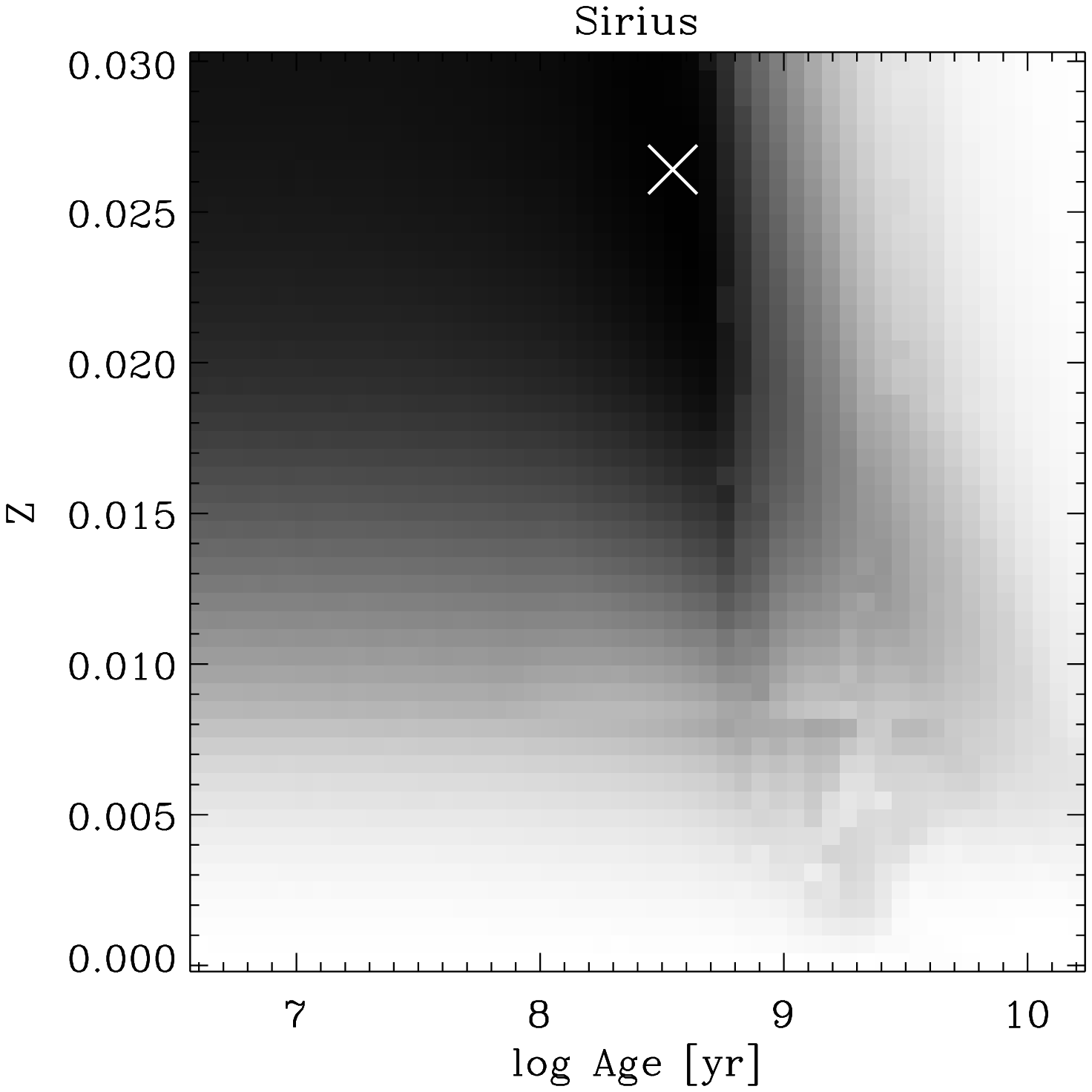}
\includegraphics[width=0.3\textwidth]{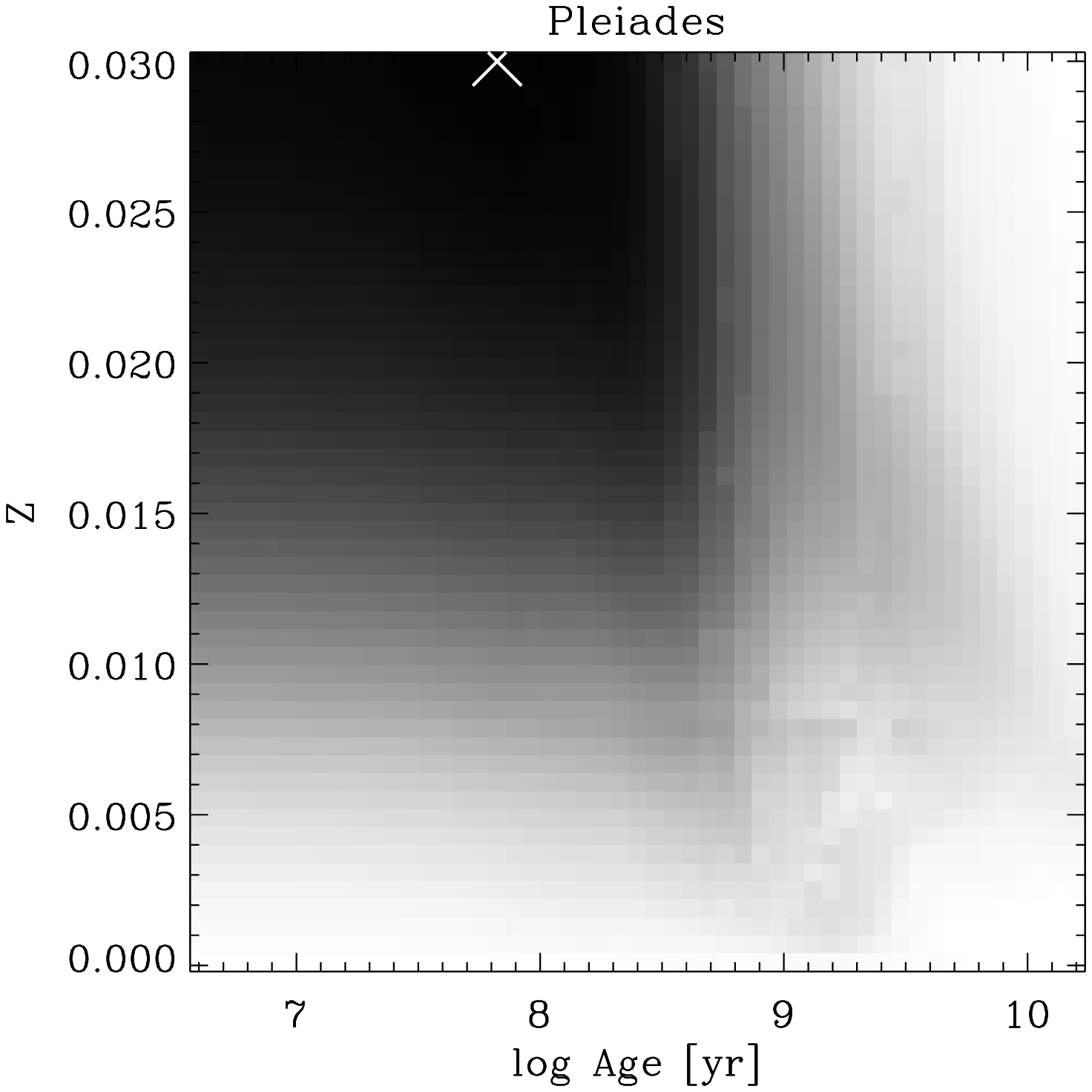}\\
\includegraphics[width=0.3\textwidth]{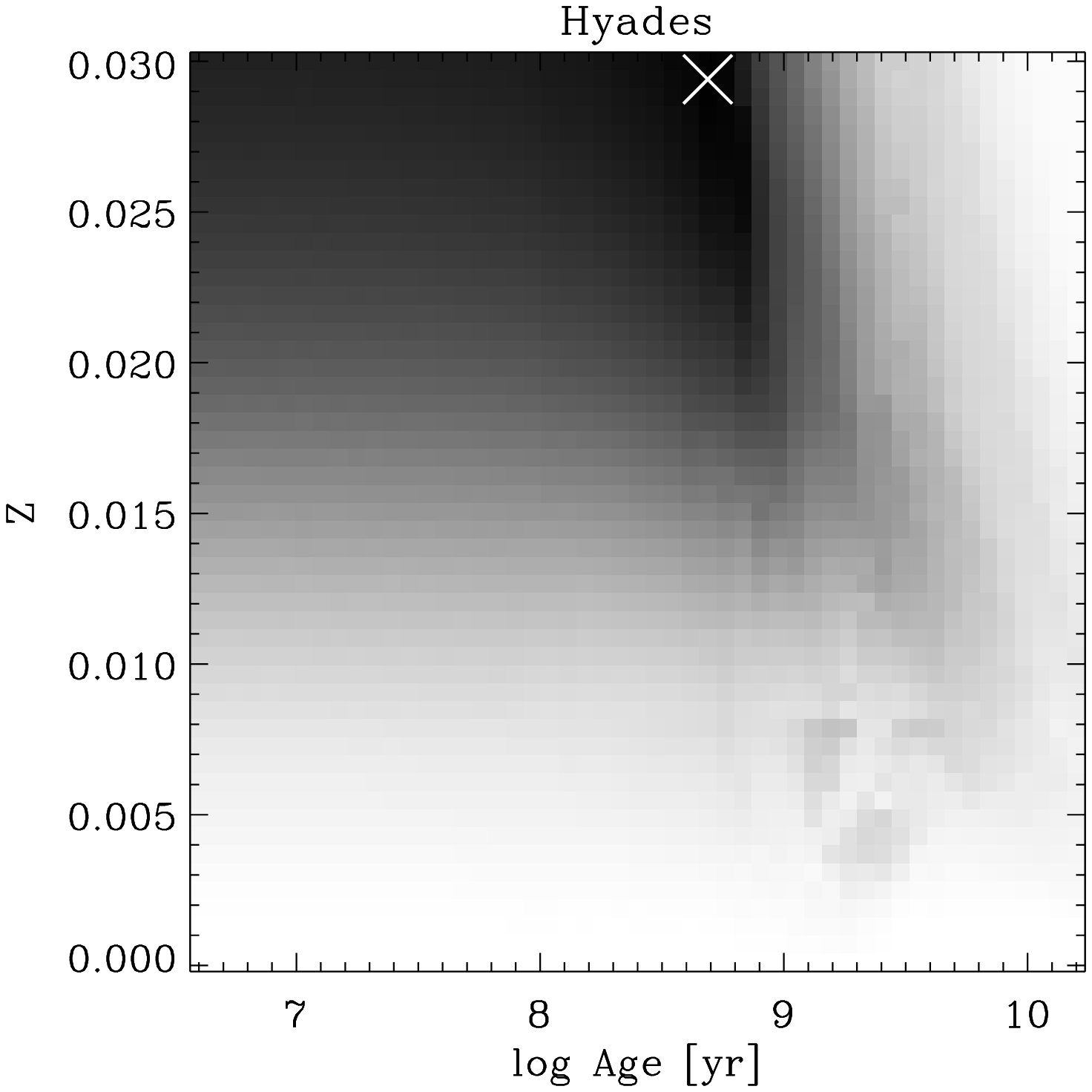}
\includegraphics[width=0.3\textwidth]{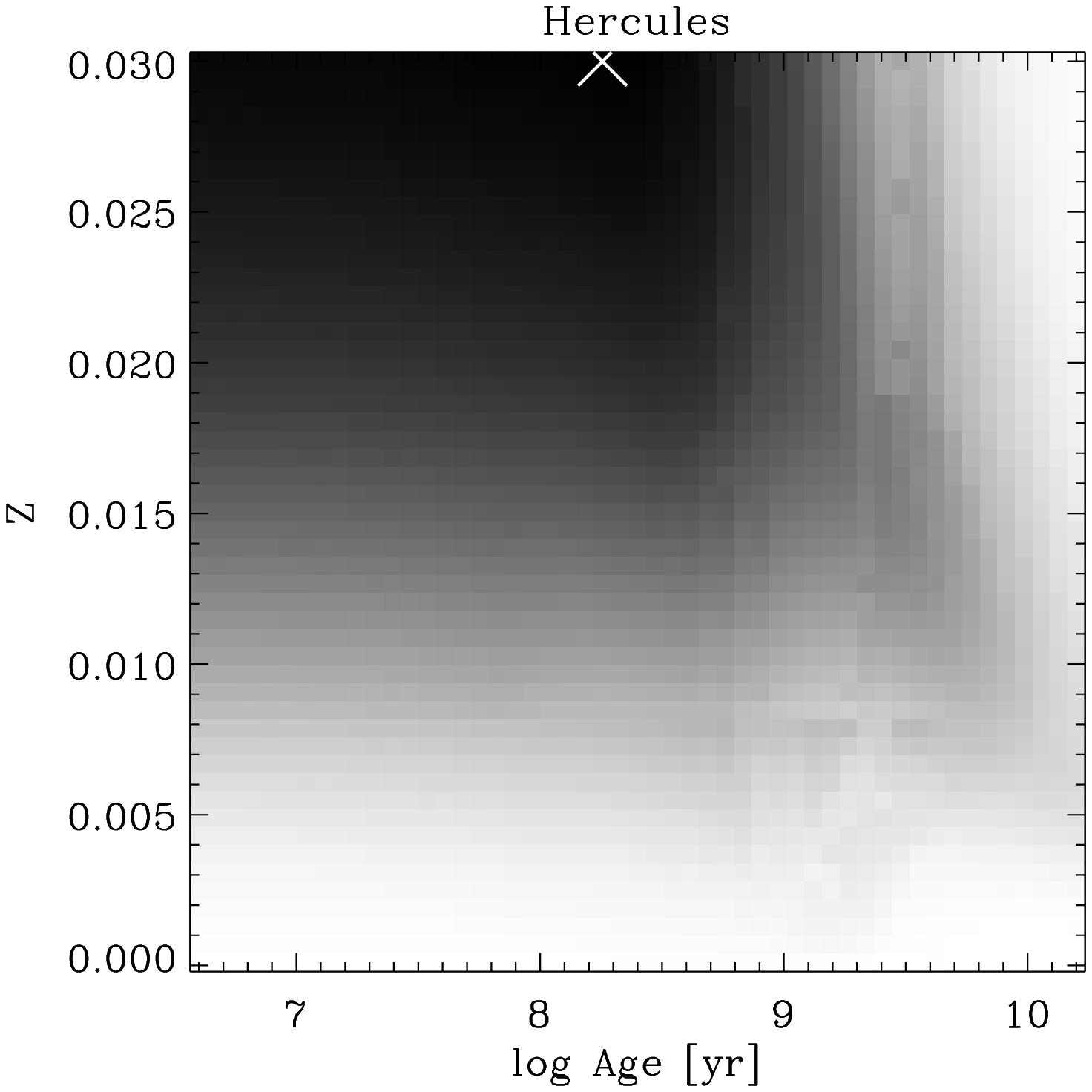}
\caption{Logarithm of the likelihood of different single-burst stellar
  population models characterized by an age and metallicity $Z$ for
  the low-velocity moving groups, with the background contamination
  level $\alpha$ for each group set to the value obtained from a
  global contamination analysis (see the text). The best-fit model
  is indicated by a white cross.}\label{fig:fit_ssp_fixalpha}
\end{center}
\end{figure}

The results of this fit when fixing $\alpha$ are shown in
\figurename~\ref{fig:fit_ssp_fixalpha} for each of the five moving
groups. The logarithm of the expression in
\eqnname~(\ref{eq:maxlikeSSP}) (up to an arbitrary normalization term)
is shown and the best-fit value of age and metallicity is
indicated. These best-fit values are given in
\tablename~\ref{table:bestfitssp} on the first line for each moving
group. For the moving groups with an associated open cluster the
best-fit ages are similar to those of the open clusters, but the
metallicities are very different. This confirms the result from
\sectionname~\ref{sec:firstlook} that the moving groups are not made
up of former open cluster members. The posterior distribution for age
and metallicity is rather broad for all of the moving groups,
indicating that there is no clear preference for a specific age and
metallicity. Given the amount of data on each moving group---each
moving group has a weight of about 10\,percent in the full velocity
distribution, translating into about 1,000 to 1,500 stars in our
sample---this is another indication that the moving groups contain
more than a single stellar population.

\begin{figure}
\begin{center}
\includegraphics[width=0.3\textwidth]{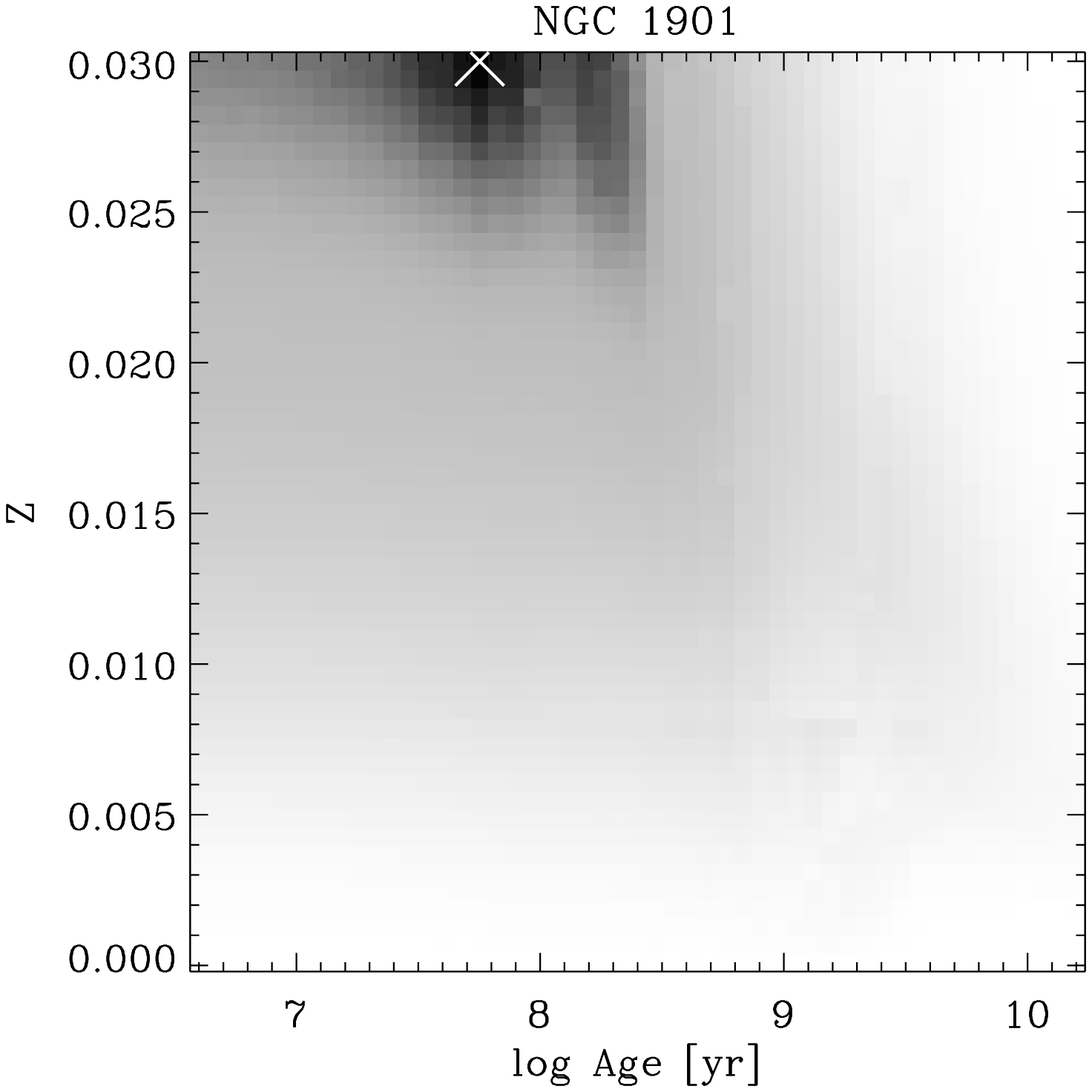}
\includegraphics[width=0.3\textwidth]{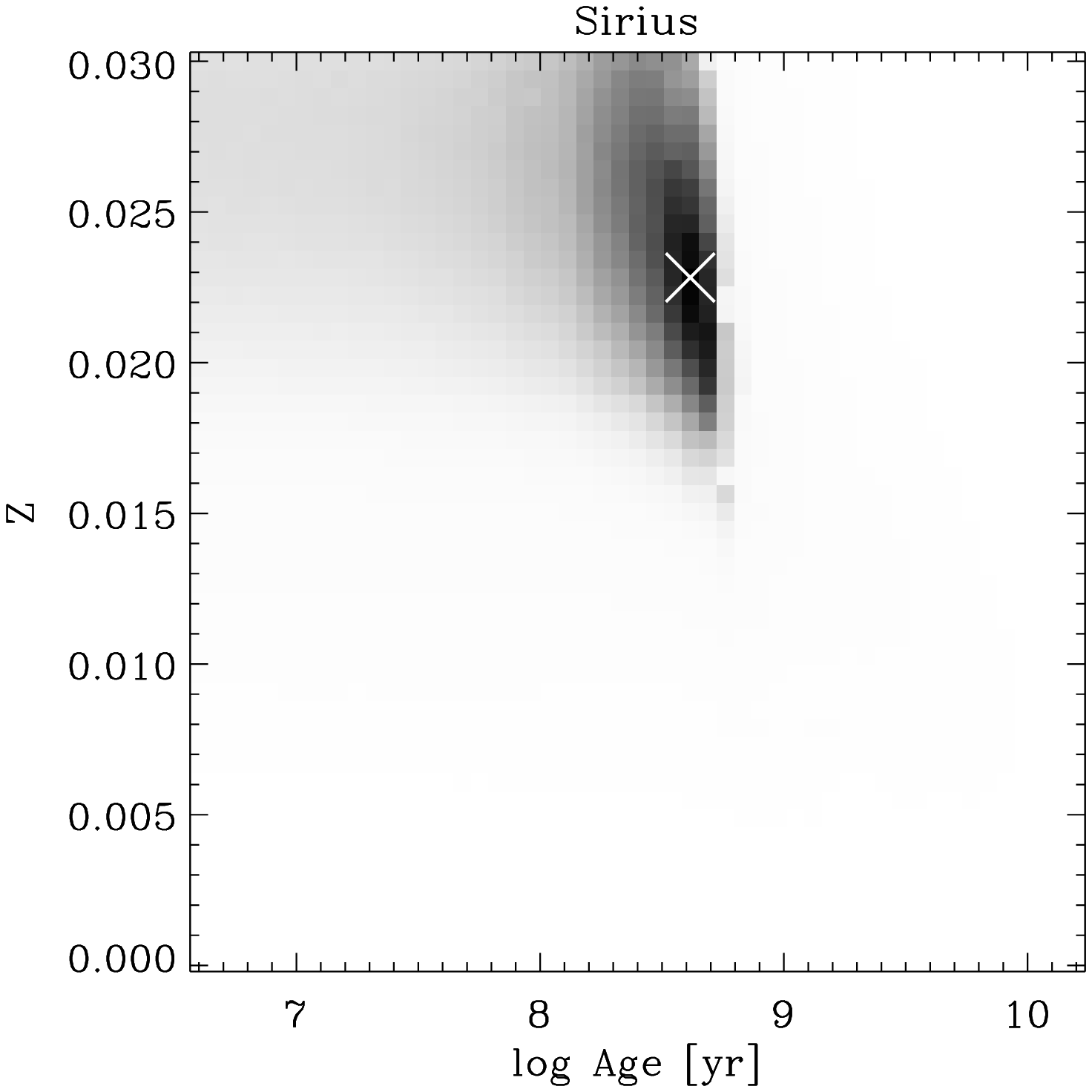}
\includegraphics[width=0.3\textwidth]{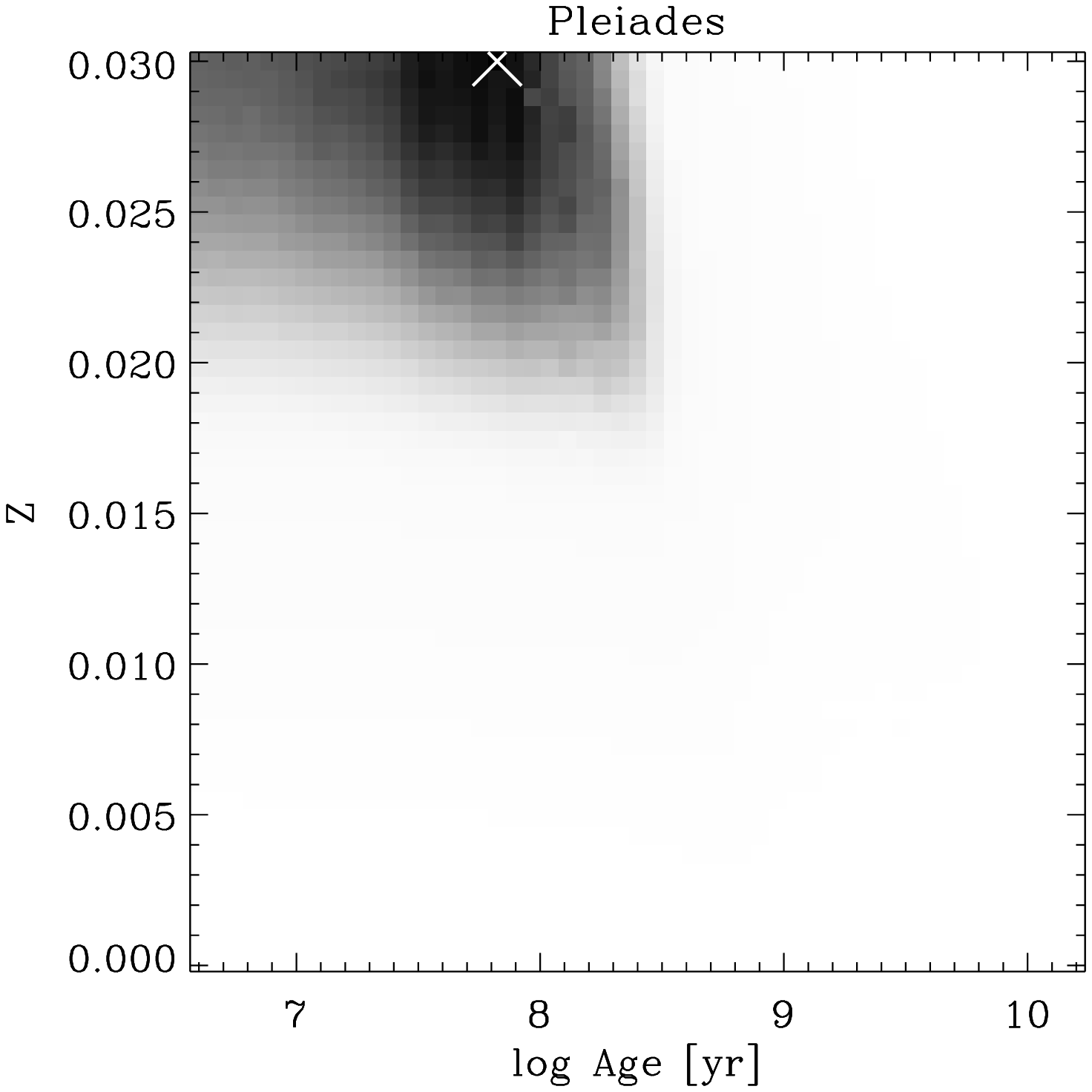}\\
\includegraphics[width=0.3\textwidth]{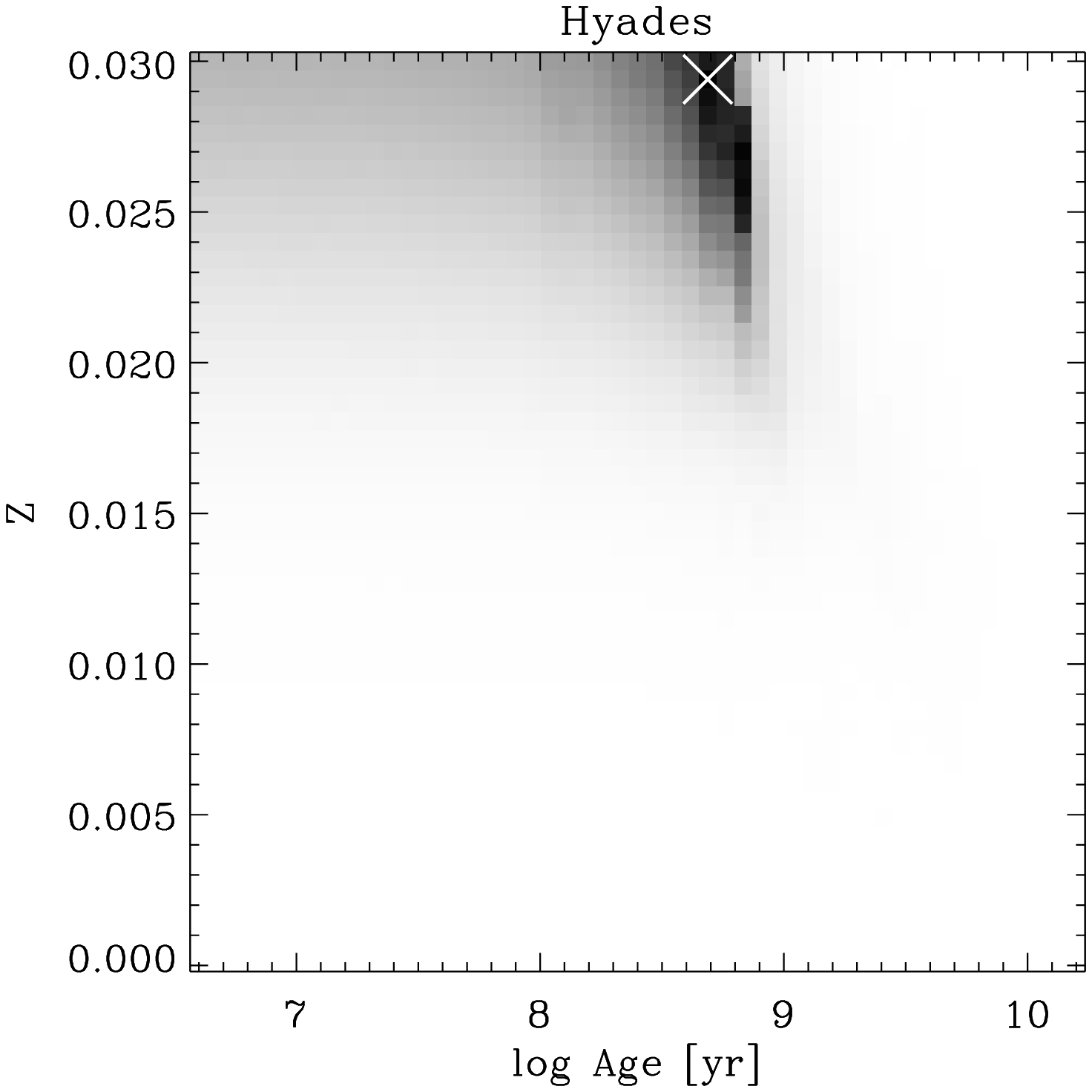}
\includegraphics[width=0.3\textwidth]{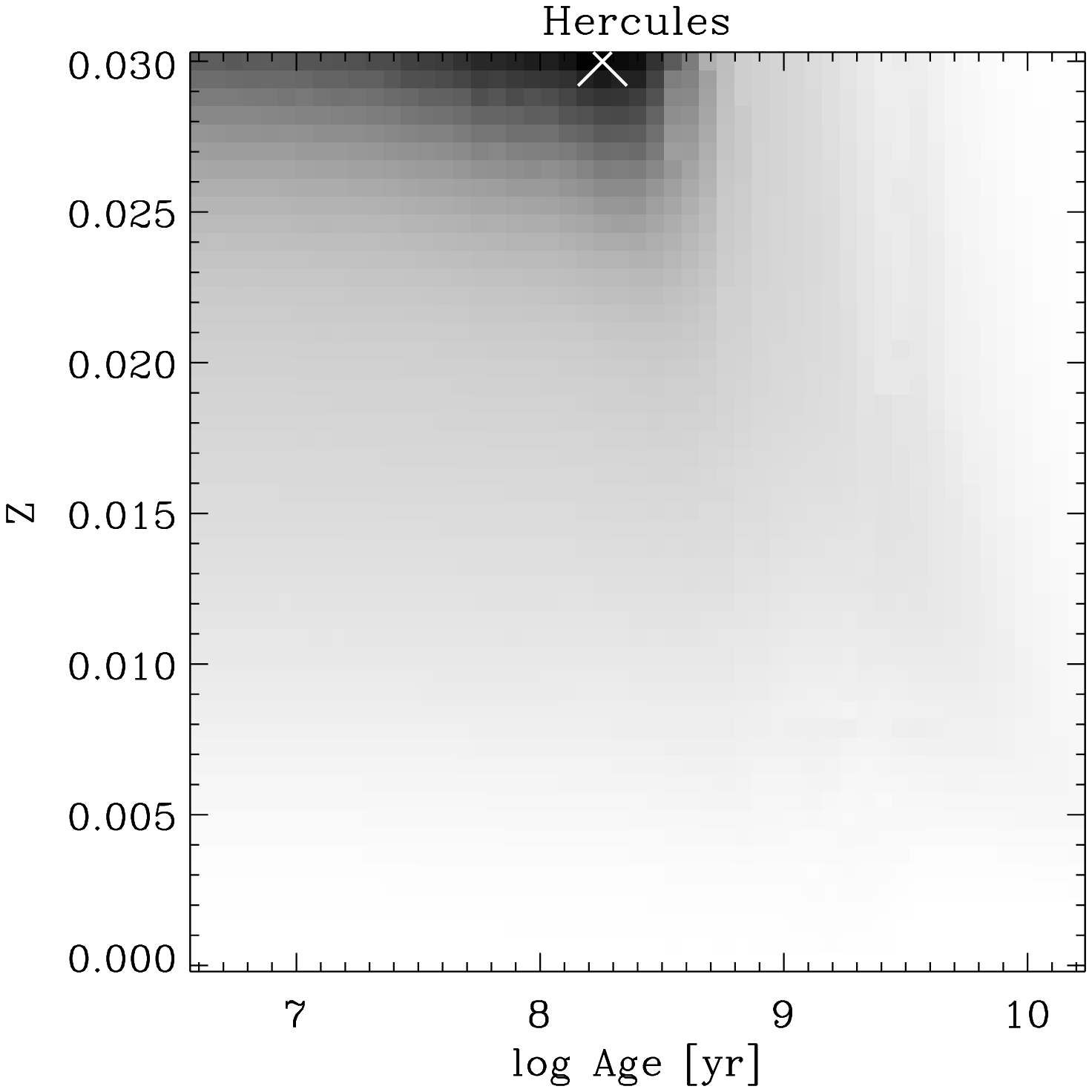}
\caption{Logarithm of the likelihood of different single stellar
population models characterized by an age and metallicity $Z$ for the
low-velocity moving groups, marginalized over the background
contamination level $\alpha$ with a uniform prior on $\alpha$. The
best-fit model is indicated by a white
cross.}\label{fig:fit_ssp}
\end{center}
\end{figure}

\begin{figure}
\begin{center}
\includegraphics[width=0.3\textwidth]{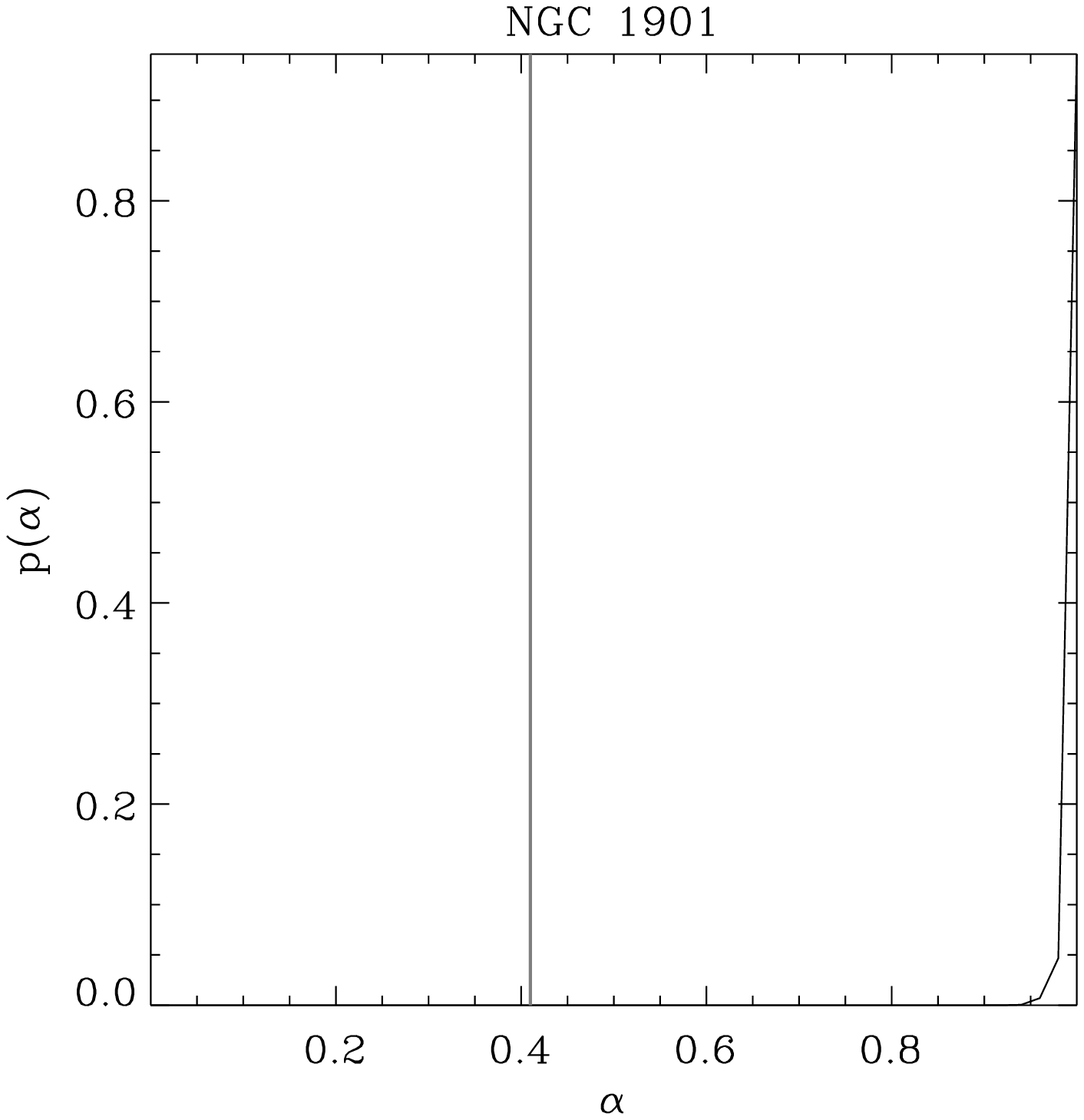}
\includegraphics[width=0.3\textwidth]{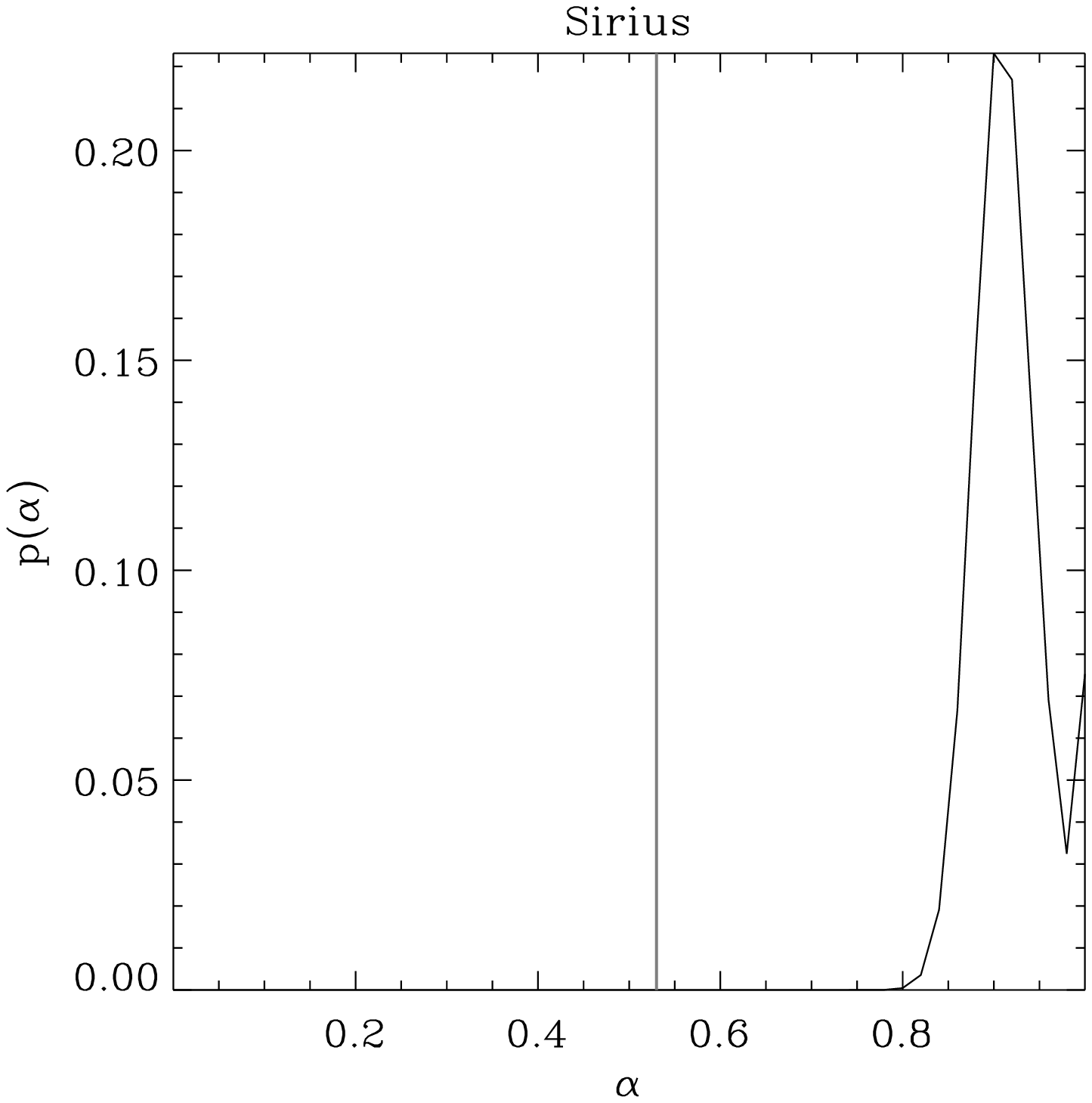}
\includegraphics[width=0.3\textwidth]{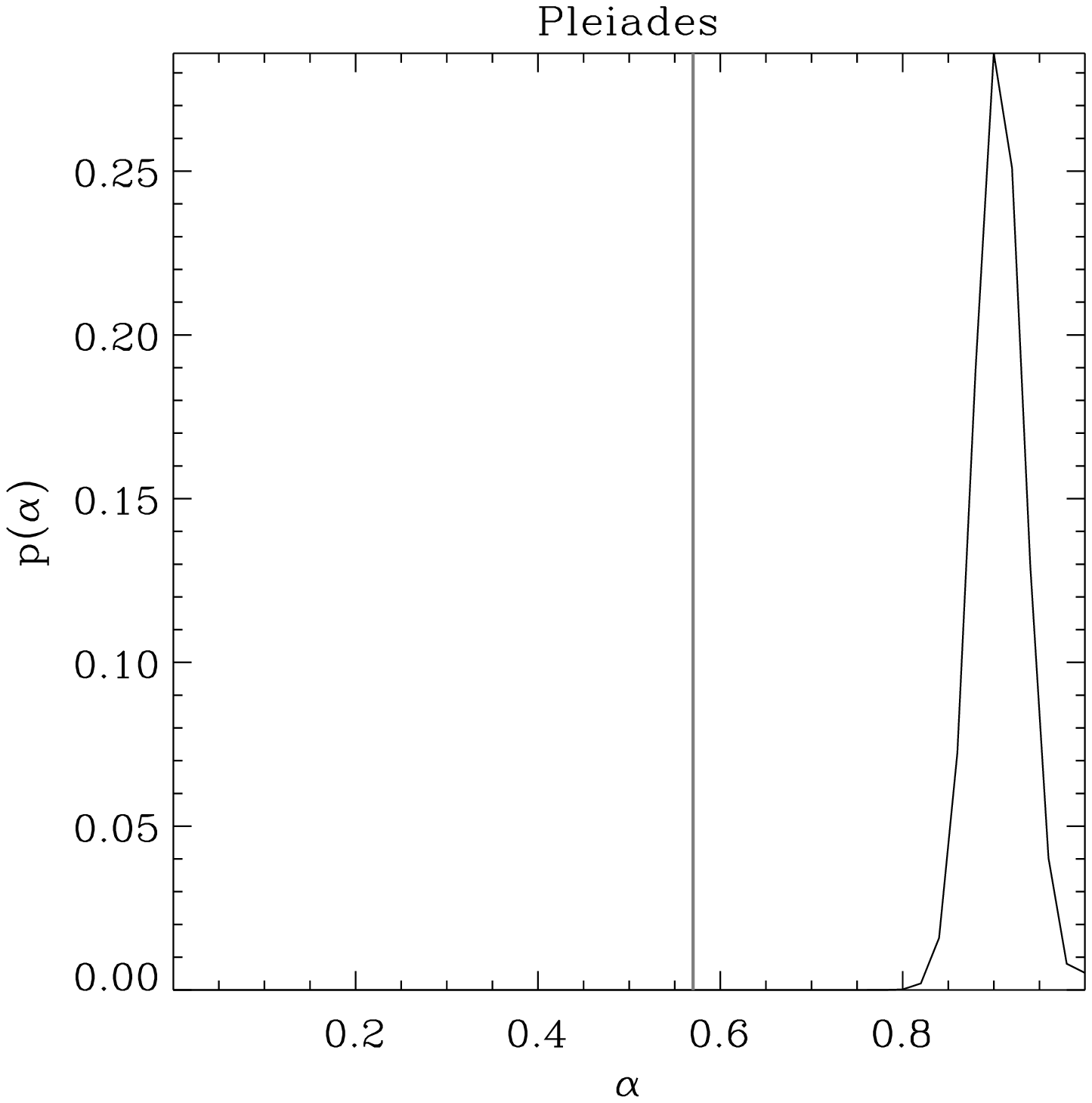}\\
\includegraphics[width=0.3\textwidth]{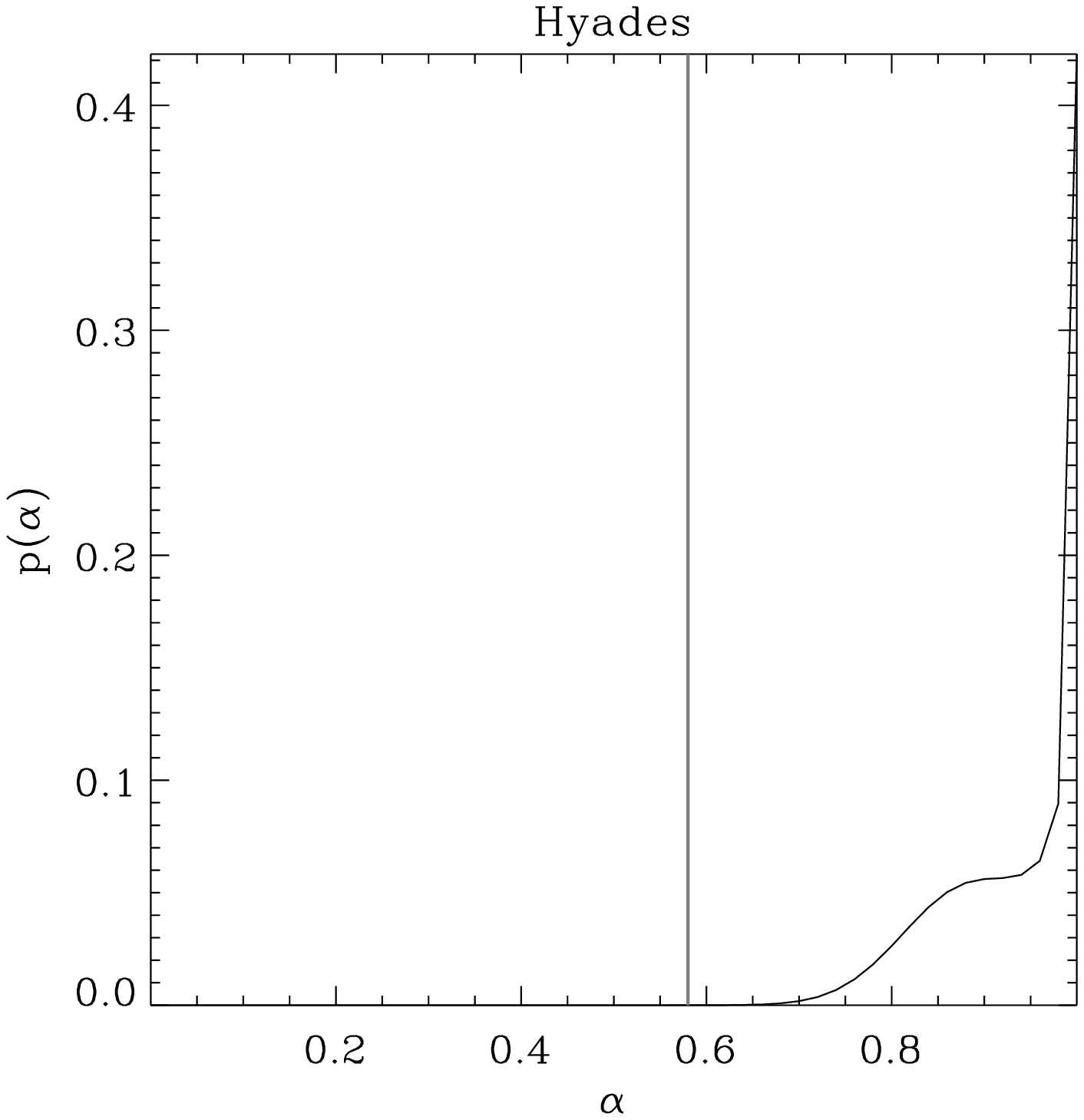}
\includegraphics[width=0.3\textwidth]{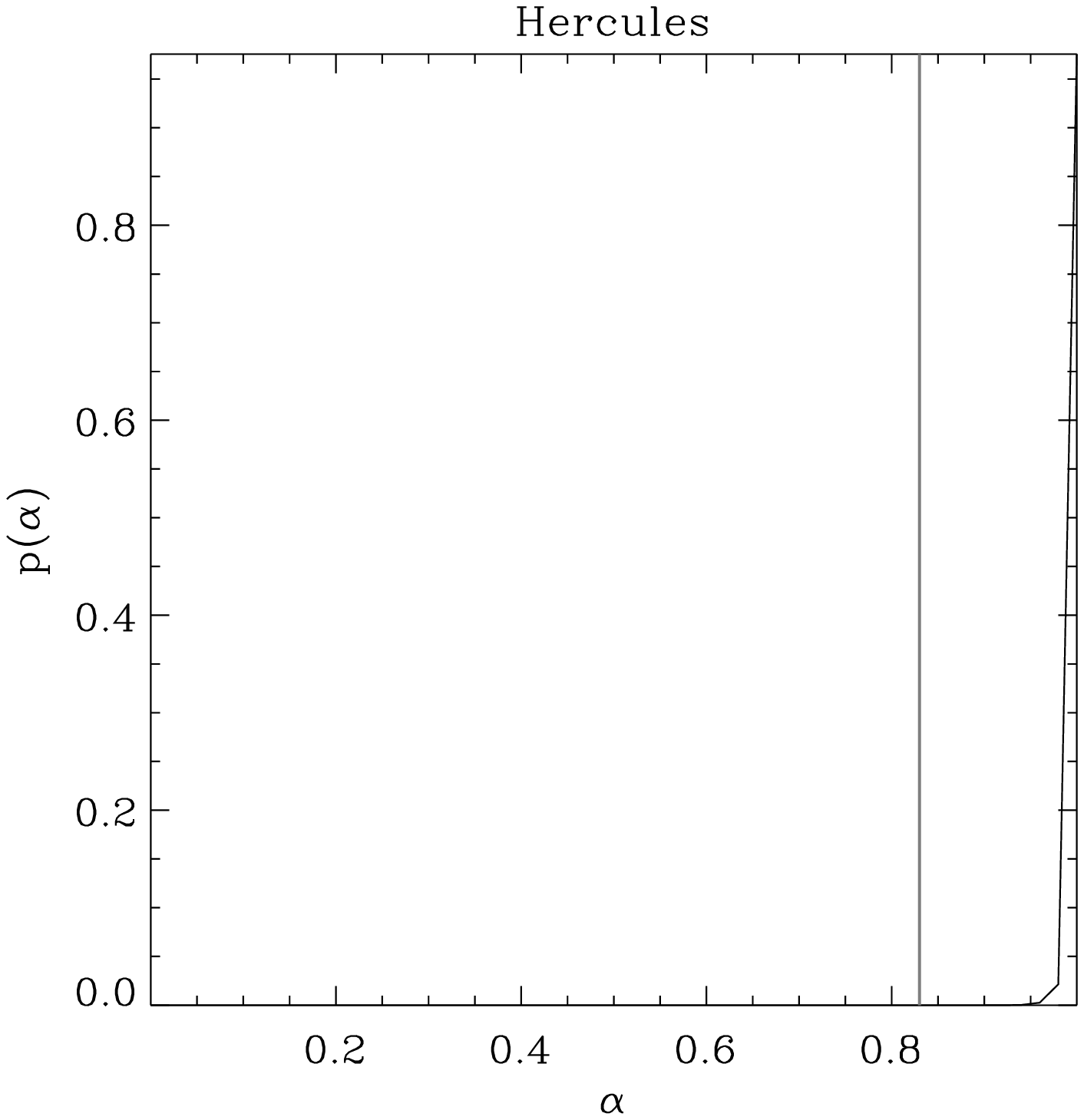}
\caption{Posterior distribution for the background contamination level
$\alpha$ for each of the moving groups, marginalized over age and
metallicity of the foreground model with uniform priors on the
metallicity and the logarithm of the age. Total
contamination---$\alpha=1$---is preferred in most cases. The value of
$\alpha$ obtained from a global contamination analysis---the value
used in \figurename~\ref{fig:fit_ssp_fixalpha}---is indicated by the
vertical line.}\label{fig:fit_ssp_alpha}
\end{center}
\end{figure}

When allowing the background contamination parameter $\alpha$ to be
fit as well, the best-fit ages and metallicities are similar to those
obtained for fixed $\alpha$, but the parameter $\alpha$ is drawn to
values close to unity. The posterior distribution for the age and the
metallicity, marginalized over $\alpha$ using a uniform prior on
$\alpha$, is shown in \figurename~\ref{fig:fit_ssp}; the posterior
distribution for $\alpha$, likewise marginalized over the logarithm of
the age and the metallicity, is shown in
\figurename~\ref{fig:fit_ssp_alpha}. The best-fit values are listed in
\tablename~\ref{table:bestfitssp} on the second line for each moving
group. It is clear from these results that the best-fit level of
background contamination is very high for each of the moving groups;
for Hercules, the best-fit value of $\alpha$ is actually equal to
unity. Especially in the marginalized distributions for $\alpha$---our
degree of belief concerning $\alpha$ given that we believe that part
of the moving group was produced in a single burst of star formation
without caring about the age and metallicity of that event---the peak
of the distribution is at large values of $\alpha$ and even at $\alpha
= 1$ for the NGC 1901 and Hyades moving groups, and in all cases much
higher than the expected level of background contamination indicated
by the vertical line. This tells us that most of each moving group, if
not all of it, is better fit by the background than by any
single-burst stellar population.

Although it is telling that the background contamination level in each
moving group, if left as a free parameter in the fit, is drawn to high
levels of contamination, we will take our hypothesis testing one step
further by examining which of the two hypotheses for each moving
group, \ie, that it is an evaporating cluster or that it is merely a
sparse sampling of the background population of stars, is better at
predicting the properties of an external data set. As this external
data set we use the stars in the \gcsabb\ sample
\citep{Nordstroem04a}, which consists of a subset of the
\Hipparcos\ data set with additional radial velocities. We select
stars that are not suspected to be giants or to be part of a binary in
exactly the same way as described in \sectionname~2.4 in \bhr. At this
point, we only take the radial velocities of this sample of
\ngcsstarsSSP\ stars, consulting the revised \Hipparcos\ catalog
\citep{vanLeeuwen07a} for all of the other properties of these
stars. This sample of stars contains stars that are in the basic
\Hipparcos\ sample that we used before to fit the age and metallicity
of the moving groups and that we used to construct the background
model as well. The trigonometric parallaxes are therefore \emph{not}
an entirely independent sample of parallaxes. But the \gcsabb\ sample
is completely external in the following sense: we can use the radial
velocities from the \gcsabb\ catalog to calculate the membership
probabilities $\pij$ for all of the stars in the \gcsabb\ sample in a
similar way as in \eqnname~(\ref{eq:pij}) but with $\RRi$ now the
projection onto the line-of-sight direction. This way of assigning
membership probabilities is independent of the way we assigned
membership probabilities before, since those were calculated using the
tangential velocities. It is in this sense that the \gcsabb\ data set
is external; in what follows we will determine whether the foreground
model trained on the basic \Hipparcos\ sample using tangential
velocities predicts the properties of the moving group members in the
\gcsabb\ sample, assigned using radial velocities, better than the
background model.

The background model predicts the distribution of the observed
parallax in \eqnname~(\ref{eq:backphotplx}). For the foreground model,
specified by an age, a metallicity, and optionally a value of the
background contamination level, the predicted distribution is given by
\eqnname~(\ref{eq:forebackground}), where the first factor is given by
\eqnname~(\ref{eq:photplxfore}) and the membership probabilities
$\pij$ are calculated using the radial velocities.

\begin{figure}
\begin{center}
\includegraphics{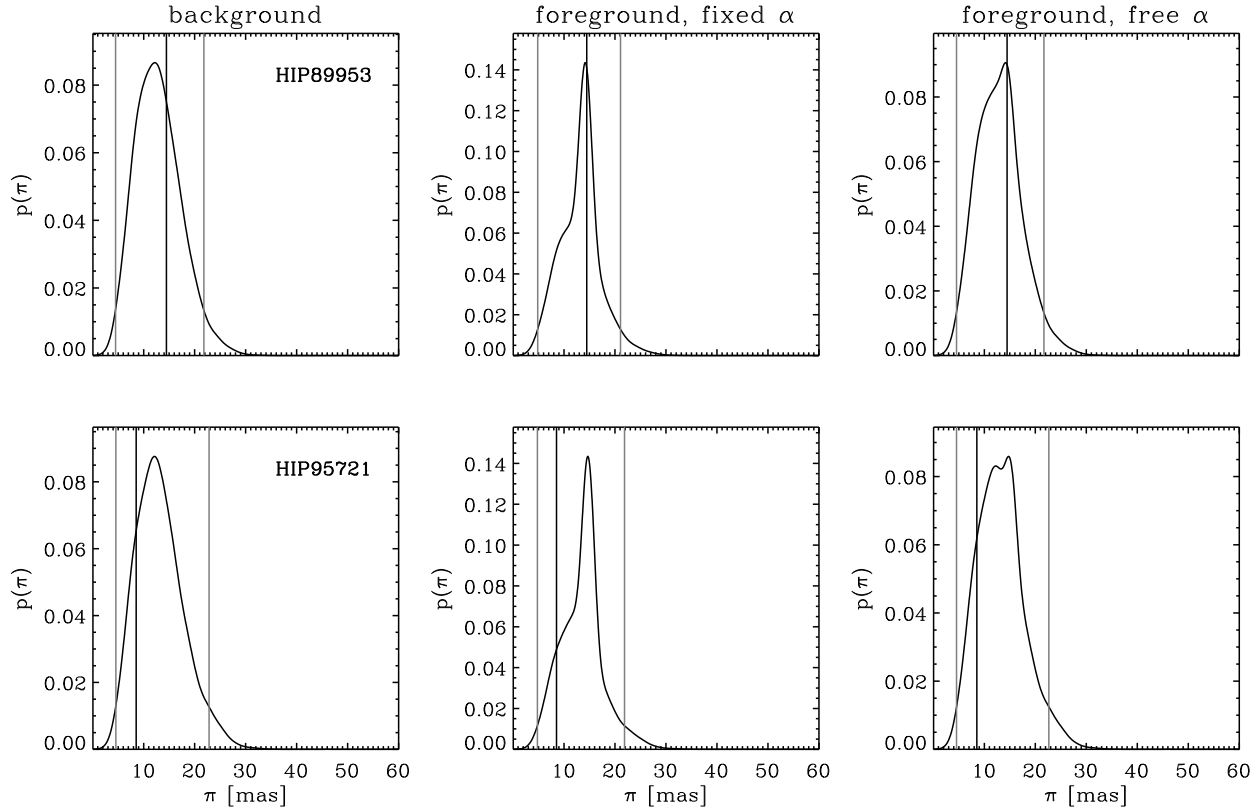}
\caption{Model selection using the \gcsabb\ sample: the background
  model prediction for two individual stellar parallaxes in the
  \gcsabb\ sample is contrasted with the best-fit foreground
  single-burst stellar population model for a fixed value of the
  background contamination $\alpha$ and the best-fit value for
  $\alpha$ for the Sirius moving group (see
  \tablename~\ref{table:bestfitssp} for the details of these best fit
  single-burst stellar populations). The foreground models are trained
  using probabilistic moving-group assignments from the \Hipparcos\
  tangential velocities, while the \gcsabb\ radial velocity is used to
  probabilistically assign the two \gcsabb\ stars featured in this
  figure to moving groups. The top row shows an example where the
  informative foreground prediction does better than the broad
  background model prediction; the bottom row shows an example where
  the narrow foreground prediction is wrong and the uninformative
  background predictions performs better. In each panel the
  probability of the parallax \parallax\ is conditional on the star's
  positional, kinematic, and photometric data (except for the observed
  trigonometric
  parallax).}\label{fig:foreground_background_contrast_example}
\end{center}
\end{figure}

In \figurename~\ref{fig:foreground_background_contrast_example} we
show figures similar in spirit to
\figurename~\ref{fig:model_selection}. For a few specially selected
stars (high probability members of the Sirius moving group) we have
calculated the background prediction for the parallax (left panel in
each row), the foreground prediction when fixing $\alpha$ at the value
determined from the global contamination analysis (middle panel in
each row), and the foreground prediction when fitting the background
contamination (right panel in each row); in making these figures we
chose the best-fit parameters for the Sirius moving group from
\tablename~\ref{table:bestfitssp}. The two stars in this figure are
chosen to illustrate the model selection and do not reflect the
general trend. The first row shows an example where the foreground
model performs well: the foreground model with fixed $\alpha$ makes a
good prediction for the parallax of this star and, by virtue of being
narrow and informative, gives a higher probability to the observed
parallax than the background model---note the difference in scale on
the $y$-axes. The second row shows the much more common situation in
which the foreground model misses completely and the observed parallax
is found in the tails of the predicted distribution; the background
model performs better by virtue of being broader.

We repeat this for all of the stars in the \gcsabb\ sample. We only
consider the \ngcsstarscolorcutSSP\ stars with colors $\bminusv < 1$
mag. We marginalize over the parameters of the foreground model to
compute the foreground probability of each parallax in the
\gcsabb\ sample and from this calculate the total probability of the
parallaxes of stars in the \gcsabb\ sample; The logarithm of this is
given in \tablename~\ref{table:sspmodelselection} for each of the
moving groups. Note that the prior distributions assumed for age,
metallicity, and $\alpha$ do matter now since these provide the
integration measure on the space of properties through which we can
integrate over these properties.

A first thing to note is that the foreground model when fixing
$\alpha$, both using only the best-fit values for the parameters as
when marginalizing over the posterior distribution for the parameters
of the foreground model, predicts the \gcsabb\ parallaxes worse than
the background model, except in the case of the Hercules moving
group. That the Hercules moving group could be considered a
single-burst stellar population is somewhat surprising, as it is
generally regarded as the best established example of a moving group
with a dynamical origin. The preference for the foreground model is
only slight and the fact that, if left free, the background
contamination parameter runs to $\alpha = 1$ is strong evidence
against it being an evaporating cluster. When we let $\alpha$ be a
free parameter, all of the foreground models perform at least as well
as the background model, although only slightly for most groups. There
might be a subsample of stars in each of the moving groups that is the
remnant of a cluster of stars. However, at the best-fit background
contamination levels in \tablename~\ref{table:bestfitssp}, hardly any
stars are assigned to the moving groups when using the relevant hard
cut on membership probability.

In the case of the Hyades, even though the foreground model only
performs as well as the background hypothesis, the best-fit foreground
model is very similar to the Hyades cluster's properties, such that a
subset of stars in the Hyades moving group may have originated from
the open cluster. This is not entirely unexpected, as there must be
\emph{some} stars that have already been lost to the open cluster but
that still share its space motion. However, this fraction is not
simply equal to the difference between the best-fit values of $\alpha$
in \tablename~\ref{table:bestfitssp} and one. In the analysis above,
we did not remove open-cluster members from our sample, and, for
example, 28 of the stars in our sample are confirmed Hyades
members---they are part of the sample from \citet{Perryman98a}
described in \sectionname~\ref{sec:firstlook}. These 28 stars make up
11\,percent of the expected 261 Hyades for this sample---they are all
high-probability members of the Hyades moving group---comparable to
the 14\,percent of non-background found in the best fit to the Hyades
moving group. These 28 stars were selected using a stringent
membership criterion and therefore we can expect the actual number of
Hyades-open-cluster members present in our sample to be even
higher. Thus, we find that only a very insignificant fraction---at
most a few percent---of a moving group can be explained by the
evaporation of a single open cluster, in disagreement with the 15 to
40\,percent, for low- and intermediate-mass stars respectively, found
by \citet{Famaey07a}.

\section{A resonant dynamical origin of the low-velocity moving groups}\label{sec:dynamics}

Now that we have firmly established that none of the moving groups can
be entirely interpreted as being an evaporating open cluster, we can
turn to investigate possible dynamical origins of the moving
groups. If not an evaporating moving cluster, the next a-priori most
likely explanation of the moving groups is that they are generated by
one of the non-axisymmetric perturbations to the Galactic potential,
\eg, by the bar or spiral arms. This is not to say that there are no
other possible explanations of the moving groups' existence---\eg,
projection effects of partially mixed phase-space structure
\citep{Tremaine99a}---but theoretical work has suggested that moving
groups naturally arise in various non-axisymmetric scenarios. As
mentioned in the Introduction, the evidence for the dynamical origin
of the moving groups has been largely circumstantial, amounting to
little more than finding that the moving groups display some variety
of ages and metallicities. The purpose of this section is to test the
hypothesis of a dynamical origin in a more specific, albeit generic,
manner.

We can broadly distinguish between two classes of dynamical origin for
the moving groups: those in which the moving groups are generated
through steady-state non-axisymmetric perturbations, and those in
which they are due to transient perturbations. This section will
mostly test the former category. Steady-state perturbations such as
those associated with the bar or spiral structure are characterized by
a pattern speed, which could potentially vary although this is not the
case in any of the dynamical scenarios considered in the literature so
far. Since orbits have associated natural frequencies---the radial and
azimuthal frequencies in the plane, or the epicycle and angular
frequencies in the epicycle approximation---strong interactions
between the non-axisymmetric perturbations and the stars occur when
these two sets of frequencies are commensurate, that is, when the
difference between the perturbation's frequency and the angular or
azimuthal frequency of the orbit is commensurate with the radial
frequency. This gives rise to the co-rotation resonance, where the
period of the perturbation is equal to the angular period of the
orbit, and the Lindblad resonances, which are associated with closed
orbits in the rotating frame of the perturbation that do not cross
themselves \citep[\eg,][]{binneytremaine}. The influence of a weak
non-axisymmetric perturbation to the overall potential is therefore
most strongly felt at these resonances
\citep[\eg,][]{LyndenBell72a}. If the moving groups' origin lies in
steady non-axisymmetric perturbations, we would expect the Sun's
present location to be near one of the resonances of the
non-axisymmetric structure to account for its strong influence on the
velocity distribution.

Simulations confirm this basic picture. Several simulations have shown
that moving-group-like structures form near the resonances associated
with the bar \citep[\eg,][]{Dehnen01a,Fux01a} or spiral structure
\citep{Quillen05a} or at the overlap of resonances of these two
\citep{Quillen03a}. Even though spiral arms are observed to start near
the end of the bar in many galaxies, the pattern speeds of these two
features are probably not strongly related in a dynamical sense, \ie,
their resonances would generically be independent of each other
because their pattern speeds are in general very different
\citep{Sellwood88a}. Note that in order to explain the Hercules moving
group as being due to the outer Lindblad resonance (OLR) of the bar
and the lower velocity moving groups as being due to resonances of the
spiral structure, the Sun would have to be in a rather special
location in the Galaxy to be at exactly the right spot with respect to
all of these. More observational evidence for either of these
scenarios is thus needed to check that the velocity distribution is
not being overfit.

It is instructive to see what happens to the orbits of stars that are
near one of the resonances to understand what properties we expect the
moving-group members to have if they are associated with a resonance
of dynamical origin. Generically, in the neighborhood of a resonance
we expect to see a bifurcation of the orbits into two families
\citep{Contopoulos75a,Weinberg94a,Kalnajs91a}. This bifurcation could
be such that one of the families is on nearly circular orbits and the
other significantly lagging with respect to the local standard of
rest, as is the case near the OLR of the bar \citep{Dehnen01a}, or it
could be such that there are no stars on nearly circular orbits any
longer and one family lags with respect to the local standard of rest,
while the other family moves faster than purely circular location (in
both cases, this is at the present location of the Sun in the
successful dynamical scenarios). At azimuths where these two families
cross, we expect to see streams in pairs in the velocity distribution.

Under the hypothesis of an OLR origin for the Hercules moving group,
there is a family of orbits that are anti-aligned with the bar and
spend most of their time inside the OLR, and there is a family of
orbits that are aligned with the bar and spend most of its time
outside the OLR \citep{Contopoulos89a}. When invoking steady-state
spiral arms to explain the existence of the Hyades/Pleiades moving
groups and the Coma Berenices (which we do not study in this paper
because it did not show up at high significance in the reconstruction
of the velocity distribution in \bhr) or Sirius moving group, the
Sun's present location is near the 4:1 ILR and there is one family of
orbits that is elongated, square-shaped, and that spends most of the
orbit outside of the ILR, and another family of orbits that is also
elongated, square-shaped, and is more typically found inside the ILR
\citep{Contopoulos86a}. Varying the parameters of the spiral
structure, moving-group-like structure also forms for other types of
orbits, but generically one family of orbits' mean radius is inside
the resonance and the other family's mean radius is outside the
resonance. This is even somewhat the case when the moving groups are
created by transient behavior of the bar---\eg, recent bar growth
\citep{Minchev09a}---although the situation is a lot messier in these
cases because of the time-dependent nature of the problem.

Thus, if the moving groups are particular manifestations of dynamics
near the resonances of the bar or spiral structure, then the previous
argument shows that the orbits of stars in the moving groups concern
different and mostly non-overlapping regions of the Galaxy: Stars that
are part of moving groups that on average lag the local circular
motion are near their apogalacticon, so their orbits will be mostly
confined to the inner Galaxy. Stars in moving groups that are ahead of
circular motion on average are near their perigalacticon, so these
stars spend most of their orbits in the outer Galaxy. Therefore, we
can expect the stellar populations of moving groups to be different
depending on their position in the $\vx-\vy$ plane, as these
populations of stars are born in different physical conditions. It is
this hypothesis that we will test in this section.

Specifically, the hypothesis we set out to test is the following. If
moving groups are associated with a family of orbits that either spend
most of the orbit inside of the Solar circle or outside of it, then,
since stars reflect the conditions of the regions in which they are
born, stars in moving groups will either have a higher than average
metallicity, or lower than average metallicity, because there is a
metallicity gradient in the Galaxy, declining outward from the
Galactic Center
\citep[\eg,][]{Shaver83a,Afflerbach97a,Nordstroem04a,Rudolph06a}. For
each moving group we can estimate this expected metallicity difference
by calculating the mean metallicity at the mean radius of each moving
group if it was moving in a simple axisymmetric potential. We can
approximate this mean radius by the radius of the circular orbit that
has the same angular momentum as the center of each moving group. The
mean radii found by assuming a flat rotation curve with a circular
velocity of 235 km s$^{-1}$ and a distance to the Galactic center of
8.2 kpc \citep{Bovy09c} are listed in
\tablename~\ref{table:metalmodelselection}. Assuming a metallicity
gradient in [Fe/H] of -0.1 dex kpc$^{-1}$
\citep[\eg,][]{Mayor76a,Nordstroem04a} these mean radii translate in
the expected metallicity differences of a few hundreds of a dex to a
tenth of a dex for Hercules in
\tablename~\ref{table:metalmodelselection}. In reality, we can expect
these differences to be larger, since the resonance-trapped orbits
will make larger excursions inward or outward than in this simple
axisymmetric argument.

There will be some spread around this mean value, but this spread will
certainly be smaller than the width of the full thin-disk metallicity
distribution, which is, as we will show below, about 0.14 dex. We can
estimate the expected width of a moving group's metallicity
distribution based on its velocity width by using the same procedure
that we used in the last paragraph. Using the velocity widths from
\bhr\, we find expected widths of a few hundreds of a dex or less;
these are given in \tablename~\ref{table:metalmodelselection}. These
expected widths are smaller than the expected metallicity offset for
each moving group except for NGC 1901. Therefore, we expect each
moving groups' metallicity distribution to be largely contained in
either the higher than average or lower than average part of the local
thin disk metallicity distribution and this effect should be
detectable.

Thus, we ask for each of the moving groups whether it is better fit by
a model with a higher or lower metallicity than the background model,
which reflects the full metallicity distribution in the Solar
neighborhood. Since the \Hipparcos\ sample that we have been using
throughout this paper does not include spectroscopic metallicity
information, we use a sample selected from the \gcsabb\ catalog
instead. We use less conservative cuts on the binary and giant
contamination of this sample to maximize the number of stars in the
sample. Giant contamination is in fact very small in this sample of F
and G dwarfs, and the presence of binaries is not really an issue
since the multiple radial velocity epochs available for all stars in
the \gcsabb\ data allow for an accurate determination of the mean
motion, although the photometric parallax technique that we use will
be slightly biased by the presence of unresolved binaries. This
affects both background and foreground models---foreground models in
this section are those with low/high metallicities---and is taken care
of in the non-parametric photometric parallax relation that we will
again establish (all models effectively use a noise-in, noise-out
approach as far as unresolved binaries are concerned).

As before we will train a non-parametric model to represent the
background hypothesis, a non-parametric photometric parallax relation
that we will establish for the \gcsabb\ stars in exactly the same way
that we trained the background model in
\sectionname~\ref{sec:background}. We cannot re-use the previous
background model, as the \gcsabb\ data are a much finer sampling in a
narrow color range of the rich color--magnitude diagram than our
previous \Hipparcos\ sample. Rather than using a parametric model for
the foreground hypothesis, we will build a non-parametric model
similar to the background model, by training it on stars that have
higher or lower metallicity than average.

\begin{figure}
\includegraphics{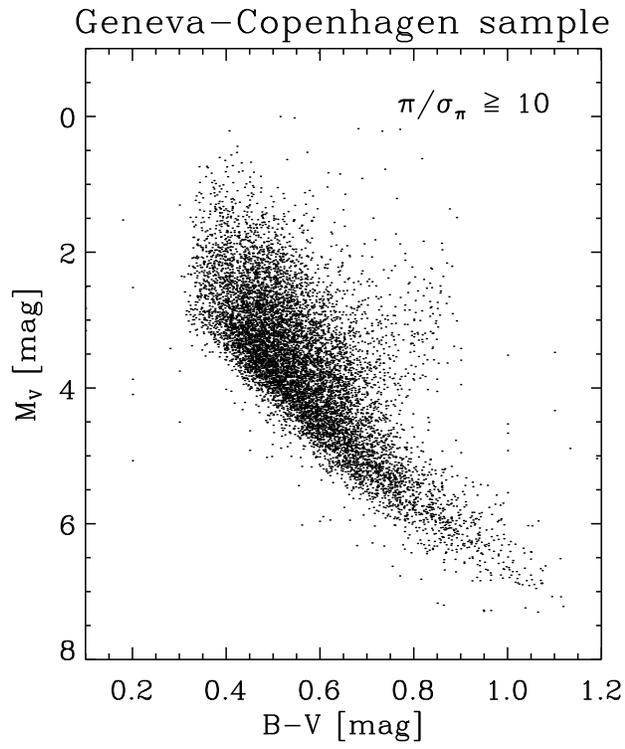}
\caption{Color--magnitude diagram of the magnitude-limited
  \gcsabb\ sample of \ngcsstars\ stars used in
  \sectionname\sectionname~\ref{sec:dynamics} and \ref{sec:sellwood}
  with relative parallax uncertainties $\lesssim$
  10\,percent.}\label{fig:gcs_cmd}
\end{figure}

We construct the \gcsabb\ sample used in this section as follows: From
the \gcsabb\ catalog we take all of the stars that have a
\Hipparcos\ counterpart and take their radial velocities with
uncertainties and their metallicities from the \gcsabb\ catalog
\citep[the latest reduction;][]{Holmberg07a,Holmberg09a}. We take the
rest of the spatial, kinematic, and photometric data from the
\Hipparcos\ catalog \citep{ESA97a,vanLeeuwen07a}. From this sample we
select those stars with accurate parallaxes
($\parallax/\sigma_\parallax \geq 10$); this leaves a sample of
\ngcsstars\ stars. The color--magnitude diagram of these stars is
shown in \figurename~\ref{fig:gcs_cmd}.

\begin{figure}
\includegraphics{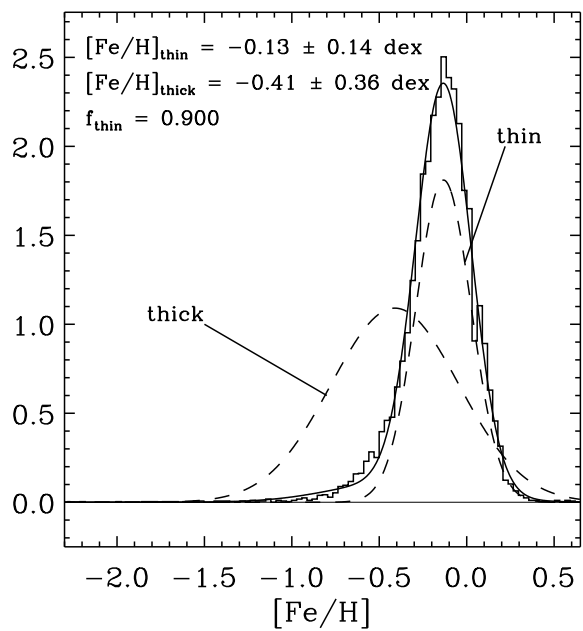}
\caption{Metallicity distribution in the Solar neighborhood: the
  distribution of metallicities of \ngcsstars\ in the
  \gcsabb\ sample. The best-fit two-Gaussian decomposition is
  overlaid: the two components as the dashed lines (the ``thin'' disk
  component has been scaled down for clarity) and the resulting
  distribution as the full line. The parameters for the best fit
  two-Gaussian distribution are given in the top-left corner as mean
  $\pm$ standard deviation of each component.}\label{fig:gcs_z}
\end{figure}

The distribution of metallicities of this sample is shown in
\figurename~\ref{fig:gcs_z}. Instead of taking a straight average of
the metallicities, we fit it as a mixture of two components,
anticipating a sizeable contribution from thick disk stars, which
could skew the inferred average thin-disk metallicity. We perform this
fit using the same deconvolution algorithm that we used to deconvolve
the velocity distribution \citep{Bovy09b}, assuming an uncertainty of
0.08 dex on the \gcsabb\ [Fe/H] values. The resulting components are
shown as the dashed curves in \figurename~\ref{fig:gcs_z}, with
arbitrary normalizations for display purposes. The best-fit
parameters---mean and width of the two component Gaussians---are given
in the top-left corner of the figure and the amplitude of the largest
component is given as well. We can identify these two components with
the thin and thick disks. The average thin disk metallicity is -0.13
dex with a spread of 0.14 dex. Note that because of the large amount
of data, this mean thin disk metallicity can be considered to be very
well determined: An estimate of the uncertainty in the mean is
$\sigma/\sqrt{f_{\mathrm{thin}}\,N}$, where $\sigma=0.16$ dex is the
width of the thin disk distribution convolved with the typical
uncertainty in the \gcsabb\ [Fe/H] values, $N=$\ngcsstars, and
$f_{\mathrm{thin}}=0.9$; the uncertainty in the mean is therefore
about 0.002 dex. Of course, this uncertainty does not include the
uncertainty in the thin-thick disk decomposition, but this is expected
to be small.

To get a first sense of the metallicities of the various moving
groups, we have computed the average metallicity of the stars in the
\gcsabb\ sample described in the previous section (before the color
cut, but the results are the same after the color cut) by weighting
the individual metallicities by the probability that the star is part
of the moving group in question, \ie,
\begin{equation}\label{eq:avgmetal}
\langle[\mathrm{Fe/H}]\rangle_j = \frac{\sum_i
  \pij\,[\mathrm{Fe/H}]_i}{\sum_i \pij}\,.
\end{equation}
In the same way we can calculate the second moment of the metallicity
distribution of each moving group.  These average metallicities and
widths are given in \tablename~\ref{table:metalmodelselection}. All of
the moving groups except for Sirius have higher metallicities than the
average thin disk metallicity, which we established above to be -0.13
dex. The Hyades moving group has a distinctively higher metallicity
than average \citep[see also][who find about the same value from a
simple cut in velocity space]{Famaey07a}; for the other moving groups
the difference is smaller and it is not clear what the significance of
this result is. The fact that the second moment of each moving group's
metallicity distribution is comparable to that of the full local
metallicity distribution indicates that the moving groups' metallicity
distributions are all very similar to that of the background. To test
the significance of the non-zero offsets from the average metallicity,
we perform a simple hypothesis test to see whether the moving groups'
metallicities are significantly different from that of the general
thin disk population.

We create two subsamples from the full sample of \ngcsstars\ stars by
taking stars with metallicities larger than the average thin disk
metallicity, and stars with metallicities lower than the average;
these samples contain \nphotgcscolorcuthighz\ and
\nphotgcscolorcutlowz\ stars, respectively. That the latter suffer
from some contamination from thick disk stars does not matter for our
purposes as we are merely interested in creating a model with lower
metallicities than the average thin disk.

We now fit a non-parametric photometric parallax relation to each of
these samples---the background model consisting of all of the stars
and the two foreground models consisting only of the low/high
metallicity subsamples of the full sample---in exactly the same way as
in \sectionname~\ref{sec:background}. In order to avoid an excessively
spiky non-parametric model, we focus on the color region 0.35 mag $<$
\bminusv\ $<$ 0.95 mag in the color--magnitude diagram (see
\figurename~\ref{fig:gcs_cmd}). This cut only removes a very small
number of stars (245 out of \ngcsstars), but makes sure that the
optimal smoothing scale is not unduly affected by the sparse sampling
of certain color regions. The width parameter of the Tricube kernel is
again set by leave-one-out cross validation.

As before, we can now calculate the total probability of the
moving-group stars' parallaxes under the assumption that they are
merely background stars, or under the assumption that they have higher
than average or lower than average metallicities. The probability of
an observed, trigonometric parallax based on the star's color,
apparent magnitude and the full \gcsabb\ training sample is again
given by the expression in \eqnname~(\ref{eq:backphotplx}), where the
sum is now over stars in the \gcsabb\ sample (with the color cut
discussed above). The probability of the observed parallax under the
foreground model is again a mix between that of assuming that the star
has higher/lower probability by virtue of being part of the moving
group, and that of the background model, since we can only
probabilistically assign membership. Thus, for each star the
foreground probability is given by
\begin{equation}
p(\parallax_{\mathrm{obs},i} | \mathrm{foreground}) = p(\parallax_{\mathrm{obs},i} | \mathrm{high/low} Z)^{\pij}\,p(\parallax_{\mathrm{obs},i} | \mathrm{background})^{1-\pij}\,,
\end{equation}
where we now make use of the full kinematical information for the
\gcsabb\ stars, since we have all three components of the velocity to
assign moving-group membership for this sample. The logarithm of the
total probability under both foreground models and the background
model thus calculated is tabulated in
\tablename~\ref{table:metalmodelselection}. If a moving group shows
clear signs of a higher or lower metallicity, and thus of a resonant
origin in a steady-state non-axisymmetric potential, we would expect
the moving group's properties to be better fit by the higher/lower
metallicity subsample than by the background model. As is clear from
\tablename~\ref{table:metalmodelselection}, no moving group shows
convincing evidence that this is the case.

The only moving group that shows weak evidence that it has a different
metallicity than the background of Solar-neighborhood stars is the
Hyades moving group, confirming the result for the Hyades found above
by calculating a weighted average of the metallicities of Hyades
members. That the Hyades moving group has a slight preference for a
higher metallicity could indicate that it is associated with a family
of orbits whose mean radii are within the Solar circle, although the
evidence is very weak. This may seem like a large factor, but one
needs to consider that this is for a sample $\alpha_{\mathrm{Hyades}}
N = 0.017 \times \nphotgcscolorcut \approx 159$ stars. Nevertheless,
since the two competing models are qualitatively similar, we conclude
that there is some weak evidence here that the Hyades is made up of
stars with higher than average metallicity.

The moving group which has the lowest likelihood of having either
higher or lower metallicity than average is the NGC 1901 moving
group. This is hardly surprising. The NGC 1901 moving group sits right
on top of the bulk of the thin disk velocity distribution and it is
therefore very hard to identify its members; its weight in the mixture
of Gaussian decomposition of the velocity distribution is also rather
large for a moving group, such that it was suggested in \bhr\ that a
large part of this Gaussian component might simply be part of the
background distribution. The analysis here confirms this intuition.

As for the other moving groups, they are all best-fit by the
background model, but we can nevertheless ask which of the two
foreground models is preferred (ignoring the background model). The
Sirius moving group prefers the low-metallicity foreground model over
the high-metallicity model, confirming what we found above. Taken
together with the result that the Hyades moving group has higher than
average metallicity, this could be interpreted as tentative evidence
in favor of the scenario in which these two moving groups arise
through spiral perturbations near the ILR (as discussed above;
\citealt{Quillen05a}). The evidence in favor of this explanation is
not strong, but given the difficulty with which group membership is
established through kinematic association, it may be compelling.

As for the Pleiades moving group, it has often been assumed that its
origin is strongly linked to that of the Hyades moving group, since
some reconstructions of the velocity distribution did not resolve the
difference between the moving groups \citep{Famaey05a} and because
they are part of the same branch in the Galactic-plane part of the
velocity distribution \citep{Skuljan99a}. Although in the metallicity
test performed here the Pleiades moving group is best fit by the
background model, the runner-up is the foreground model with lower
metallicity than average, as opposed to the best-fitting model for the
Hyades, which has higher than average metallicity---note that this
result is not confirmed by the calculation of the average
metallicities above. This argues against a common origin of the
Pleiades and Hyades moving groups. Recent bar growth has recently been
proposed as a scenario in which the Hyades and Pleiades have a
different origin \citep{Minchev09a}. Because of the transient nature
of the effect of bar growth on the local velocity distribution, it is
quite possible that the stars that make up the Pleiades moving group
do not have a preference for a single metallicity at the present
epoch.

Finally, we observe that the Hercules moving group, long thought to be
a signature of the OLR of the bar and the moving group with the
largest and most significant expected metallicity anomaly, is not
preferentially fit by a model with higher metallicity than average, as
would be the case if the resonant origin were correct, but is instead
better fit by the background model. The background model is preferred
by quite a large margin. Focusing only on the preferred foreground
model, however, we see that the higher metallicity model is strongly
preferred over the lower metallicity model. If one prefers to think
that the overall preference for the background model is due to the
difficulty of assigning group membership, or perhaps due to the slight
offset between the Gaussian component identified in the reconstruction
of the velocity distribution used here and other reconstructions of
the velocity distribution, then this strong preference for the higher
metallicity model over the lower metallicity model could be taken as
evidence for the resonant origin of the Hercules moving group.

Because of the many assumptions and simplifications made in estimating
the expected metallicity distribution of the various moving groups,
the test we performed in this section is largely
qualitative. Nevertheless, the fact that most of the moving groups are
best fit by the background model as opposed to the higher or lower
metallicity models challenges the explanation that the moving groups
are associated with resonances related to the bar and/or spiral
structure. If the moving groups nevertheless have a dynamical origin,
then the dynamical effect is probably transient and less cleanly
described in terms of supporting orbits. The predictions of the
transient spiral or bar models in which moving groups arise in the
literature \citep[\eg,][]{deSimone04a,Minchev09a} do not contain very
definite descriptions of the expected stellar content of the moving
groups generated through transient perturbations. Therefore, at this
time it is hard to say whether these models are preferred by the data,
but they do gain in likelihood if only because of the relative drop in
likelihood of the resonant models due to their not being strongly
supported by the data here.

\section{Hints of recurrent spiral structure}\label{sec:sellwood}

If the solar neighborhood is currently near the ILR of the current
cycle in the recurrent spiral-structure scenario of
\citet{Sellwood91a} scenario described in
\sectionname~\ref{sec:intro}, we would expect to see a feature in the
local energy-angular-momentum distribution corresponding to stars
being scattered at the ILR. Some tentative signs of this have been
detected in the distribution of \Hipparcos\ stars \citep{Sellwood00a},
although this analysis made use of the reconstruction of the local
velocity distribution derived from tangential velocities alone by
\citet{Dehnen98b}. With the full kinematical information in the
\gcsabb\ catalog, we can construct the energy--angular-momentum
distribution without making any symmetry assumptions, and we can ask
whether any of the moving groups are actually a manifestation of the
groove feature in the angular-momentum distribution that drives spiral
structure.

\begin{figure}
\includegraphics{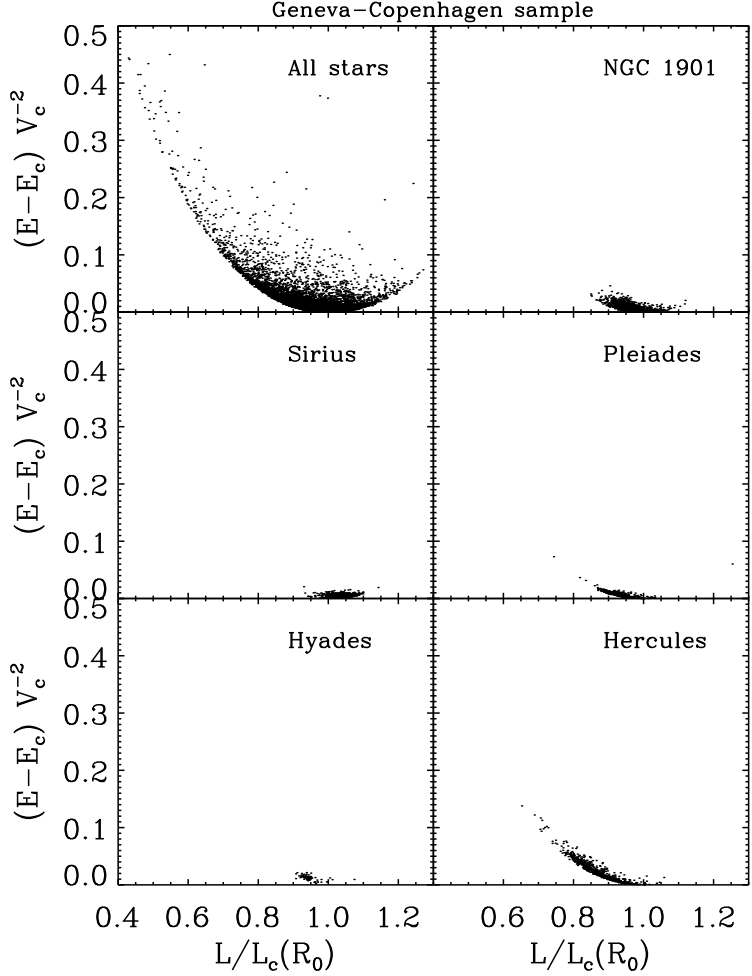}
\caption{Lindblad diagram: Distribution of the \gcsabb\ stars in
  energy--angular momentum space assuming a Mestel disk model for the
  Galaxy with circular velocity of 235 km s$^{-1}$ and $R_0 = 8.2$
  kpc; $E_c \equiv E_c(L)$ is the energy of a circular orbit with
  angular momentum $L$, $L_c(R_0)$ is the angular momentum of the
  circular orbit going through the Sun's present location. The
  location of high probability ($\pij > 0.5$) members of the moving
  groups in this diagram is shown in the remaining panels. None of the
  moving groups stand out as a feature in this
  diagram.}\label{fig:sellwood}
\end{figure}

In order to calculate the integrals of the motion of the stars in the
\gcsabb\ sample we need to assume a Galactic disk potential to convert
positions and velocities into energy and angular momentum. We use a
simple model for the disk potential, a Mestel disk
\citep{Mestel63a,binneytremaine}, which has a flat rotation curve and
is uniquely characterized by the circular velocity; we assume a
circular velocity at the Sun of $V_c = 235$ km s$^{-1}$ and calculate
Galactocentric distances using $R_0 = 8.2$ kpc
\citep[\eg,][]{Bovy09c}. The resulting distribution in energy and
angular momentum of the \ngcsstars\ \gcsabb\ stars that was used in
the previous section is shown in \figurename~\ref{fig:sellwood}. The
lower cutoff in energy as a function of angular momentum is a
selection effect: since the stars are all within about 100 pc from the
Sun, stars on nearly circular orbits with angular momenta different
from that at the Solar circle do not make large enough excursions to
make it into our sample.

We have also indicated the locations of the moving groups in this
diagram, by making hard assignments of stars to moving groups using
$\pij > 0.5$, where $\pij$ is again calculated using the full
three-dimensional velocity vector and the level of the hard cut is set
to the value that gives an overall fraction of stars in moving groups
of about 40\,percent. There does not seem to be a clear scattering
feature in this distribution. The Hercules moving group is,
unsurprisingly, the only moving group that could potentially be
associated with a scattering feature, but since it lies very close to
the selection cutoff, it is hard to tell whether the Hercules moving
group corresponds to a genuine scattering feature in this diagram or
whether this is just the selection cutoff.

\begin{figure}
\includegraphics{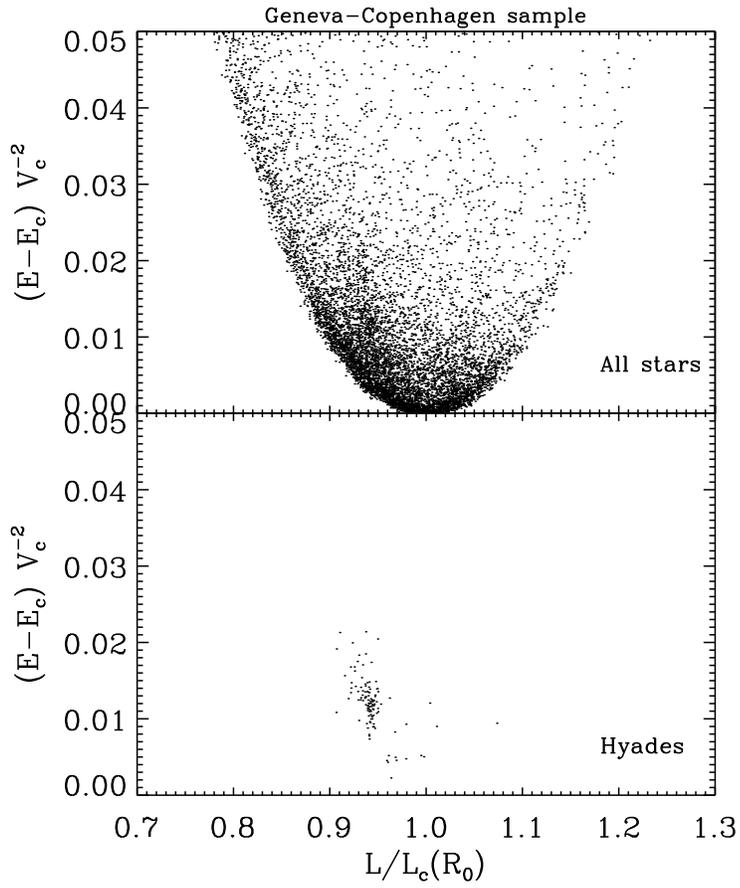}
\caption{Zoom of \figurename~\ref{fig:sellwood} for all stars and for
  the Hyades moving group. The Hyades members occupy a narrow range in
  angular momentum that corresponds to a feature in the distribution
  for all stars in the top panel.}\label{fig:sellwood_hyades}
\end{figure}

Recently, \citet{sellwood10a} has argued that the Hyades moving group
rather than the Hercules moving group corresponds to the
inner-Lindblad scattering feature. This feature is not apparent in
\figurename~\ref{fig:sellwood}, since it concerns stars with an order
of magnitude less random energy. For ease of comparison with
\citet{sellwood10a}, \figurename~\ref{fig:sellwood_hyades} shows the
\gcsabb\ stars from \figurename~\ref{fig:sellwood} with the smallest
random motions, as well as the Hyades moving-group members. It is
clear from this figure that the Hyades members do indeed correspond to
the weak feature apparent in the top panel, confirming that the Hyades
moving group might be a telltale of the recurrent nature of the Milky
Way's spiral structure. This explanation does leave a few questions
unanswered. The other low-velocity moving groups do not stand out in
the energy--angular-momentum space. Ignoring the Hercules moving
group, which can potentially be explained by the bar, how are the
other moving groups formed if they are not the result of inhomogeneous
star formation? Since the recurrent spiral structure is supposed to
move inward, with the next spiral pattern's corotation radius near the
inner Lindblad radius of the previous pattern, it is unlikely that the
other moving groups are the result of scattering features associated
with previous patterns since these features should be at larger values
of the angular momentum and random energy. The result that the Hyades
moving group is created by the scattering of stars at the ILR is also
slightly at odds with the higher metallicity preference for the Hyades
moving group found in \sectionname~\ref{sec:dynamics}: Since stars are
scattered inward at the ILR, the Hyades stars originate at greater
Galactocentric radii and should therefore be, if anything, less
metal-rich than average.

\section{Discussion and future work}\label{sec:discussion}

\begin{figure}
\includegraphics{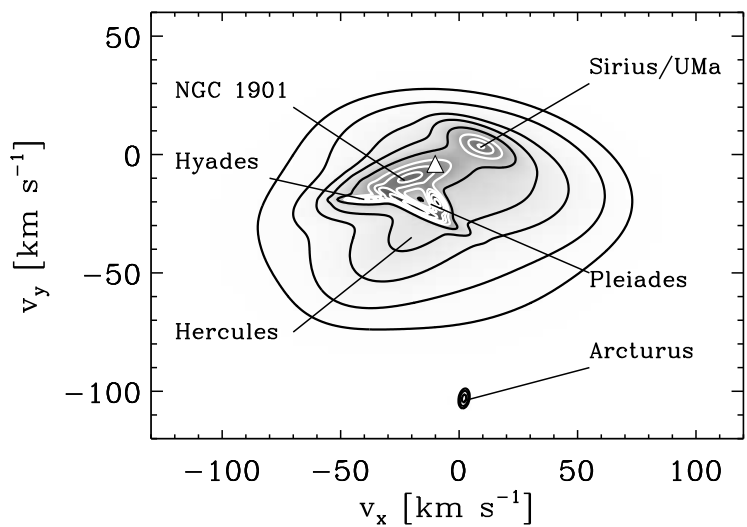}
\caption{Velocity distribution in the Solar neighborhood (from
  \citealt{Bovy09a}) in the Galactic plane with the moving groups
  studied in this work indicated. The density grayscale is linear and
  contours contain, from the inside outward, 2, 6, 12, 21, 33, 50, 68,
  80, 90, and 95 percent of the distribution. The first
  five of these contours are white and somewhat blended together; 50
  percent of the distribution is contained within the innermost dark
  contour. The origin in each of these plots is at the Solar velocity;
  the velocity of the Local Standard of Rest \citep{Hogg05a} is
  indicated by a triangle.}\label{fig:veldist}
\end{figure}

The tests and discussions above have all focused on determining the
nature of the moving groups identified in
\figurename~\ref{fig:veldist}, and we have been able to rule out and
provide support for some possible scenarios through which these moving
groups may have formed. However, the groups shown in
\figurename~\ref{fig:veldist} have been determined as Gaussian
components in a deconvolution of the observed velocity distribution
using 10 Gaussians. \bhr\ found that the best-fitting
mixture-of-Gaussians model contained only 10 components: When using
more components, the velocity distribution was overfit as it became
clear by testing its predictions of the external \gcsabb\
radial-velocity data set. This does \emph{not}, however, constitute an
endorsement that the individual components have any physical
interpretation: only the mixture itself, that is, the full
distribution, can be considered real, the individual components are
just positioned in such a way as to best describe the overall velocity
distribution. It is therefore fair to ask whether the results in this
paper have not been unduly influenced by our identification of moving
groups with individual components of the mixture.

In \bhr, we argued that moving groups can be associated with
individual components of the mixture for a few different reasons. The
overall reconstructed velocity distribution contains a number of
distinct peaks (see \figurename~\ref{fig:veldist}). These peaks can be
unambiguously identified with specific components of the mixture, and
therefore we can cross-correlate structures in the velocity
distribution with the Gaussian components. Peaks, or overdensities, in
the velocity distribution are what are generally called moving
groups. Thus, since peaks in the velocity distribution are what define
moving groups, and these peaks can be identified as individual
components in the mixture, we can associate individual components with
moving groups. Furthermore, the peaks in the velocity distribution
compare favorably with the fiducial locations of the classical moving
groups that are studied in this paper, although there are some small
differences, such as that the Pleiades is resolved as two components,
and that the Hercules moving group is both more smoothly connected to
the bulk of the distribution than is generally thought to be the case
and is located at slightly lower velocities than usual.

The generally accepted kinematic properties of the moving groups
amount to not much more than a rough location and an even rougher
estimate of the size and orientation of the moving group. The shape of
the moving groups in the direction out of the plane is rarely
discussed, although all of the moving groups' vertical velocities are
presumably as well mixed as those of the general background
population, because of the efficiency of phase mixing in the vertical
direction. Similarly, until \bhr, the weight of the individual moving
groups in the velocity distribution, or even the total weight of
substructure in the distribution had never been quantitatively
determined. It is hard to make quantitative estimates of group
membership for individual stars, especially if not all of the velocity
components of the stars are measured. The locations, shapes, and
relative importance that we used in this paper allow for an objective
way to estimate membership probabilities for a large sample of stars
for all of the moving groups. While one can argue over whether these
locations, shapes, and relative weights are exactly right for the
moving groups, the objective, probabilistic procedure that we followed
in this paper should be preferred over ad hoc choices on which to base
membership assignments.

We also do not expect small biases in the parameters of the moving
groups to affect the conclusions of this paper very much. If the
moving groups are actually located at slightly different locations in
velocity space, if their profiles deviate from Gaussians in the wings,
or if their relative weights are slightly higher or slightly lower
than that which was assumed here, the computed membership
probabilities will be somewhat wrong, but not by large factors. That
is, high probability members based on the parameters that we assumed
for the moving groups will remain high probability members even for
slightly different parameters. If the moving groups had shown a clear
preference for an explanation of their existence over the others in
the previous sections, \eg, if they were much better fit by a
single-burst stellar population than by the background distribution,
this conclusion would have stood out at high significance even if we
computed membership probabilities slightly wrong. Thus, the main
conclusion of this paper---that no moving group shows clear evidence
of having originated through one of the scenarios discussed
here---holds whatever you believe about our parameterization. The more
tentative conclusions reached here, however, should be interpreted
with care.

Another caveat has to do with the possibility of radial mixing playing
an important role in the chemical evolution of the Galactic
disk. Radial mixing \citep{sellwood02a} is the process in which stars
can migrate radially from their birthplaces over large distances while
remaining on nearly circular orbits. Such mixing causes a wider range
of birth radii to be present at any Galactocentric radius and can
therefore weaken expected correlations between, for example,
metallicity and Galactocentric radius or metallicity and age
\citep[\eg,][]{roskar08a}. Radial mixing occurs naturally in galactic
disks with transient spiral structure---only stars scattered at
corotation can be scattered without increasing their random motion, so
a large range of frequencies needs to be present for radial mixing to
occur throughout the disk---but recently it has been shown that the
coupling between a steady-state bar and steady-state spiral arms can
also lead to significant radial migration \citep{Minchev09b}. In this
scenario, stars from a wide range of birth radii and metallicities can
migrate radially and be trapped into the bar's and spiral structure's
resonances, leading to a potentially significant dilution of the
metallicity-offset effect we searched for in
\sectionname~\ref{sec:dynamics}. More work is necessary to test
whether the resonance-overlap radial mixing is consistent with
observations of the Solar neighborhood \citep[\cf][]{schoenrich09a}
and whether the metallicity distributions of the moving groups created
by the resonances are consistent with the results from
\sectionname~\ref{sec:dynamics}.

The scenarios discussed and explicitly tested here do not constitute
an exhaustive set of the possible origins of the moving groups. We
have only tested some of the simplest explanations for the existence
of moving groups, but these simple explanations do command a
considerable amount of weight in the discussion on the origin of the
moving groups. Our tests considered all of the main classes of
explanations for the origin of the moving groups, however, within
these classes we did not test whether the moving groups are related to
transient non-axisymmetric perturbations to the Galactic potential,
nor did our test of the evaporating-cluster scenario include the
possibility that the moving groups are the remnants of \emph{several}
open clusters. All of these alternative explanations provide \emph{a
priori} reasonable explanations of the moving groups' existence and
should therefore be tested. Testing these explanations will be harder
because the stellar content of the moving groups will have to be
determined in greater detail than what has been done here. Theoretical
work and simulations will also have to establish the nature of the
moving groups in the scenarios where they are due to transient
perturbations to allow the data on the stellar content of the moving
groups to be interpreted in terms of these models.

Future work to elucidate the origin of the moving groups could go
beyond the simple tests performed here by fitting more complicated
models for the chemical composition and star-formation history of each
moving group. This ``chemical tagging'' \citep{Freeman02a} could lead
to greater insight into the kind of stars or orbits that make up the
moving groups. Fitting these more general models will be considerably
more complicated than what has been done here. Nevertheless, the
probabilistic approach followed here in which all stars in the sample
are carried through the analysis of each moving group with appropriate
membership-probability weights---a weak cut could be done for
computational efficiency---will be essential in these more
sophisticated analyses to study the kinematic structures that are the
moving groups.

\section{Conclusions}

A summary of our results is the following:

$\bullet$ We use large samples of stars extracted from the
\Hipparcos\ and \gcsabb\ catalogs to study the properties of the five
most prominent low-velocity moving groups: the NGC 1901 group, the
Sirius group, the Pleiades group, the Hyades group, and the Hercules
group. Using membership probabilities calculated in a probabilistic
manner based on the tangential velocities of the stars, the radial
velocities, or both, and by propagating these membership probabilities
through our whole analysis, we are able to use the maximum number of
stars in the study of each moving group---an order of magnitude
improvement for most of the moving groups---and avoid any possible
biases that could result from making hard cuts on membership
probabilities in analyses of this kind.

$\bullet$ For the four moving groups in our sample with an associated
open cluster, we asked whether the moving groups could consist of
stars that have evaporated from these open clusters. By comparing the
parallaxes of the stars that we predict if the stars in the moving
groups have the same age and metallicity as the open cluster that the
moving group is associated with the observed trigonometric parallax,
we establish that a large part of each moving group is poorly fit by
the assumption that it has the same stellar population as the open
cluster. This establishes beyond any reasonable doubt that the moving
groups are \emph{not} fundamentally associated with their eponymous
open clusters.

$\bullet$ Next we studied whether each moving group could conceivably
be associated with \emph{any} open cluster, not necessarily the one
normally associated with it. We constructed a background model in
which the moving group is nothing more than a sparse sampling of the
local disk population of stars and single-burst stellar population
foreground models parameterized by an age, a metallicity, and a level
of background contamination. For reasonable values of the background
contamination we find that only the Hercules moving group displays
marginal evidence that it could be a remnant of a past star formation
event. However, letting the level of background contamination run
free, all of the moving groups prefer very large values of the
contamination, reaching values close to complete contamination by the
background, especially in the case of the Hercules moving
group. Therefore, we can confidently conclude that none of the moving
groups is a remnant of a single open cluster.

$\bullet$ To test scenarios in which moving groups are formed as a
consequence of resonances associated with the bar and/or spiral
structure, we asked whether the moving groups are better fit by a
model with higher than average---or lower than average---metallicity,
such as would generically be the case in resonant models for the
moving groups. We find that of all the moving groups only the Hyades
moving group shows a metallicity preference, toward higher
metallicity. All of the other moving groups are best represented by
the background population of stars, although the Sirius moving group
prefers a lower than average metallicity over higher than average,
which, together with the higher than average metallicity of the Hyades
could be an indication of a spiral-structure-associated resonance
origin for the Hyades and Sirius moving groups. The Pleiades moving
group is preferably fit by a lower than average metallicity rather
than a higher than average metallicity, arguing against a common
origin for the Hyades and Pleiades moving groups. The Hercules moving
group has a preference toward higher metallicity, consistent with it
being associated with the OLR of the bar. We stress that all of this
evidence is very tentative and the background model is the preferred
model in most cases, raising the likelihood of transient
non-axisymmetric perturubation scenarios for the origin of the moving
groups.

$\bullet$ We confirm the result of \citet{sellwood10a} that the Hyades
moving groups might be associated with features---grooves---in the
angular momentum distribution as would be expected in some models of
recurrent spiral structure.

\acknowledgments It is a pleasure to thank the anonymous referee for
valuable comments and Michael Aumer, Mike Blanton, Iain Murray and Sam
Roweis for helpful discussions and assistance.  Financial support for
this project was provided by the National Aeronautics and Space
Administration (grant NNX08AJ48G) and the National Science Foundation
(grant AST-0908357). DWH is a research fellow of the Alexander von
Humboldt Foundation.

\clearpage
\begin{deluxetable}{lrr@{.}lr@{.}l}
\tablecaption{Best fit single-stellar-population models for the low-velocity moving groups\label{table:bestfitssp}}
\tablecolumns{6}
\tablewidth{0in}
\tablehead{\colhead{Group} & \colhead{Age}& \multicolumn{2}{c}{$Z$}{} &\multicolumn{2}{c}{$\alpha$\tablenotemark{\protect{\ref{alpha}}}}\\
\colhead{} & \colhead{(Myr)} & \multicolumn{2}{c}{} & \multicolumn{2}{c}{}}
\startdata
NGC1901........ & 180  & 0&030 & 0 & 41\\
NGC1901........ & 56  & 0&030 & 0 & 98\\
\\
Sirius............... & 350  & 0&026 & 0&53\\
Sirius............... & 413 & 0&023 & 0&90\\
\\
Pleiades........... & 67 & 0&030 & 0&57\\
Pleiades........... & 67 & 0&030 & 0&90\\
\\
Hyades............. & 488 & 0&029 & 0&58\\
Hyades............. & 679 & 0&027 & 0&86\\
\\
Hercules........... & 180 & 0&030 & 0&83\\
Hercules........... & 180 & 0&030 & 1&00
\enddata
\tablecomments{The first line for each group lists the best-fit age and metallicity for the fixed value for $\alpha$ in the last column---this value was obtained from a global contamination analysis (see the text)---the second line lists the overall best fit age, metallicity, and $\alpha$.}
\setcounter{tableone}{1}
\makeatletter
\let\@currentlabel\oldlabel
\newcommand{\@currentlabel}{\thetableone}
\makeatother
\renewcommand{\thetableone}{\alph{tableone}}
\tablenotetext{\thetableone}{\label{alpha}
Background contamination level.\stepcounter{tableone}}
\end{deluxetable}

\clearpage
\begin{deluxetable}{lrrrrr}
\tablecaption{Model selection using the \gcsabb\ sample: is the single-stellar-population model for the moving groups preferred?\label{table:sspmodelselection}}
\tabletypesize{\small}
\tablecolumns{6}
\tablewidth{0in}
\tablehead{\colhead{Group} & \colhead{Best-fit \ssp,} &\colhead{Marginalized \ssp,} & \colhead{Best fit \ssp,} & \colhead{Marginalized \ssp,}\\
\colhead{} & \colhead{fixed $\alpha$} & \colhead{fixed $\alpha$} & \colhead{free $\alpha$} & \colhead{free $\alpha$}}
\startdata
NGC1901......... &       -262 & -262 & 17 & 17\\
Sirius............... &  -61 & -61 & 0 & 1\\
Pleiades........... &    -70 & -70 & 1 & 5\\
Hyades............. &    -8 & -8 & 0 & 0\\
Hercules........... &    2 & 2 & 0 & 15
\enddata
\tablecomments{The difference between the logarithm of the probability
  of the parallaxes of the \ngcscolorcut\ stars in the \gcsabb\ sample
  used in \sectionname~\ref{sec:ssp} (\bminusv $<$ 1 mag) under the
  various foreground models and that under the background model
  is given for each moving group. The logarithm of the likelihood of
  the background model is -23,155. ``Marginalized'' probabilities have
  the uncertainties in the best-fit values integrated out by
  marginalizing over the posterior distribution for age, metallicity,
  and, if applicable, background contamination level $\alpha$.}
\end{deluxetable}

\clearpage
\begin{deluxetable}{lr@{.}lr@{.}lr@{.}lr@{.}lr@{.}lrrrrr}
\tablecaption{High/low metallicity model selection: do the moving groups have higher or lower metallicities than the background disk population?\label{table:metalmodelselection}}
\tabletypesize{\small}
\rotate
\tablecolumns{14}
\tablewidth{0in}
\tablehead{\colhead{Group} & \multicolumn{2}{c}{$R_c(L)$\tablenotemark{\protect{\ref{rl}}}} & \multicolumn{2}{c}{Expected $\Delta$[Fe/H]\tablenotemark{\protect{\ref{fehmodel}}}} & \multicolumn{2}{c}{Expected $\sigma$[Fe/H]\tablenotemark{\protect{\ref{fehmodel}}}} & \multicolumn{2}{c}{$\langle\Delta$[Fe/H]$\rangle$\tablenotemark{\protect{\ref{feh}}}} & \multicolumn{2}{c}{$\sigma$[Fe/H]\tablenotemark{\protect{\ref{feh}}}} & \colhead{High Metallicity} &\colhead{Low Metallicity}\\
\colhead{} & \multicolumn{2}{c}{(kpc)} & \multicolumn{2}{c}{(dex)} & \multicolumn{2}{c}{(dex)} & \multicolumn{2}{c}{(dex)} & \multicolumn{2}{c}{(dex)} & \colhead{} &\colhead{}}
\startdata
NGC1901.........     &  8&0 &  0&02 &  0&025 &  0&02  & 0&16 &    -203  & -144 \\
Sirius...............&  8&5 & -0&03 &  0&015 & -0&03  & 0&15 &    -108  & -8 \\
Pleiades...........  &  7&6 &  0&05 &  0&015 &  0&02  & 0&16 &    -43  & -37 \\
Hyades.............  &  7&6 &  0&05 &  0&003 &  0&11  & 0&14 &    3  & -14 \\
Hercules...........  &  7&2 &  0&10 &  0&040 &  0&01  & 0&17 &   -40  & -106
\enddata
\tablecomments{The difference between the logarithm of the probability
  of the parallaxes of the \nphotgcscolorcut\ stars in the
  \gcsabb\ sample (0.35 mag $<$\bminusv $<$ 0.95 mag) under the
  higher/lower metallicity foreground models and that under the
  background model is given for each moving group. The logarithm of
  the likelihood of the background model is -27882.}
\setcounter{tabletwo}{1} \makeatletter \let\@currentlabel\oldlabel
\newcommand{\@currentlabel}{\thetabletwo} \makeatother
\renewcommand{\thetabletwo}{\alph{tabletwo}}
\tablenotetext{\thetabletwo}{\label{rl} Galactocentric radius of the circular orbit with the same angular momentum as the center of the moving group. \stepcounter{tabletwo}}
\tablenotetext{\thetabletwo}{\label{fehmodel} Expected metallicity anomaly and spread based on the mean radius ($\approx R_c(L)$), the velocity width of the moving group and a metallicity gradient of -0.1 dex kpc${-1}$. \stepcounter{tabletwo}}
\tablenotetext{\thetabletwo}{\label{feh} Average metallicity and spread of each
  moving group, computed by weighting the metallicities in the
  \gcsabb\ sample with the membership probabilities (see
  \eqnname~(\ref{eq:avgmetal})). The width includes the measurement uncertainty, which is about 0.08 dex.\stepcounter{tabletwo}}
\end{deluxetable}

\clearpage

\clearpage

\clearpage

\clearpage

\clearpage

\clearpage

\clearpage

\clearpage

\clearpage

\clearpage

\clearpage

\clearpage

\clearpage

\clearpage

\clearpage

\clearpage

\clearpage

\clearpage

\end{document}